\documentclass[11pt,a4paper]{article}

\usepackage{jheppub}
\usepackage{graphicx}
\usepackage{dcolumn}
\usepackage{bm}
\usepackage{amsmath}
\usepackage{braket}
\usepackage{slashed}    
\usepackage{epstopdf}
\usepackage{placeins}
\usepackage{multirow}
\usepackage{makecell}
\usepackage[shortlabels]{enumitem}
\usepackage{cleveref}
\usepackage{xcolor}
\usepackage[font=small, labelfont=bf, textfont={small,it}]{caption}

\newcommand{\beq}{\begin{eqnarray}}
\newcommand{\eeq}{\end{eqnarray}}
\newcommand{\beqnn}{\begin{eqnarray*}}
\newcommand{\eeqnn}{\end{eqnarray*}}

\newcommand{\DW}{D_{\scriptscriptstyle{\rm W}}}
\newcommand{\QW}{Q_{\scriptscriptstyle{\rm W}}}
\newcommand{\SW}{S_{\scriptscriptstyle{\rm W}}}
\newcommand{\effsup}{{\scriptscriptstyle{(\mathrm{eff})}}}
\newcommand{\DWp}{\widetilde{D}_{\scriptscriptstyle{\rm W}}}

\newcommand{\ZS}{Z_{\scriptscriptstyle{\rm S}}}
\newcommand{\ZP}{Z_{\scriptscriptstyle{\rm P}}}
\newcommand{\ZA}{Z_{\scriptscriptstyle{\rm A}}}
\newcommand{\PCAC}{{\scriptscriptstyle{\rm PCAC}}}
\newcommand{\TEK}{{\scriptscriptstyle{\rm TEK}}}
\newcommand{\A}{{\scriptscriptstyle{\rm A}}}
\newcommand{\s}{{\scriptscriptstyle{\rm s}}}
\newcommand{\R}{{\scriptscriptstyle{\rm R}}}
\newcommand{\chir}{{\scriptscriptstyle{(\chi)}}}
\newcommand{\SU}{\mathrm{SU}}
\newcommand{\MSbar}{\overline{\mathrm{MS}}}
\newcommand{\mutwoGeV}{\mu=2\text{ GeV}}
\newcommand{\dd}{\mathrm{d}}
\newcommand{\ee}{\mathrm{e}}
\newcommand{\ii}{\mathrm{i}}
\newcommand{\Tr}{\mathrm{Tr}}
\newcommand{\Nf}{N_{\scriptscriptstyle{\rm f}}}
\newcommand{\Vphyseff}{\mathcal{V}_{\scriptscriptstyle{\rm eff}}}

\usepackage[normalem]{ulem}

\begin{document}

\title{\centering Non-perturbative determination of meson masses and low-energy constants in large-$N$ QCD}

\author[a]{Claudio Bonanno,}
\author[a]{Margarita Garc\'ia P\'erez,}
\author[a,b]{Antonio Gonz\'alez-Arroyo,}
\author[c,d]{\\Ken-Ichi Ishikawa,}
\author[d]{Masanori Okawa}

\affiliation[a]{Instituto de F\'isica Te\'orica UAM-CSIC, Calle Nicol\'as Cabrera 13-15,\\Universidad Aut\'onoma de Madrid, Cantoblanco, E-28049 Madrid, Spain}

\affiliation[b]{Departamento de F\'isica Te\'orica, Universidad Aut\'onoma de Madrid,\\M\'odulo 15, Cantoblanco, E-28049 Madrid, Spain}

\affiliation[c]{Core of Research for the Energetic Universe,\\Graduate School of Advanced Science and Engineering,\\Hiroshima University, Higashi-Hiroshima, Hiroshima 739-8526, Japan}

\affiliation[d]{Graduate School of Advanced Science and Engineering, Hiroshima University,\\Higashi-Hiroshima, Hiroshima 739-8526, Japan}

\emailAdd{claudio.bonanno@csic.es}
\emailAdd{margarita.garcia@csic.es}
\emailAdd{A.Gonzalez-Arroyo@pm.me}
\emailAdd{ishikawa@theo.phys.sci.hiroshima-u.ac.jp}
\emailAdd{okawa@hiroshima-u.ac.jp}

\abstract{We provide first-principles non-perturbative determinations of the low-lying meson mass spectrum of large-$N$ QCD in the 't Hooft limit $N_{\scriptscriptstyle{\rm f}}/N\to 0$, as well as of three low-energy constants appearing in the QCD chiral expansion: the quark condensate $\Sigma$, the pion decay constant $F_\pi$, and the next-to-leading-order coupling $\bar{\ell}_4$. Using the excited state masses in the $\pi$ and $\rho$ channels, we are able to investigate the behavior of their radial Regge trajectories. Concerning QCD low-energy constants, we are able to assess the magnitude of sub-leading corrections in $1/N$ by combining our $N=\infty$ results with previous finite-$N$ determinations. Our calculation exploits large-$N$ twisted volume reduction to efficiently perform numerical Monte Carlo simulations of the large-$N$ lattice discretized theory. We employ several values of $N$ up to $N=841$, 5 values of the lattice spacing, and several values of the quark mass, to achieve controlled continuum and chiral extrapolations.}


\maketitle

\vspace*{-0.31\baselineskip}

\section{Introduction}

Although Quantum Chromo-Dynamics (QCD), the theory of strong interactions among quarks and gluons within the Standard Model, is mathematically described by a $\SU(N)$ gauge theory with number of colors $N=3$, there is plenty of theoretical and phenomenological interest in studying QCD in the large-$N$ limit $N\to\infty$.

If the number of colors is sent to infinity keeping the number of quark flavors $\Nf$, the quark masses and the 't Hooft coupling $\lambda=g^2 N$ constant, the diagrammatic expansion of large-$N$ QCD exhibits remarkable simplifications, and organizes itself into an expansion in powers of $1/N$~\cite{tHooft:1973alw}, where the leading-order contribution is given by planar diagrams only. Moreover, each fermion loop suppresses the diagram by a factor of $\Nf/N$, so that quark contribution is sub-leading compared to the gluon one, leading to the following general expected scaling for any observable with a finite large-$N$ limit:
\beq
\braket{\mathcal{O}}(N,\Nf) = A_0 + A_1\frac{\Nf}{N} + (A_2+A_2^\prime \Nf^2) \frac{1}{N^2} + \mathcal{O}\left(\frac{1}{N^3}\right),
\eeq
where the large-$N$ limit $A_0$ is independent of the number of flavors $\Nf$. Given that the large-$N$ $1/N$ expansion is purely non-perturbative in terms of the coupling $\lambda$, it provides a powerful theoretical tool to investigate certain phenomenologically-relevant features of the non-perturbative domain of QCD. For example, it allows to clarify how the chiral anomaly~\cite{tHooft:1976rip} and the non-trivial dependence of QCD on the topological parameter $\theta$ can explain the $\eta^\prime$ meson mass~\cite{Witten:1978bc,Witten:1979vv,Veneziano:1979ec,Kawarabayashi:1980dp,Witten:1980sp,DiVecchia:1980yfw}.

Since only a limited set of non-perturbative aspects of large-$N$ gauge theories can be fully grasped with analytical tools, one has in general to resort on numerical Monte Carlo simulations of the lattice-discretized theory to systematically investigate QCD in the large-$N$ regime. The standard approach to study the large-$N$ limit of gauge theories on the lattice consists in performing simulations at finite values of $N$, which are eventually extrapolated towards $N=\infty$ by fitting the first few terms of the $1/N$ expansion to the data. This approach has been extensively applied both to pure Yang--Mills theories~\cite{DelDebbio:2001sj,Lucini:2001ej,DelDebbio:2002xa,Lucini:2004my,DelDebbio:2006yuf,Vicari:2008jw,Allton:2008ty,Lucini:2010nv,Lucini:2012gg,Bali:2013kia,Bonati:2016tvi,Ce:2016awn,Bennett:2020hqd,Bonanno:2020hht,Athenodorou:2021qvs,Bennett:2022gdz,Bonanno:2022yjr,Bonanno:2024ggk,Athenodorou:2024loq,Sharifian:2025fyl} and to simulations with dynamical quarks~\cite{DeGrand:2016pur,Hernandez:2019qed,DeGrand:2020utq,Hernandez:2020tbc,DeGrand:2021zjw,Baeza-Ballesteros:2022azb,DeGrand:2023hzz,DeGrand:2024lvp,DeGrand:2024frm,Baeza-Ballesteros:2025iee,Butti:2025rnu}, with the typical values of $N$ employed being, respectively, of the order of 8--10 and 5--6 at most.

The approach followed in this study to investigate large-$N$ QCD is instead different, and based on the concept of \emph{large-$N$ volume reduction}~\cite{PhysRevLett.48.1063,BHANOT198247,Gross:1982at,GONZALEZARROYO1983174,PhysRevD.27.2397,Aldazabal:1983ec,Kiskis:2002gr,Narayanan:2003fc,Kovtun:2007py,Unsal:2008ch,Gonzalez-Arroyo:2010omx,Neuberger:2020wpx}. As first shown in the seminal paper by Eguchi and Kawai~\cite{PhysRevLett.48.1063}, at $N=\infty$ lattice Yang--Mills theories enjoy a dynamical equivalence of space-time and color degrees of freedom, meaning that these degrees of freedom appear intertwined at large $N$. This in practice permits the numerical study of lattice large-$N$ gauge theories in the thermodynamic limit reducing the lattice volume down to a single space-time point $V=1$~\cite{Gonzalez-Arroyo:1983cyv,Das:1984jh,Das:1985xc,Narayanan:2004cp,Gonzalez-Arroyo:2005dgf,Kiskis:2009rf,Hietanen:2009ex,Hietanen:2010fx,Bringoltz:2011by,Hietanen:2012ma,Gonzalez-Arroyo:2012euf,Gonzalez-Arroyo:2013bta,Lohmayer:2013spa,Gonzalez-Arroyo:2014dua,GarciaPerez:2014azn,GarciaPerez:2015rda,Perez:2015ssa,Gonzalez-Arroyo:2015bya,Perez:2017jyq,GarciaPerez:2020gnf,Perez:2020vbn,Butti:2023hfp,Bonanno:2023ypf,Butti:2022sgy,Bonanno:2024bqg,Bonanno:2024onr}. The clear advantage of this strategy is that it allows to reach much larger values of $N$, up the order of $10^2$--$10^3$, compared to the standard approach, thus allowing to practically work already at $N=\infty$. On the other hand, as we will also discuss later, finite-$N$ corrections in reduced models are of different nature, and essentially show up as finite-volume effects, thus reduced models only allow the computation of the leading-order term in the $1/N$ expansion. Instead, reliably computing sub-leading terms in the $1/N$ expansion requires the combination of $N=\infty$ results obtained from reduced models with standard finite-$N$ ones. For this reason, these two approaches should be regarded as \emph{complementary}.

Large-$N$ volume reduction works only provided that center symmetry is unbroken, which is clearly not true if standard periodic boundary conditions are taken on a single-site box, due to the well-known spontaneous center-symmetry breaking occurring in the deconfined phase of Yang--Mills theories~\cite{Fingberg:1992ju,Beinlich:1997ia,Campostrini:1998zd,Lucini:2001ej,Lucini:2002ku,Lucini:2003zr,Lucini:2004yh,Lucini:2005vg,Lucini:2012wq,Borsanyi:2022xml,Lucini:2023irm,Cohen:2023hbq}. However, this is not an obstruction, as there are several ways to bypass this issue and enforce center symmetry in a one-point box. In this study we will use twisted boundary conditions to this end, and adopt the so-called \emph{Twisted Eguchi--Kawai} (TEK) model~\cite{GONZALEZARROYO1983174,PhysRevD.27.2397,Gonzalez-Arroyo:2010omx} to perform our numerical calculations (for other strategies to enforce center symmetry on a small box to achieve large-$N$ volume independence see, e.g., Refs.~\cite{BHANOT198247,Narayanan:2003fc,Kovtun:2007py,Unsal:2008ch}).

The TEK model offers a very powerful framework to efficiently study gauge theories directly at $N=\infty$, allowing also to exploit the several simplifications that occur in this limit. In recent times it has allowed the investigation of many intriguing physical quantities of several $4d$ large-$N$ lattice gauge theories, including large-$N$ QCD~\cite{Gonzalez-Arroyo:2012euf,GarciaPerez:2014azn,Gonzalez-Arroyo:2015bya,Perez:2020vbn,Bonanno:2023ypf,Butti:2023hfp}, large-$N$ $\mathcal{N}=1$ Supersymmetric Yang--Mills theory~\cite{Butti:2022sgy,Bonanno:2024bqg,Bonanno:2024onr}, and large-$N$ adjoint QCD~\cite{Gonzalez-Arroyo:2013bta,GarciaPerez:2015rda,Hamada:2025whg}. The present study can be placed within this context, as it is about the computation of the low-lying meson spectrum of large-$N$ QCD, and of the two low-energy constants appearing in the low-energy effective QCD chiral Lagrangian at leading order, namely, the pion decay constant $F_\pi$ and the quark condensate $\Sigma$. On general theoretical grounds, one expects meson masses and decay constants to have a well-defined and finite large-$N$ limit, and chiral symmetry at $N=\infty$ to be spontaneously broken~\cite{Coleman:1980mx} in a similar fashion as standard $N=3$ QCD, with $\Sigma/N$ and $F_\pi / \sqrt{N}$ also having a well-defined and finite large-$N$ limit. These topics were addressed in a certain number of lattice studies in the past with a variety of approaches. The large-$N$ limit of meson masses and decay constant was studied from the scaling of finite-$N$ determinations obtained both from pure Yang--Mills theories~\cite{DelDebbio:2007wk,Bali:2013kia} and full QCD~\cite{DeGrand:2016pur,Hernandez:2019qed} simulations, and even exploiting large-$N$ volume reduction~\cite{Narayanan:2005gh,Hietanen:2009tu} (with a different approach with respect to the one followed here, the so-called ``continuum reduction''). However, no study of the continuum limit is presented in these papers, with the exception of the recent study~\cite{DeGrand:2023hzz}, where the continuum limit of $\Sigma/N$ and $F_\pi/\sqrt{N}$ is computed for $N=3,4,5$ colors.

The goal of the present study is to provide solid non-perturbative determinations of the large-$N$ limit of meson masses and decay constants with controlled chiral-continuum extrapolations and controlled finite-volume effects from the TEK model. The computation of these observables from the TEK model was already addressed in our previous studies~\cite{Perez:2020vbn,Bonanno:2023ypf}, but in the present manuscript we will present updated and extended calculations which will allow us to largely improve on our old results. This improvement was possible thanks to the employment of several new ensembles generated with larger values of $N$ with respect to our previous studies: in particular, we have been able to simulate $\SU(N)$ gauge theories with $N=529, 841$, to be compared with $N=169, 289$ and $361$ used in~\cite{Perez:2020vbn,Bonanno:2023ypf}. Since in the TEK model the effective lattice size is given by $\ell=a\sqrt{N}$ (with $a$ the lattice spacing), reaching larger effective volumes was helpful in several respects.\\
First, it helped us in taming finite-volume effects affecting the determination of meson masses for the smallest values of the quark mass considered, as we could reach larger values of the effective lattice size in units of the pion mass $m_\pi \ell$, thus allowing a better control of chiral extrapolations towards the massless quark limit. This is particularly crucial for heavier mesons, as finite size effects result in a spurious constant contribution at large time separations in the correlation functions employed to extract masses from lattice simulations. Since the exponential decay for heavier states is much faster, it becomes very hard to disentangle the physical signal from this finite-volume artifact if $m_\pi\ell$ is not large enough, possibly biasing the mass extraction.\\
Secondly, it allowed us to include a fifth, finer lattice spacing in our analysis while keeping a large enough value of the physical effective volume $\Vphyseff = a^4 N^2$, which allowed to improve the study of the continuum limit of the considered observables.\\
Finally, this allowed us to go higher up in the spectrum and obtain more reliable determinations of the excited state masses for the $\pi$ and the $\rho$ channels, as more points along the temporal direction were available to study the large-time-separation decay of correlation functions.

In the end, for what concerns ground-state meson masses and the two leading-order QCD low-energy constants, we are able to provide $N=\infty$ results as a function of the pion mass down to the chiral limit which are perfectly compatible with, but much more accurate and solid with respect to, our previous determinations~\cite{Perez:2020vbn,Bonanno:2023ypf}. When possible, we also compared these results with previous large-$N$ determinations given in the literature, finding always good agreement. In addition to these achievements, in this study we are also able to provide a series of brand new results. First, we provide large-$N$ determinations of the first and second excited $\pi$ and $\rho$ masses, which allowed us to test the linearity and universality of their corresponding radial Regge trajectories. Second, we report the first large-$N$ determination of the low-energy constant $\bar{\ell}_4$, appearing at next-to-leading-order in Chiral Perturbation Theory ($\chi$PT) and parameterizing the leading-order pion-mass dependence of $F_\pi$. Finally, we are able to assess the magnitude of sub-leading $1/N$ corrections to $\Sigma$, $F_\pi$ and $\bar{\ell}_4$ by combining our TEK results with those provided for $N=3,4,5$ colors in~\cite{DeGrand:2023hzz}.

This manuscript is organized as follows. In Sec.~\ref{sec:setup} we describe our numerical setup; in Sec.~\ref{sec:res} we present our results; in Sec.~\ref{sec:conclu} we draw our conclusions and discuss possible future perspectives of this study.

\section{Numerical setup}\label{sec:setup}

In this section we will summarize our lattice discretization and the numerical strategies followed to compute meson masses and QCD low-energy constants.

\subsection{Twisted-reduced model for large-\texorpdfstring{$N$}{N} QCD}

In this paper we study QCD in the 't Hooft large-$N$ limit, where the number of quark flavors is kept constant and finite as $N$ tends to infinity, so that $\Nf/N \to 0$. In the 't~Hooft limit, the contribution of dynamical quarks is sub-leading and suppressed by a factor of $1/N$ with respect to that of the gluons. Therefore, large-$N$ QCD is a theory of dynamical gluons and quenched quarks. At the practical level, this implies the exclusion of the fermion determinant from the lattice functional integral that is sampled using Monte Carlo methods, so that gluon dynamics does not ``feel'' the quark back-reaction. Quark fields thus only appear in fermionic observables, which are always computed in the background of the dynamical gluon fields sampled from the Monte Carlo.

As already explained in the introduction, we will exploit twisted large-$N$ volume reduction to simulate gluon dynamics on the lattice at large-$N$. This means that our theory can be thought of as a standard lattice Yang--Mills theory defined on a single-site lattice with twisted boundary conditions. Consequently, our lattice Yang--Mills TEK model only involves $d=4$ $\SU(N)$ matrices with no space-time index. However, it is of the utmost importance to bear in mind that this does \emph{not} mean that we are really working on a space-time torus of vanishing extent. In fact, for our choice of twisted boundary conditions, our particle fields actually propagate over an \emph{effective} volume $\Vphyseff = a^4N^2\equiv \ell^4$, $\ell=a\sqrt{N}$, and finite-volume effects are traded for finite-$N$ effects.

Due to the argument previously outlined, the partition function of the $d=4$ single-site TEK model describing large-$N$ QCD contains only the pure-gauge Yang--Mills action:
\beq\label{eq:part_func}
Z_\TEK \equiv \int [\dd U]\, \ee^{-\SW[U]}, \quad [\dd U] \equiv \left[\prod_{\mu\,=\,1}^{d}\dd U_\mu\right]
\eeq
with $[\dd U]$ the $\SU(N)$ invariant Haar measure and
\beq\label{eq:TEK_Wilson_action}
\SW[U] = -N b \sum_{\mu\,=\,1}^{d}\sum_{\nu \,\ne\, \mu} z_{\nu\mu} \Tr\left\{U_\mu U_\nu U_\mu^\dagger U_\nu^\dagger\right\}
\eeq
the TEK Wilson plaquette action. More details on the Monte Carlo algorithm employed to sample the gluon fields according to the functional integral~\eqref{eq:part_func} can be found in Ref.~\cite{Perez:2015ssa}.

In the action~\eqref{eq:TEK_Wilson_action}, the quantity $b=1/(Ng^2)$ is the inverse of the bare 't Hooft coupling, $U_\mu$ are the $d=4$ $\SU(N)$ gauge link matrices describing the lattice gluon fields, and $z_{\nu \mu}$ is the twist factor implementing twisted boundary conditions, chosen to be a $N^{\text{th}}$-root of 1:
\beq\label{eq:twist_def}
z_{\nu\mu} = z_{\mu\nu}^* = \exp\left\{\frac{2\pi \ii}{N} n_{\nu\mu} \right\}.
\eeq

In this study we work with the so-called \emph{symmetric} twist, meaning that $N$ is taken to be a perfect square $N=L^2$, and the integer-valued anti-symmetric twist tensor $n_{\nu \mu}$ appearing in Eq.~\eqref{eq:twist_def} is chosen to be:
\beq
n_{\nu \mu} = - n_{\mu \nu} = k(L) L, \quad \text{($\nu>\mu$)},
\eeq
with $k(L)$ a co-prime integer with $L$. As explained in Refs.~\cite{Gonzalez-Arroyo:2010omx,Chamizo:2016msz,GarciaPerez:2018fkj}, the flux parameter $k(L)$ has to be scaled with $L$ in order to avoid center-symmetry breaking~\cite{Ishikawa:2003,Bietenholz:2006cz,Teper:2006sp,Azeyanagi:2007su} that would spoil large-$N$ volume reduction. Moreover, appropriate choices of the value of $k$ allow to reduce non-planar finite-volume (i.e., finite-$N$) corrections, see Refs.~\cite{Perez:2017jyq,Bribian:2019ybc}. With this setup, the effective torus size is given by:
\beq
\ell = a L = a \sqrt{N}, \qquad \quad \Vphyseff = \ell^4 = a^4 L^4 = a^4 N^2,
\eeq
with $a$ the lattice spacing. The choices for $N$ and $k$ adopted in this study are summarized in Tab.~\ref{tab:N_values}.

\begin{table}[!t]
\begin{center}
\begin{tabular}{|c|c|c|}
\hline
$N$ & $L=\sqrt{N}$ & $k$\\
\hline
289 & 17 & 5 \\
361 & 19 & 7 \\
529 & 23 & 7 \\
841 & 29 & 9 \\
\hline
\end{tabular}
\end{center}
\caption{Values of the number of colors $N$, of the effective lattice size $L=\sqrt{N}$, and of the parameter $k$ used in this study.}
\label{tab:N_values}
\end{table}

Concerning the fermion sector, in our study we consider $\Nf=2$ degenerate quenched quark flavors, which are discretized according to the Wilson formulation. In space-time coordinates, the lattice Wilson Dirac operator in our setup reads~\cite{Gonzalez-Arroyo:2015bya}:
\beq
\DW = \frac{1}{2\kappa} - \frac{1}{2} \sum_{\mu \, =\, 1}^{d}\left[(\mathbb{I}+\gamma_\mu) \otimes \mathcal{W}_\mu + (\mathbb{I}-\gamma_\mu)\otimes \mathcal{W}_\mu^\dagger \right].
\eeq
Here $\mathbb{I}$ is the identity matrix, $1/(2\kappa)=am_0$ is the lattice bare quark mass, $\gamma_\mu$ are the Dirac matrices, while $\mathcal{W}_\mu$ is an $N^2 \times N^2$ matrix defined as:
\beq
\mathcal{W}_\mu = U_\mu \otimes \Gamma_\mu^*.
\eeq
The links $U_\mu$ are the gauge links sampled via the Monte Carlo according to the functional integral~\eqref{eq:part_func}, while the $\Gamma_\mu$ are known as \emph{twist eaters}, and are $\SU(N)$ matrices satisfying the equation:
\beq\label{eq:master_eq_twist_eaters}
\Gamma_\mu \Gamma_\nu = z_{\nu\mu}^* \Gamma_\nu \Gamma_\mu,
\eeq
with $z_{\nu\mu}$ the same twist factor~\eqref{eq:twist_def} appearing in the TEK Wilson action. The solution of Eq.~\eqref{eq:master_eq_twist_eaters} with a symmetric twist is unique up to a similarity transformation and up to the multiplication by a center element~\cite{Gonzalez-Arroyo:1997ugn}. Since we only consider observables which are traces of link matrices, and since our lattice action is center-symmetric, both these redundancies are irrelevant for our purposes. Thus, we are free to pick one specific solution of Eq.~\eqref{eq:master_eq_twist_eaters} and keep it fixed throughout our calculations to avoid any ambiguity.

\subsection{Lattice determination of meson masses}\label{sec:mesonmass_setup}

Meson masses are extracted from the large Euclidean time decay of the zero-momentum meson lattice correlation functions, following standard numerical techniques used in all customary lattice studies of the QCD spectrum. We briefly summarize all the relevant steps in this section, and refer the reader to Refs.~\cite{Gonzalez-Arroyo:2015bya,Perez:2020vbn} for further details.

The zero-momentum Euclidean time correlation function for the meson A is:
\beq
C_{\A}(\tau) = \left\langle\mathcal{O}_{\A}^\dagger(\tau)\mathcal{O}_{\A}(0)\right\rangle, \qquad \mathcal{O}_{\A}(\tau) = \sum_{\vec{x}} \overline{\psi}(\vec{x},\tau)\gamma_{\A} \psi(\vec{x},0),
\eeq
with $\Tr$ the trace over color and spinor indices, $\psi$ the quark field and $\mathcal{O}_{\A}(\tau)$ the appropriate interpolating operator projecting onto the zero-momentum meson state A, cf.~Tab.~\ref{tab:interp_ops}. The correlation function can thus be rewritten as follows, up to an irrelevant pre-factor:
\beq
C_{\A}(\tau) = \sum_{\vec{x}} \left\langle\Tr\left\{\gamma_{\A} \DW^{-1}(\vec{x},\tau; \vec{0},0) \gamma_{\A} \DW^{-1}(\vec{0},0;\vec{x},\tau)\right\}\right\rangle,
\eeq
where $\DW^{-1}(\vec{x}_1,\tau_1; \vec{x}_2, \tau_2)$ is the quark propagator between the space-time points $x_1$ and $x_2$, equal to the inverse Dirac operator. Finally, one can extract the meson masses in the A channel from the large-time exponential decay of $C_{\A}$:
\beq\label{eq:timedecay_tcorr}
C_{\A}(\tau) \underset{\tau \to \infty}{\sim} \vert c_{\A} \vert^2 \, \exp\left\{-m_\A \tau\right\} + \vert c_{\A^*} \vert^2 \, \exp\left\{-m_{\A^*} \tau\right\} + \dots, \\
c_\A= \bra{\mathrm{A},\vec{p}=\vec{0}}\mathcal{O}_{\A}(0)\ket{0}, \quad c_{\A^*}= \bra{\mathrm{A}^*,\vec{p}=\vec{0}}\mathcal{O}_{\A}(0)\ket{0}, \dots,
\eeq
with $\mathrm{A}^{*}$ denoting the first excited state, and the dots standing for neglected heavier excited states.

In the TEK model, however, one cannot directly compute meson correlation functions in coordinate space, due to lattice volume reduction. The solution to bypass this issue was given in Ref.~\cite{Gonzalez-Arroyo:2015bya}. In a few words, the idea is to let quarks live on an extended lattice, and let them ``feel'' a periodic potential obtained by replicating the gauge fields $\sqrt{N}$ times, in a similar fashion to the Bloch description of electron propagation in a crystal. At the practical level, this approach allows to obtain zero-momentum time correlators in coordinate space from the inverse Fourier transform of the correlation function in momentum space:
\beq \label{eq:inv_fourier}
C_{\A}(\tau) = \sum_{p_0} \ee^{-\ii p_0 \tau} \widetilde{C}_{\A}(p_0),
\eeq
with
\beq
\widetilde{C}_{\A}(p_0) = \left\langle\Tr\left\{\gamma_{\A} \DWp^{-1}(0,\vec{0}) \gamma_{\A} \DWp^{-1}(p_0,\vec{0})\right\}\right\rangle
\eeq
the correlator in momentum space (at vanishing spatial momentum $\vec{p}=\vec{0}$), and
\beq
\DWp(p_0,\vec{p}) &=& \frac{1}{2\kappa} - \frac{1}{2} \sum_{\mu \, =\, 1}^{d}\left[(\mathbb{I}+\gamma_\mu) \otimes \widetilde{\mathcal{W}}_\mu + (\mathbb{I}-\gamma_\mu)\otimes \widetilde{\mathcal{W}}^\dagger_\mu\right],\\
\widetilde{\mathcal{W}}_\mu &=& \ee^{\ii a p_\mu} \, \mathcal{W}_\mu, \qquad p_\mu = (p_0, \vec{p}),
\eeq
the TEK Wilson Dirac operator in momentum space. Here we choose the total time extent to be $T=2L = 2\sqrt{N}$, thus, the time-component of the momentum assumes the discrete values:
\beq
a p_0 = \frac{\pi}{\sqrt{N}}j, \quad j\in\mathbb{N}, \quad j=1, \dots, T.
\eeq

\begin{table}[!t]
\begin{center}
\begin{tabular}{|c||c|c|c|c|c|}
\hline
Meson & $\pi$ & $\rho$ & $a_0$ & $a_1$ & $b_1$ \\
\hline
$\gamma_\A$ & $\gamma_5$ & $\gamma_i$ & $\mathbb{I}$ & $\gamma_5\gamma_i$ & $\frac{1}{2}\varepsilon_{ijk}\gamma_j\gamma_k$\\
\hline
$J^{\rm PC}$ & $0^{-+}$ & $1^{--}$ & $0^{++}$ & $1^{++}$ & $1^{+-}$\\
\hline
\end{tabular}
\end{center}
\caption{Quantum numbers $J^{\rm PC}$ and interpolating operators $\gamma_\A$ for the meson channels $\mathrm{A}=\pi,\rho,a_0,a_1,b_1$ considered in this study.}
\label{tab:interp_ops}
\end{table}

Another important ingredient of the numerical recipe to extract masses from lattice correlation functions is the \emph{Generalized EigenValue Problem} (GEVP), which is used to build an interpolating operator with better projection onto the desired states. Indeed, building the correlation function from the GEVP helps in reducing higher excited state contamination which, due to the finite time extent in actual calculations, could introduce unwanted systematics in the determination of lower-lying ones. The idea is to consider an extended basis of interpolating operators $\mathcal{B}=\left\{\gamma_{\A}^{\ell}\right\}$ for the channel A, and solve the following eigenproblem:
\beq\label{eq:GEVP}
C^{\ell\ell^\prime}_{\A}(\tau) v_{\ell^\prime} = \lambda(\tau,\tau_0) C^{\ell\ell^\prime}_{\A}(\tau_0) v_{\ell^\prime}.
\eeq
In this expression $C^{\ell\ell^\prime}_{\A}(\tau)$ is the inverse Fourier transform, c.f.~Eq.~\eqref{eq:inv_fourier}, of:
\beq\label{eq:tcorr_smeared}
\widetilde{C}^{\ell\ell^\prime}_{\A}(p_0) = \left\langle\Tr\left\{\gamma_{\A}^{\ell} \DWp^{-1}(0,\vec{0}) \gamma_{\A}^{\ell^\prime} \DWp^{-1}(p_0,\vec{0})\right\}\right\rangle.
\eeq
In our study, we build the extended operator basis using a smearing procedure:
\beq
\gamma_{\A}^{\ell} \equiv D_{\s}^{\ell} \gamma_{\A}, \qquad D_{\s} \equiv \frac{1}{1+6c}\left[1+c \sum_{i\,=\,1}^{3}\left(\overline{U}_i\otimes\Gamma_i^{*}+\overline{U}^{\dagger}_i\otimes\Gamma_i^{\rm t}\right)\right],
\eeq
where the exponent $\ell$ is the smearing level, and where $\overline{U}_i$ stands for APE-smeared spatial gauge links obtained after $n_{\scriptscriptstyle{\rm APE}}$ steps. The smeared link $\overline{U}_i$ is iteratively obtained according to:
\beq
U^{(n+1)}_i = \mathrm{Proj}_{\scriptscriptstyle{\SU(N)}}\left[(1-f)U^{(n)}_i + \frac{f}{4}\sum_{j \, \ne \, i}\left(z_{ij} U_j U_i U_j^\dagger + z_{ji} U_j^\dagger U_i U_j\right)\right],
\eeq
where $n=1,\dots,n_{\scriptscriptstyle{\rm APE}}$ counts the number of APE-smearing steps. Concerning the smearing parameters we used $n_{\scriptscriptstyle{\rm APE}}=10$, $f=0.081$, $c=0.4$, and $\ell$ up to 100, with the typical number of operators in our GEVP basis being $4-8$.

Once the GEVP is solved, for given $\tau$ and $\tau_0$, the optimal operator achieving the largest overlap with the ground state of the desired channel is computed as:
\beq
C^{\scriptscriptstyle{(\rm best)}}_{\A} = \left(\bar{v}^{\scriptscriptstyle{(1)}}_{\ell^\prime}\right)^\dagger C^{\ell^\prime\ell}_{\A}\bar{v}^{\scriptscriptstyle{(1)}}_{\ell},
\eeq
with $\bar{v}^{\scriptscriptstyle{(1)}}$ the eigenvector related to the first largest eigenvalue $\bar{\lambda}_1(\tau,\tau_0)$. Practically, to define the optimal operator, we solved the GEVP for fixed times $\tau_0$ and $\tau$ (with $\tau=\tau_0+\delta$, $\delta>0$), and extracted the meson mass in the channel A by fitting the optimal correlator to the following fit function:
\beq\label{eq:fit_opt_tcorr}
C^{\scriptscriptstyle{(\rm best)}}_{\A}(\tau; \tau_0, \delta) = A \cosh\left[m_{\A}\left(\tau-\frac{aT}{2}\right)\right], \quad \tau \in [\tau_1,\tau_2],
\eeq
where the cosh takes into account the finiteness of the time extent, and where the fit is performed in a finite range $\tau \in [\tau_1,\tau_2]$. We typically used $\delta/a=1-2$ and $\tau_0/a=2-5$ for this purpose.

In order to ascertain the precision of the obtained ground state mass determinations, the GEVP was also solved for several values of $\tau$ (i.e., for several values of $\delta$). Indeed, this procedure allows to cross-check the value of $m_\A$ obtained from the fit to Eq.~\eqref{eq:fit_opt_tcorr} by verifying that the effective mass, obtained from the largest eigenvalue of Eq.~\eqref{eq:GEVP} as,
\beq\label{eq:effmass_def}
m_{\A}^{\scriptscriptstyle{(\rm eff)}}(\tau) = \frac{-\log\left[\bar{\lambda}_1(\tau,\tau_0)\right]}{\tau-\tau_0},
\eeq
exhibits a plateau around a value compatible with $m_\A$ in the same range $[\tau_1,\tau_2]$ where the cosh fit is performed. Excited state masses are also extracted from the plateau of the effective mass~\eqref{eq:effmass_def} by employing the $n^{\text{th}}$ largest GEVP eigenvalue $\bar{\lambda}_n(\tau,\tau_0)$ ($n>1$) instead of $\bar{\lambda}_1(\tau,\tau_0)$.

Finally, at the practical level, the traces appearing in Eq.~\eqref{eq:tcorr_smeared} were computed stochastically averaging over random sources:
\beq\label{eq:stoc_trace_def}
\Tr\left\{\gamma_{\A}^{\ell} \DWp^{-1}(0,\vec{0}) \gamma_{\A}^{\ell^\prime} \DWp^{-1}(p_0,\vec{0})\right\} \longrightarrow \left\langle \mathrm{tr}\left\{ z^\dagger\left[\gamma_{\A}^{\ell} \DWp^{-1}(0,\vec{0}) \gamma_{\A}^{\ell^\prime} \DWp^{-1}(p_0,\vec{0})\right] z \right\}\right\rangle_{s}.
\eeq
Here, $\Tr$ includes the trace over spinor, color ($U_{\mu}$), and space-time ($\Gamma_\mu$) indices, while $\mathrm{tr}$ includes only the trace over the spinor index, whereas the color and space-time traces are evaluated via the random $\mathbb{Z}_4$ source method: $\braket{z^\dagger \mathcal{M} z}_s= \frac{1}{N_{\rm src}} \sum_{i\,=\,1}^{N_{\rm src}}z^\dagger_i \mathcal{M} z_i \equiv \frac{1}{N_{\rm src}} \sum_{i\,=\,1}^{N_{\rm src}} \mathcal{M}_{ii}$, where $z_i$ is a random vector whose elements (for both color and space-time indices) are $\mathbb{Z}_4$ random numbers, and where $N_{\rm src}$ is the number of random sources employed (we typically used $N_{\rm src}=5-10$). For any given random vector $z_i$, the trace appearing in Eq.~\eqref{eq:stoc_trace_def} can be rewritten as:
\begin{align}
\mathcal{M}_{ii} &= \mathrm{tr}\left\{ z_i^{\dagger} \gamma_{\A} D_{\s}^{\ell} \DWp^{-1}(0,\vec{0})\gamma_{\A} D_s^{\ell^\prime} \DWp^{-1}(p_0,\vec{0}) z_i\right\} 
\notag\\
&=
\mathrm{tr}\left\{z_i^{\dagger}\gamma_{\A} D_{\s}^{\ell} \gamma_5 \left(\DWp^{-1}(0,\vec{0})\right)^{\dagger}\gamma_5\gamma_{\A} D_{\s}^{\ell^\prime} \DWp^{-1}(p_0,\vec{0}) z_i\right\}
\notag\\
&=
\mathrm{tr}\left\{ \gamma_{\A} \gamma_5 \left[z_i^{\dagger} D_{\s}^{\ell} \left(\DWp^{-1}(0,\vec{0})\right)^{\dagger}\right] \gamma_5\gamma_{\A} \left[D_{\s}^{\ell^\prime}\DWp^{-1}(p_0,\vec{0})z_i\right]\right\}
\notag\\
&=
\mathrm{tr}\left\{ \gamma_{\A} \gamma_5 \left[\DWp^{-1}(0,\vec{0}) D_{\s}^{\ell} z_i \right]^{\dagger} \gamma_5\gamma_{\A} \left[D_{\s}^{\ell^\prime}\DWp^{-1}(p_0,\vec{0})z_i\right]\right\}
\notag  \\
&=
\mathrm{tr}\left\{ \left [ \gamma_{\A} \gamma_5 \left(S^{\mathrm{src}-\ell}_{i}\right)^{\dagger} \right] \cdot \left[\gamma_5\gamma_{\A} S^{\mathrm{snk}-\ell^\prime}_{i} \right] \right\},
\end{align}
where it is understood that both the color and space-time indices are contracted in the inner product $\left(S^{\mathrm{src}-\ell}_{i}\right)^\dagger \cdot S^{\mathrm{snk}-\ell^\prime}_{i}$. The source and sink vectors $S^{\mathrm{src}-\ell}_{i}$ and $S^{\mathrm{snk}-\ell^\prime}_{i}$ are solutions of the following linear equations:
\begin{align}
\label{eq:source}
\DWp(0,\vec{0}) S^{\mathrm{src}-\ell}_{i} & = D_{\s}^{\ell} z_i, \\
\label{eq:sink}
\DWp(p_0,\vec{0}) G &= z_i,
\end{align}
with
\beq
S^{\mathrm{snk}-\ell^\prime}_{i} = D_s^{\ell^\prime} G.
\eeq
These equations are solved using a mixed precision solver, which is based on the defect correction Richardson iteration algorithm: the outer defect correction iteration is evaluated in double precision arithmetic, while
the inner iteration is evaluated with the BiCGStab algorithm in single precision arithmetic, which is a popular choice for the non-Hermitian Wilson--Dirac quark solver in QCD. To improve the numerical performances, especially for larger values of $N$, we employed temporal gauge fixing within the solver and GPU accelerators. When multiple GPU accelerators are available in a computational node, we distributed the linear equations with different right-hand side vectors (i.e., with different spin components and $\mathbb{Z}_4$ sources) to the available GPU cards enabling concurrently solving independent equations.

\subsection{Lattice definitions of the quark mass}\label{sec:quarkmass_lattice_setup}
Clearly, all the masses of the meson tower extracted with the method outlined in the previous section will depend on the chosen quark mass $m$, which can be changed by varying the hopping parameter $\kappa$ appearing in the Wilson Dirac operator. Calculating this dependence is not only important because it contains physical information, but is also necessary to provide extrapolated results of meson masses towards the chiral limit $m=0$.

In this paper we will consider two possible lattice definitions of the bare quark mass. One, the so called ``subtracted mass'', exhibits both multiplicative and additive renormalizations:
\beq
\ZS m_{\R} = (m_0 - m_c) = \frac{1}{a} \left(\frac{1}{2\kappa}-\frac{1}{2\kappa_c}\right),
\eeq
where $\ZS$ is the renormalization constant of the quark scalar density, and $\kappa_c$ is the critical $\kappa$ value for which the renormalized quark mass $m_\R$ vanishes. Another popular choice is instead based on the axial Ward identity and the partially conserved axial current (PCAC):
\begin{equation}
m_{\PCAC} = \frac{\braket{\Tr\left[\partial_0 A_0(\tau) P(0)\right]}}{2\braket{\Tr\left[P(\tau) P(0)\right]}},
\end{equation}
with $A_0(\tau)$ the space-integrated time component of the quark axial vector current, and $P$ the space-integrated pseudo-scalar quark density. This definition indeed only gets a multiplicative renormalization:
\beq
m_{\PCAC} = \frac{\ZP}{\ZA}m_\R,
\eeq
with $\ZP$ and $\ZA$ the renormalization constants of the pseudo-scalar and axial quark bilinears. Thus, $m_\PCAC=0$ automatically implies $m_\R=0$. Therefore, these two definitions of the quark mass offer two possible ways to define the chiral point $m_\R=0$, i.e., the critical $\kappa$: either $\kappa=\kappa^{(\pi)}_c$ when $m_\pi=0$, or $\kappa=\kappa_c^{(\PCAC)}$ when $m_\PCAC=0$. These two determinations of $\kappa_c$ should be equal up to lattice artifacts, as we will show in Sec.~\ref{sec:kappac_determination}.

At the practical level, the determination of the PCAC quark mass is obtained from a constant fit to the following ratio of correlation functions:
\beq
a m_{\PCAC} = \frac{C_{055}(\tau+a)-C_{055}(\tau-a)}{4C_{55}(\tau)}.
\eeq
In momentum space, the correlation functions entering the lattice definition of $m_\PCAC$ are:
\beq
\widetilde{C}_{055}(p_0) &=& \left\langle\Tr\left\{\gamma_0 \gamma_5 \DWp^{-1} (p_0)\gamma_5^{\scriptscriptstyle{(\rm opt)}} \DWp^{-1}(0)\right\}\right\rangle,\\
\widetilde{C}_{55}(p_0) &=& \left\langle\Tr\left\{\gamma_5 \DWp^{-1} (p_0)\gamma_5^{\scriptscriptstyle{(\rm opt)}} \DWp^{-1}(0)\right\}\right\rangle,\\
\gamma_5^{\scriptscriptstyle{(\rm opt)}}&=& \sum_{\ell} \overline{v}^{(\pi)}_\ell \gamma_5^{\ell},
\eeq
with $\overline{v}^{(\pi)}$ the eigenvector related to the largest eigenvalue of the GEVP that is solved for the pion correlation function, and where all traces are again computed using stochastic sources.

\subsection{Lattice determination of the QCD low-energy constants \texorpdfstring{$B$}{B}, \texorpdfstring{$\Sigma$}{Sigma}, \texorpdfstring{$F_\pi$}{Fpi}, \texorpdfstring{$\bar{\ell}_4$}{L4}}\label{sec:LEC_lattice_setup}

At leading order (LO) in the momentum $p^2$ and in the quark mass $m$, only two Low-Energy Constants (LECs) are necessary to write the chiral effective Lagrangian describing QCD with $\Nf=2$ degenerate light quark flavors~\cite{Gasser:1983yg} with mass $m$. One is the scheme- and scale-dependent quark condensate:
\beq
\Sigma = - \lim_{m\,\to\,0} \braket{\overline{\psi}\psi}.
\eeq
The other one is the pion decay constant:
\beq\label{eq:Fpi_def}
\sqrt{2}F_\pi = \lim_{m\,\to\,0}\frac{1}{m_\pi} \bra{0} \overline{\psi} \gamma_0 \gamma_5 \psi\ket{\pi,\vec{p}=\vec{0}}.
\eeq
Note that we work with the normalization that, in standard $N=3$ QCD, would correspond to $F_\pi \simeq 92$ MeV at the $\Nf=2$ isospin-symmetric physical point. Both these LECs are divergent in the large-$N$ limit, and scale as
\beq
\Sigma \sim \mathcal{O}(N), \qquad F_\pi\sim\mathcal{O}(\sqrt{N}),
\eeq
thus in this study we will compute $\Sigma/N$ and $F_\pi/\sqrt{N}$, which instead are expected to have a well-defined and finite large-$N$ limit.

These two quantities enter the celebrated Gell-Mann--Oakes--Renner (GMOR) relation, expressing the quark mass dependence of the pion mass:
\beq\label{eq:GMOR_def}
m^2_\pi = 2\frac{\Sigma_\R}{F_\pi^2} m_\R = 2 B_\R m_\R,
\eeq
with the subscript ``R'' denoting renormalized quantities, through the combination
\beq
B_\R\equiv\frac{\Sigma_\R}{F_\pi^2}\sim\mathcal{O}(N^0),
\eeq
which is sometimes employed in place of $\Sigma$. Note that the product $\Sigma_\R m_\R$ (or, equivalently, $B_\R m_\R$) is renormalization group invariant (RGI), thus we do not need to specify any particular renormalization scheme or scale. Another fundamental relation where the chiral condensate appears is the renowned Banks--Casher (BC) formula:
\beq\label{eq:BC_def}
\frac{\Sigma}{\pi} = \lim_{\lambda\,\to\, 0} \lim_{m\,\to\,0} \lim_{V\,\to\,\infty} \rho(\lambda,m), \qquad (\slashed{D}+m)v_\lambda = (\ii\lambda+m)v_\lambda,
\eeq
relating the quark condensate to the value of the chiral limit of the spectral density of the Dirac operator in the origin.

In this study we will exploit both the GMOR equation~\eqref{eq:GMOR_def} to compute $B$ and the BC relation~\eqref{eq:BC_def} to compute $\Sigma$, which combined will thus allow the determination of $F_\pi$. This result will be then cross-checked with a direct calculation of $F_\pi$ from a suitable discretization of its defining formula~\eqref{eq:Fpi_def}. In this section we will summarize all the necessary ingredients for these tasks.

Let us start from the LEC $B_\R$. This is easily extracted from the chiral behavior of the pion mass as a function of the Wilson hopping parameter:
\beq
m_\pi^2(\kappa) = 2 B_\R m_\R(\kappa) = 2\frac{B_\R}{a\ZS}\left(\frac{1}{2\kappa}-\frac{1}{2\kappa_c}\right).
\eeq
Fitting lattice data of $m_\pi^2$ as a function of $\kappa$ according to this equation permits the determination of $\kappa_c^{(\pi)}$ and of the bare quantity $B=B_\R/\ZS$, which will be renormalized using the large-$N$ non-perturbative determinations of $\ZS$ in the $\MSbar$ scheme at the conventional scale $\mutwoGeV$ of~\cite{Castagnini:2015ejr}.

The chiral condensate $\Sigma$ alone will instead be extracted from the BC relation using the Giusti--L\"uscher spectral method~\cite{Giusti:2008vb}. The idea behind this numerical strategy is to consider the \emph{mode number}, i.e., the integral of the Dirac spectral density,
\beq
\braket{\nu(M,m)} &=& \braket{\# \, \lambda \text{ such that } |\ii\lambda+m|^2 \le M^2}\\
&=& V \int_{-M_0}^{M_0} \dd\lambda \, \rho(\lambda,m), \qquad M_0^2=M^2-m^2,
\eeq
with $V$ the space-time volume, and build an ``effective'' (i.e., mass-dependent) condensate,
\beq
\Sigma^{\effsup}(m) = \frac{\pi}{2}\frac{\braket{\nu(M,m)}}{VM_0},
\eeq
which, in the chiral limit, will tend to the actual condensate $\Sigma$:
\beq\label{eq:chircorrs_Sigmaeff}
\Sigma^{\effsup}(m) = \Sigma \left[1+ C m + \mathcal{O}(m^2)\right].
\eeq
The linear relation between $\Sigma$ and $\nu$, valid for sufficiently small quark masses and for sufficiently small values of $M_0$, of course stems directly from the BC relation.

As shown in Ref.~\cite{Bonanno:2023ypf}, the Giusti--L\"uscher method can be applied to our reduced model once the substitution $V\longrightarrow \Vphyseff=a^4N^2$ is performed. At the practical level, we solved the eigenproblem
\beq\label{eq:eigenproblem_modenumber}
\QW u_\alpha = \alpha \, u_\alpha, \qquad \alpha \in \mathbb{R}, \qquad \QW \equiv \gamma_5 \DW,
\eeq
and simply computed $\braket{\nu(M,m)}$ as:
\beq
\braket{\nu(M,m)} = \braket{\# \, \alpha \text{ such that } \alpha^2 \le M^2}.
\eeq
Then, the renormalized effective condensate is computed in terms of the renormalized \emph{slope} $\mathcal{S}_\R$ of the mode number as a function of $M_\R$, computed in a point $\overline{M}_\R$ sufficiently close to $M_\R = m_\R$:
\beq\label{eq:effective_cond_def}
\frac{1}{N}\Sigma^{\effsup}_\R(m) = \frac{\pi}{2\Vphyseff}\sqrt{1-\left(\frac{m_\R}{\overline{M_\R}}\right)^2} \, \mathcal{S}_\R, \qquad \mathcal{S}_\R = \frac{\partial}{\partial M_\R} \frac{\braket{\nu_\R(M,m)}}{N}\bigg\vert_{M_\R = \overline{M}_\R}.
\eeq
Finally, the actual quark condensate is obtained extrapolating $\Sigma_\R^{\effsup}$ assuming linear corrections in the quark mass, cf.~Eq.~\eqref{eq:chircorrs_Sigmaeff}.

The renormalization of Eq.~\eqref{eq:effective_cond_def} has been worked out in Ref.~\cite{Giusti:2008vb}:
\beq
\braket{\nu} = \braket{\nu_\R}, \qquad M =\ZP M_\R.
\eeq
Since the slope is practically obtained from a linear best fit of $\braket{\nu}$ as a function of $M$, the Giusti--L\"uscher method allows the computation of the bare condensate $\Sigma^{\effsup} = \Sigma^{\effsup}_\R/\ZP$. The renormalization of this quantity is performed as follows:
\beq\label{eq:renormalization_Sigmaeff}
\frac{1}{N}\Sigma^{\effsup}_\R = \frac{\Sigma^{\effsup}_\R}{N\ZP}\times \ZS \times \frac{\ZP}{\ZS}.
\eeq
The non-perturbative determinations of $\ZS$ come again from Ref.~\cite{Castagnini:2015ejr}, while non-perturbative determinations of $\ZP/\ZS$ can be obtained from the very same eigenvectors of the eigenproblem solved to obtain the mode number, cf.~Eq.~\eqref{eq:eigenproblem_modenumber}, as follows~\cite{Giusti:2008vb}:
\beq\label{eq:specsum_ZPZS}
\left(\frac{\ZP}{\ZS}\right)^2 = \lim_{m \, \to \, 0} \frac{\braket{s_{\scriptscriptstyle{\rm P}}(M,m)}}{\braket{\nu(M,m)}}, \qquad \braket{s_{\scriptscriptstyle{\rm P}}(M,m)} = \sum_{\vert\alpha\vert \le M} \sum_{\vert\alpha^\prime\vert\le M} \vert u_{\alpha}^\dagger \gamma_5 u_{\alpha^\prime}\vert^2.
\eeq
Note that the ratio of spectral sums defining $\ZP/\ZS$ is expected to exhibit a plateau as a function of $M$. Also, let us stress that the large-$N$ non-perturbative spectral computation of $\ZP/\ZS$ is a new result first presented in this paper.

Clearly, since the GMOR and the BC formulas allow the computation of $B_\R$ and $\Sigma_\R/N$, it is straightforward to obtain $F_\pi/\sqrt{N}$ from the inversion of the defining equation for $B_\R$. However, one can also directly compute $F_\pi$ using a suitable discretization of Eq.~\eqref{eq:Fpi_def}. This latter route will serve as a cross-check of our determinations of $B_\R$ and $\Sigma_\R$. In practice, one performs a fit to a constant of the following ratio of correlation functions:
\beq
\frac{aF_\pi}{\ZA} = \frac{C_{055}(\tau)}{\sqrt{m_\pi C_\pi(\tau)}},
\eeq
with $C_{055}(\tau)$ and $C_{\pi}(\tau)$ the correlators already introduced in Sec.~\ref{sec:mesonmass_setup}. In this paper, we will follow a different strategy to renormalize the bare pion decay constant with respect to the previous TEK study~\cite{Perez:2020vbn}. In particular, it will be performed as follows:
\beq
\frac{F_\pi}{\sqrt{N}} = \frac{F_\pi}{\ZA \sqrt{N}} \times \frac{\ZP}{\ZS} \times \left(\frac{\ZP}{\ZS \ZA}\right)^{-1}.
\eeq
The ratio $\ZP/\ZS$ is obtained from the Dirac spectrum as earlier explained, while the ratio $\ZP/(\ZS \ZA)$ can be computed from a linear best fit of the chiral behavior of the PCAC quark mass as a function of the hopping parameter:
\beq
m_\PCAC(\kappa) = \frac{\ZP}{\ZA}m_\R(\kappa) = \frac{\ZP}{\ZS\ZA}\frac{1}{a}\left(\frac{1}{2\kappa}-\frac{1}{2\kappa_c}\right).
\eeq
Of course, this best fit procedure also allows the determination of $\kappa_c^{(\PCAC)}$.

Finally, in this study we will also be able to extract the RGI dimensionless LEC $\bar{\ell}_4$ appearing at next-to-leading-order (NLO) in the chiral expansion of QCD~\cite{Gasser:1983yg}. Indeed, this quantity appears in the coefficient parameterizing the LO pion mass dependence of $F_\pi$:
\beq\label{eq:ell4_def}
F_\pi(m_\pi) = F_\pi\left[1+\frac{\bar{\ell}_4}{16\pi^2F_\pi^2} m_\pi^2 +\mathcal{O}(m_\pi^4)\right], \qquad \quad \frac{\bar{\ell}_4}{N} \sim \mathcal{O}(N^0).
\eeq

\section{Results}\label{sec:res}

This section is devoted to the discussion of our numerical results. First, we present our determination of the scale setting and of the chiral point, necessary ingredients to fix the overall scale and to take the continuum and chiral limits. Then, we present our results for the low-lying meson spectrum, and we study the radial Regge trajectories in the $\pi$ and $\rho$ channels. Finally, we discuss our large-$N$ determination of the QCD LECs $B$, $\Sigma$, $F_\pi$ and $\bar{\ell}_4$, and how to compute their $N$-dependence by combining them with finite-$N$ results of Ref.~\cite{DeGrand:2023hzz}.

\subsection{Scale setting}\label{sec:scale_set}

As it is customary in lattice calculations, one has to restore the appropriate energy dimension in any observable computed in lattice units by dividing for the appropriate power of the lattice spacing. This in turn requires the determination of the lattice spacing in terms of a reference observable of choice, in units of which any other quantity will be expressed. This procedure of \emph{scale setting} is customarily done in terms of quantities which can be accurately measured with a relatively cheap numerical effort.

Since in the literature several choices have been employed, we will also consider a few alternatives, so that the comparison with previously available results will present no difficulties. In particular, we will consider the string tension $\sigma$, whose determination in the TEK model was tackled in Ref.~\cite{Gonzalez-Arroyo:2012euf}, as well as the well-known gradient flow scale $t_0$. Moreover, in this study we will also consider another reference gradient flow scale, $t_1$. The detailed discussion of the calculation of these gradient flow quantities will be presented in a separate forthcoming paper~\cite{tonypietrowip} (see also~\cite{Butti:2023oep,Butti:2023hfp}, where the first dedicated TEK studies of $t_0$ and $t_1$ have been presented). In this section, we will just limit to briefly recall how gradient flow scale setting works, how the scales $t_0$ and $t_1$ are defined, and how one can pass from one scale to the other.

Gradient flow~\cite{Narayanan:2006rf,Luscher:2009eq,Lohmayer:2011si} is a smoothening technique rooted on renormalization group flow and consists in evolving the gauge fields according to the differential equation:
\beq
\partial_t B_\mu (t) = D_\nu G_{\mu \nu} (t), \qquad B_\mu (t = 0) = A_\mu.
\eeq
The flow equation introduces a new length scale in the game, $\sqrt{8t}$, which can be physically interpreted as the length scale below which ultraviolet fluctuations are smoothened away. This length scale can be used for the purpose of scale setting by fixing the value of $t^2 \braket{E(t)}$, with
\beq\label{eq:t0_N3_def}
E(t) = \frac{1}{6 V} \sum_{x} \sum_{\nu>\mu}\frac{1}{2} \Tr \left \{G_{\mu \nu} (x;t)G_{\mu \nu} (x;t)\right\},
\eeq
the flowed energy density, to some reference value which is kept fixed across all lattice spacings. The typical choice for $t_0$ at $N=3$ is:
\beq\label{eq:t0_N3}
t^2 \braket{E(t)}\bigg\vert_{t\,=\,t_0} = 0.3.
\eeq
Since the energy density scales with $N$ as
\beq
t^2 \braket{E(t)} \underset{N\,\to\,\infty}{\sim} \frac{N^2-1}{N},
\eeq
the definition~\eqref{eq:t0_N3_def} is typically extended to larger SU$(N)$ gauge groups as:
\beq\label{eq:t0_Ninf_Giusti}
\frac{N}{N^2-1} t^2 \braket{E(t)}\bigg\vert_{t\,=\,t_0} = c = 0.1125,
\eeq
since $0.1125 \times 8/3 = 0.3$. However, in the large-$N$ limit, it is also possible to consider another definition $t^\prime_0$ which still reproduces the $N=3$ definition in Eq.~\eqref{eq:t0_N3}, and still has a finite large-$N$ limit:
\beq\label{eq:t0_Ninf_our}
\frac{1}{N} t^2 \braket{E(t)}\bigg\vert_{t\,=\,t^\prime_0} = c = 0.1,
\eeq

\begin{table}[!t]
\begin{center}
\begin{tabular}{|c|c|c|c||c|}
\hline
$b$ & $a/\sqrt{8t^\prime_0}$ $(c=0.1)$ & $a/\sqrt{8t_1}$ $(c=0.05)$ & $a\sqrt{\sigma}$ & $a/\sqrt{8t_0}$ $(c=0.1125)$\\
\hline
0.355  &  0.22962(74) &  0.37707(57) &  0.2410(30) & 0.21506(69)\\
0.360  &  0.19486(53) &  0.31776(50) &  0.2058(25) & 0.18250(50)\\
0.365  &  0.16725(90) &  0.27181(74) &  0.1783(17) & 0.15664(84)\\
0.370  &  0.14652(64) &  0.23607(61) &  0.1573(19) & 0.13723(60)\\
0.375  &  0.12863(61) &  0.20576(55) &  0.1361(17) & 0.12047(57)\\
\hline
\end{tabular}
\end{center}
\caption{Summary of the different quantities that will be used for scale setting in this study. The values of $c$ in the gradient flow scales refer to the values of the quantity $t^2\braket{E(t)}/N = c$ used to define them. Results for $\sigma$ come from Ref.~\cite{Gonzalez-Arroyo:2012euf}, while those for $t^\prime_0$ and $t_1$ will be presented in a separate forthcoming publication~\cite{tonypietrowip} (see Refs.~\cite{Butti:2023oep,Butti:2023hfp} for preliminary determinations). Finally, $t_0$ is obtained from $t_0^\prime$ using Eq.~\eqref{eq:conv_t0p_to}.}
\label{tab:scale_setting}
\end{table}

\noindent Since, for $N=\infty$, Eq.~\eqref{eq:t0_Ninf_Giusti} becomes:
\beq
\frac{1}{N} t^2 \braket{E(t)}\bigg\vert_{t\,=\,t_0} = c = 0.1125,
\eeq
it is clear that $t_0^\prime$ and $t_0$ will only be equal for $N=3$, and differ for $4\le N \le \infty$. Since we have TEK $N=\infty$ determinations of $t^\prime_0$ at our disposal, but previous finite-$N$ lattice studies use $t_0$, in order to compare our findings with the existing literature it is important to know how to convert from one scale to another. Given that $c=0.1$ and $c=0.1125$ are numerically very close, and given that the quantity $f(t)\equiv t^2 \braket{E(t)}/N$ is almost linear around $t\sim t_0$, performing a Taylor expansion of this function it is possible to reliably estimate the ratio $t_0/t_0^\prime$ from a best fit of $f(t)$ to a second-order polynomial:
\beq\label{eq:conv_t0p_to}
\frac{t_0}{t_0^\prime} = 1.140(3).
\eeq
Finally, we will also consider the scale obtained cutting the energy density in a different value of $c$:
\beq
\frac{1}{N} t^2 \braket{E(t)}\bigg\vert_{t\,=\,t_1} = c = 0.05.
\eeq
This alternative definition is very useful because it generally yields more precise results compared to $t_0$, and suffers for smaller finite-volume effects, at the price of slightly larger lattice artifacts. Thus, this scale will be used for intermediate calculations, and then final results will be converted to either $\sigma$ or $t_0$.

\begin{figure}[!t]
\centering
\includegraphics[scale=0.4]{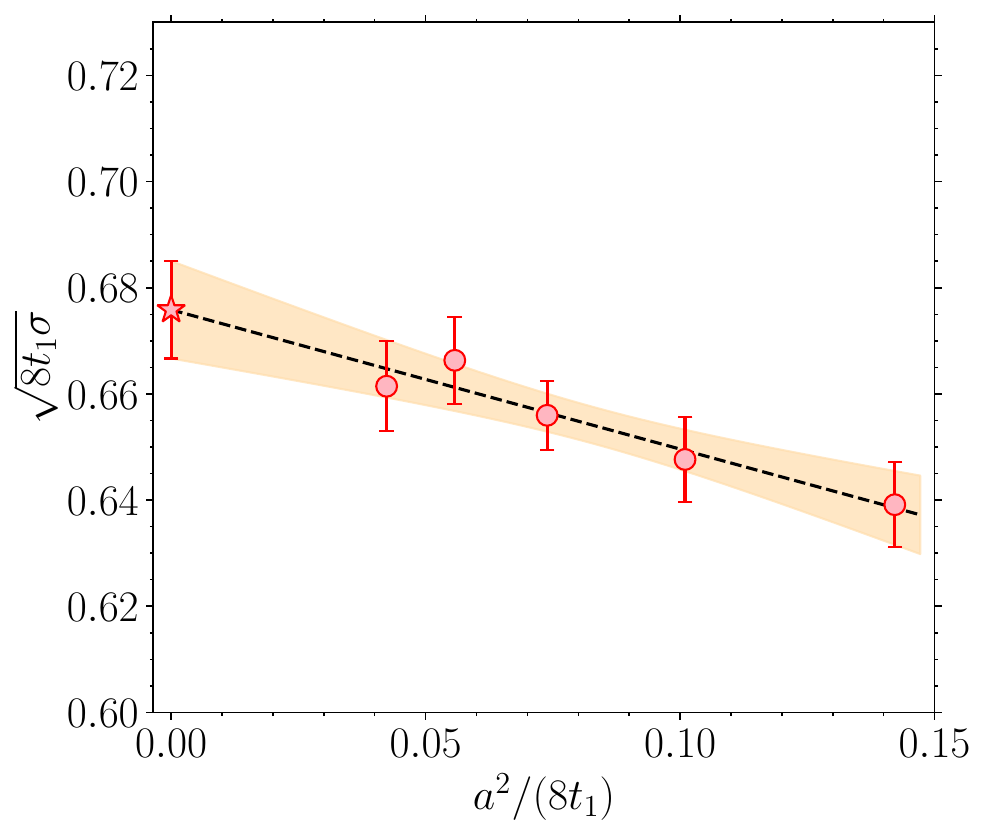}
\includegraphics[scale=0.4]{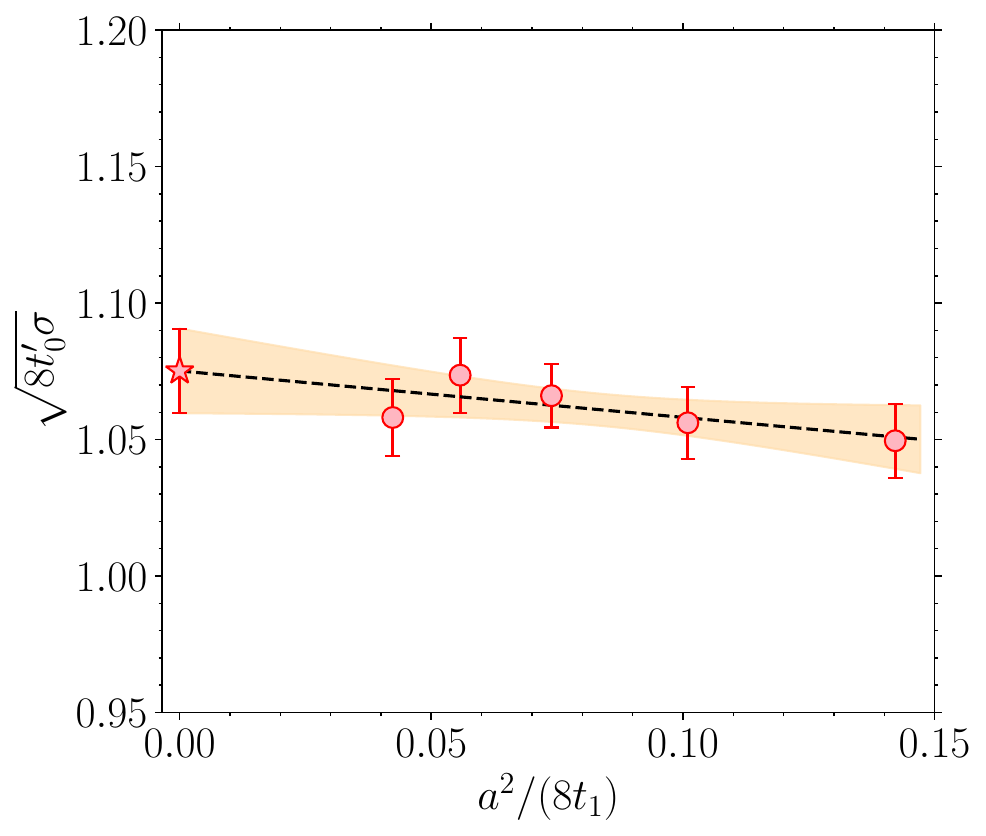}
\includegraphics[scale=0.4]{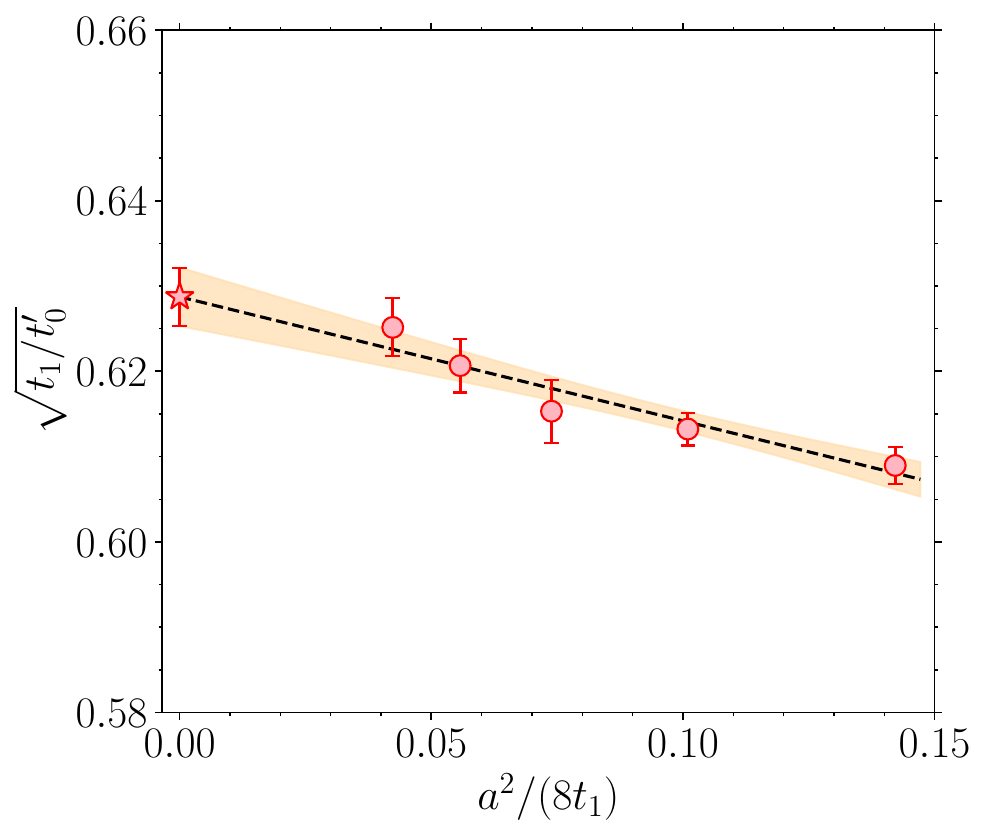}
\caption{Continuum limit of ratios of quantities used for scale setting assuming standard $\mathcal{O}(a^2)$ lattice artifacts for gluonic quantities.}
\label{fig:scales}
\end{figure}

Our results for the scale setting quantities are summarized in Tab.~\ref{tab:scale_setting}. In Fig.~\ref{fig:scales} we also show the continuum extrapolations for the ratios of two of these quantities assuming standard $\mathcal{O}(a^2)$ lattice artifacts for gluonic quantities. As it can be seen, we observe scaling in all cases, meaning that our determinations of the scale are perfectly consistent among them. The final continuum results for the ratio of scales are:\footnote{Quoted errors on ratios of scale setting quantities are just statistical. Possible systematic errors related to residual finite-$N$ effects have been estimated to be very small: $\sim \mathcal{O}(0.5\%)$ for $\sqrt{8t_1\sigma}$, and $\sim \mathcal{O}(1.5\%)$ for $\sqrt{t_1/t_0^\prime}$. They have also been verified to be negligible when using $\sqrt{t_1/t_0^\prime}$ to convert physical quantities from $t_1$ to $t_0$ or $\sigma$ units.}
\beq
\label{eq:conv_t1_sigma}
\sqrt{8t_1\sigma} &=& 0.6759(92),\\
\label{eq:conv_t0p_sigma}
\sqrt{8t_0^\prime\sigma} &=& 1.075(16),\\
\label{eq:conv_t1_t0p}
\sqrt{t_1/t_0^\prime} &=& 0.6287(34),
\eeq
Using Eq.~\eqref{eq:conv_t0p_to}, we obtain for the large-$N$ limit of the scale $t_0$ defined in Eq.~\eqref{eq:t0_Ninf_Giusti}:
\beq\label{eq:conv_t0_sigma}
\sqrt{8t_0\sigma}=1.148(17).
\eeq

\subsection{Determination of the chiral point}\label{sec:kappac_determination}

Given that we aim at studying the chiral behavior of meson masses, eventually computing their value in the chiral limit, it is of utmost importance to define the chiral point.

As discussed in Sec.~\ref{sec:setup}, this can be achieved either by looking at the critical value of the hopping parameter $\kappa$ corresponding to $m_\PCAC=0$, or at the one corresponding to $m_\pi=0$. The determinations of $\kappa_c$ from the PCAC or the pion mass typically do not coincide exactly at coarse lattice spacing due to lattice artifacts, but they should get closer as the lattice spacing gets finer, eventually converging to the same value in the continuum limit.

Our determinations for the pion and PCAC mass are reported in Tab.~\ref{tab:raw_mpi_mPCAC} in Appendix~\ref{app:rawdata}. In this study, thanks to the employed values of $N$ we are able to simulate effective volumes which satisfy $m_\pi \ell = a m_\pi \sqrt{N} \gtrsim 4.5$. This threshold, which is of the same order of that customarily chosen in standard $N=3$ QCD simulations, allows for a solid control over finite-volume effects. For completeness, in Tab.~\ref{tab:raw_mpi_mPCAC} we also report the values of the effective volumes in physical units $\ell\sqrt{\sigma} = a L \sqrt{\sigma} = a \sqrt{N} \sqrt{\sigma}$, with typical values $\ell\sqrt{\sigma}\gtrsim 4$, corresponding to the new ensembles with $N=529$ and $841$. In a few cases, for heavier pions we also included a few data points obtained with $N=289$ coming from the previous work~\cite{Perez:2020vbn}, as they also had sufficiently large values of $m_\pi \ell$.

\begin{table*}[!t]
\begin{center}
\begin{tabular}{|c|c|c|c|c|}
\hline
$b$ & $aB_\R/\ZS$ & $\kappa_c$ ($m_\pi$) & $\ZP/(\ZS \ZA)$ & $\kappa_c$ ($m_\PCAC$)\\
\hline
0.355 & 1.350(27)  & 0.162882(74)  & 0.8415(63) & 0.163044(34)  \\
0.360 & 1.254(33)  & 0.160046(54)  & 0.862(10)  & 0.160321(33)  \\
0.365 & 1.091(33)  & 0.158051(80)  & 0.844(11)  & 0.158251(37)  \\
0.370 & 0.971(20)  & 0.156415(38)  & 0.8649(74) & 0.156544(16)  \\
0.375 & 0.848(23)  & 0.155009(39)  & 0.897(14)  & 0.155025(23)  \\
\hline
\end{tabular}
\end{center}
\caption{Summary of the best fit parameters extracted from the chiral behavior of $m_\pi$ and of $m_\PCAC$, cf.~Eq.~\eqref{eq:mpi_chir_fit} and Eq.~\eqref{eq:mpcac_chir_fit}.}
\label{tab:kappac_res}
\end{table*}

\begin{figure}[!t]
\centering
\includegraphics[scale=0.34]{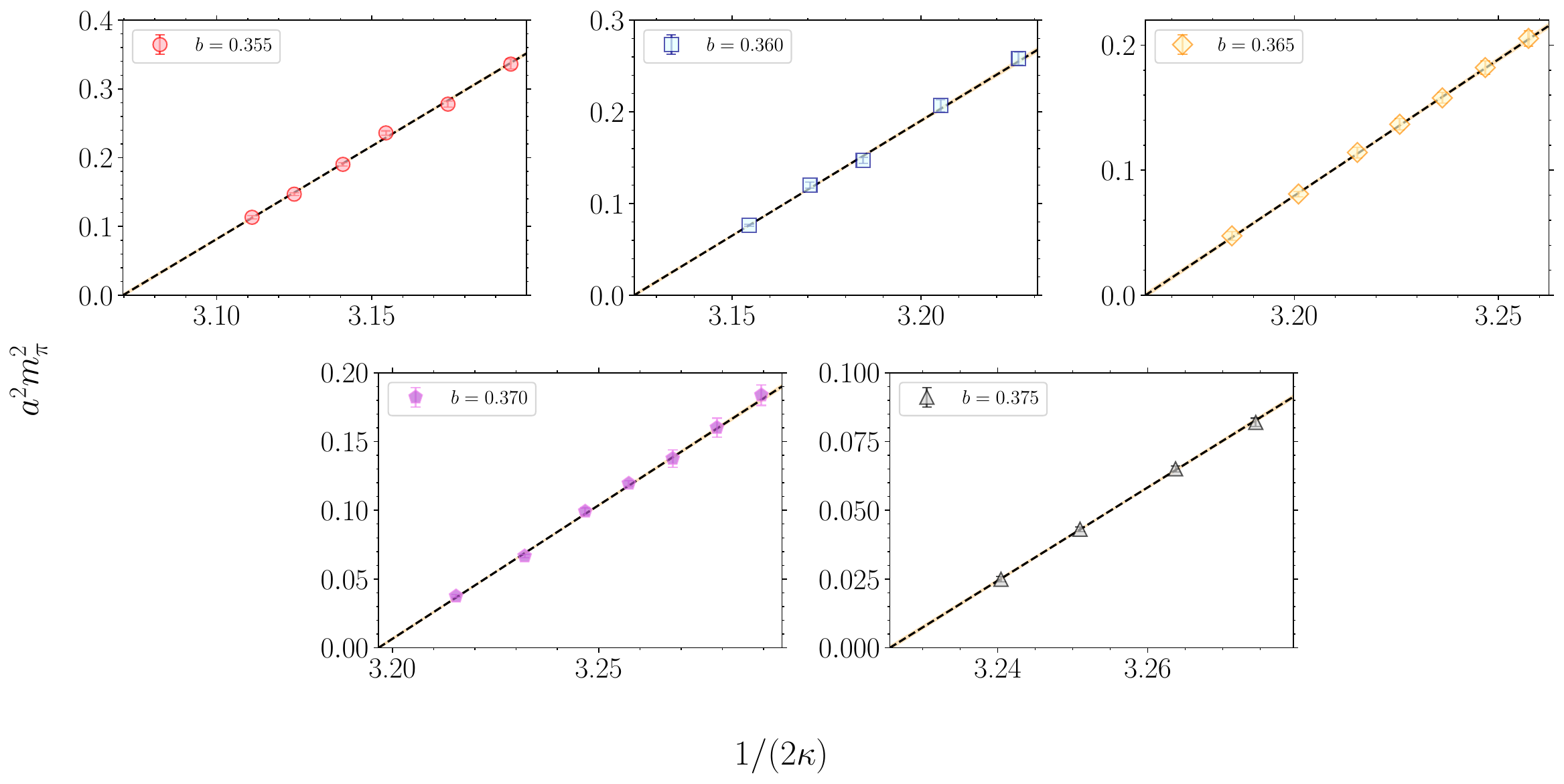}
\includegraphics[scale=0.34]{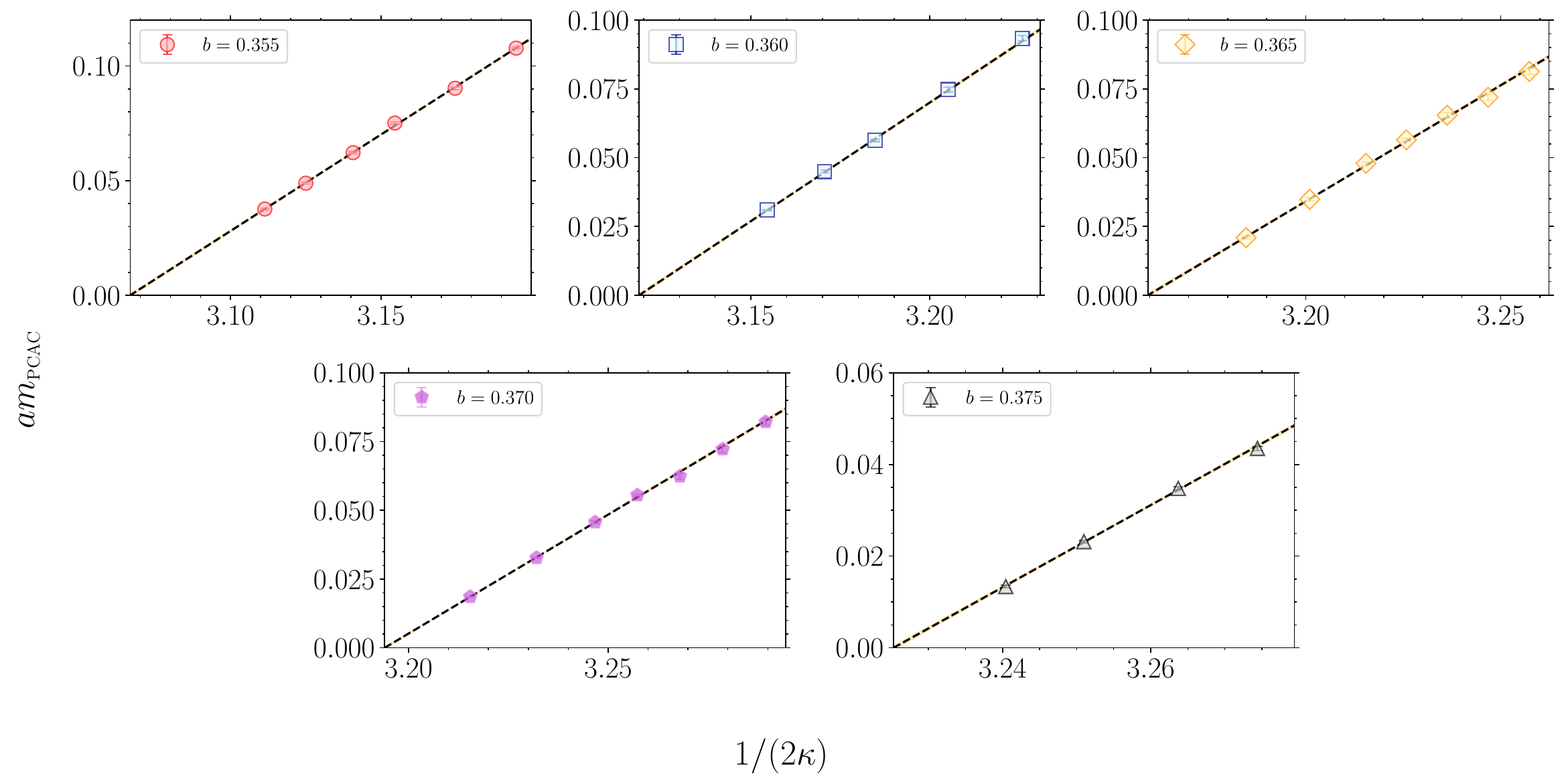}
\caption{Chiral behavior of the pion mass (top panel) and of the PCAC quark mass (bottom panel) as a function of the Wilson hopping parameter $\kappa$.}
\label{fig:kappacrit_det}
\end{figure}

In Fig.~\ref{fig:kappacrit_det} we display the chiral behavior for each value of $b$ of both $m_\pi$ and $m_\PCAC$. Best fits are performed according to the following fit functions:
\beq
\label{eq:mpi_chir_fit}
a^2 m_\pi^2 = \frac{a B_\R}{\ZS} \left( \frac{1}{2\kappa} - \frac{1}{2\kappa_c}\right),\\
\label{eq:mpcac_chir_fit}
a m_\PCAC = \frac{\ZP}{\ZS\ZA} \left( \frac{1}{2\kappa} - \frac{1}{2\kappa_c}\right).
\eeq
Best fit results are reported in Tab.~\ref{tab:kappac_res}. As it can be appreciated, while results for $\kappa_c$ obtained from these two fit procedures are not exactly compatible for the smallest values of $b$ explored due to lattice artifacts, they come together as the lattice spacing gets finer, and no difference can be observed at the largest value of $b$ simulated, confirming our general expectations.

\FloatBarrier

\subsection{Results for meson masses}\label{sec:mesonmass_res}

This section is devoted to collect our results for large-$N$ meson masses. Our goal is to compute these masses in the continuum limit, and we aim at reconstructing their value as a function of the quark mass $m$ (or, equivalently, as a function of $m_\pi^2$) down to the chiral limit. To extract the ground state mass of the $\rho$, $a_0$, $a_1$ and $b_1$ mesons, we used the same ensembles adopted for the pion mass computations presented in the previous section. For all but a few of the $N=529$ and $N=841$ ensembles we were also able to extract the mass of the first and the second excited states in the $\pi$ and $\rho$ channels, denoted with $^*$ and $^{**}$ respectively, whose determination represents a brand new result first presented in this paper.

\begin{figure}[!t]
\hspace*{-2.3\baselineskip}
\includegraphics[scale=0.62]{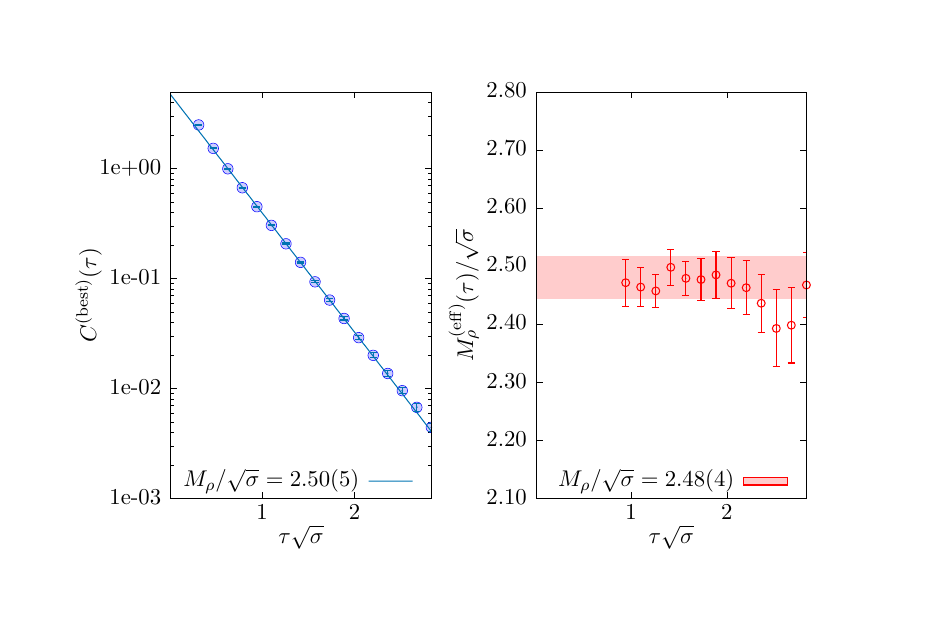}
\hspace*{-4\baselineskip}
\includegraphics[scale=0.62]{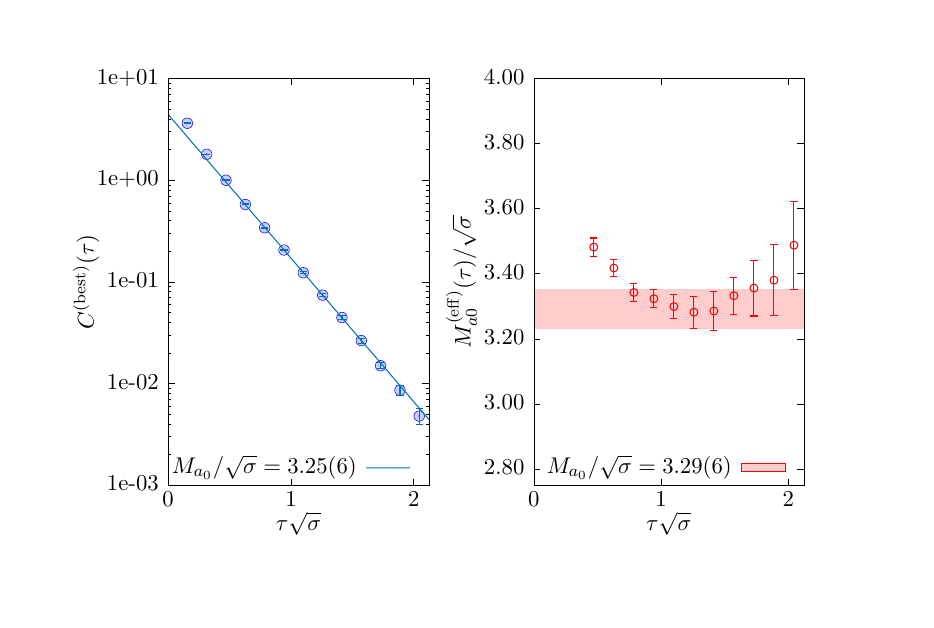}
\vspace*{-2.0\baselineskip}
\caption{Examples of exponential decays of the $\rho$ and $a_0$ optimal correlators obtained from the resolution of the GEVP, and related plateaus in the effective masses, cf.~Eq.~\eqref{eq:fit_opt_tcorr} and Eq.~\eqref{eq:effmass_def}. All plots in this figure refer to $N=841$, $b=0.370$, $\kappa=0.1540$.}
\label{fig:ex_tcorr_fit}
\end{figure}

Let us start by displaying an example of mass extraction for the $\rho$ and $a_0$ mesons for one of the $N=841$ ensembles in Fig.~\ref{fig:ex_tcorr_fit}. As expected, the meson correlation functions exhibits an exponential decay over several orders of magnitude for sufficiently large values of $\tau$, and the mass extracted from the large-$\tau$ behavior of the optimal correlator is perfectly compatible with the one that would be obtained from the plateau exhibited by the effective mass.

All meson mass results as a function of the lattice spacing and quark mass are collected in Tab.~\ref{tab:raw_other_mesons} in Appendix~\ref{app:rawdata}. Let us start our discussion from the ground state masses, and the first excited $\pi$ and $\rho$ state masses. Being these results obtained with Wilson fermions, they are affected by $\mathcal{O}(a)$ lattice artifacts. Therefore, in order to extract our final results as a function of the pion masses, we performed for all meson states a global fit involving all our determinations for different values of $b$ and $\kappa$ according to the following fit function:
\beq
\frac{1}{\sqrt{\sigma}} m_{\A}(m_\pi,a) = \frac{m^{\chir}_\A}{\sqrt{\sigma}} + C_{\A}^{(1)} \frac{m_\pi^2}{\sigma} + C_{\A}^{(2)} \frac{m_\pi^4}{\sigma^2} + k_{\A} a\sqrt{\sigma}.
\eeq
Thus, in the continuum limit, the functional form:
\beq\label{eq:chircont_formula}
\frac{1}{\sqrt{\sigma}} m_{\A}(m_\pi) = \frac{m^{\chir}_{\A}}{\sqrt{\sigma}} + C_{\A}^{(1)}\frac{m_\pi^2}{\sigma} + C_{\A}^{(2)} \frac{m_\pi^4}{\sigma^2}
\eeq
describes the pion mass dependence of the mass of the meson ``A'' down to the chiral limit in the continuum theory, with $m_\A^{\chir}$ representing the value at vanishing quark mass. Recall that chiral logs vanish at large $N$, and thus are not included in Eq.~\eqref{eq:chircont_formula}. Let us also stress that by $m_{\A}/\sqrt{\sigma}$ we actually mean:
\beq\label{eq:conv_sigma_through_t1}
\left(a m_\A\right) \times \left(\frac{\sqrt{8t_1}}{a}\right) \times \left(\frac{1}{\sqrt{8t_1\sigma}}\right),
\eeq
with $\sqrt{8t_1\sigma}$ the continuum-extrapolated value in Eq.~\eqref{eq:conv_t1_sigma}. Indeed, given the better precision of $a/\sqrt{8t_1}$ with respect to $a\sqrt{\sigma}$, this scale setting procedure allows to obtain much more accurate results, and much more constrained fits, with respect to simply using the direct determinations of $a\sqrt{\sigma}$ to restore physical units. The choice of expressing all final results in units of $\sigma$ is just done for convenience, as it will be easier to compare our findings with previous determinations in the literature.

\begin{figure}[!t]
\centering
\includegraphics[scale=0.29]{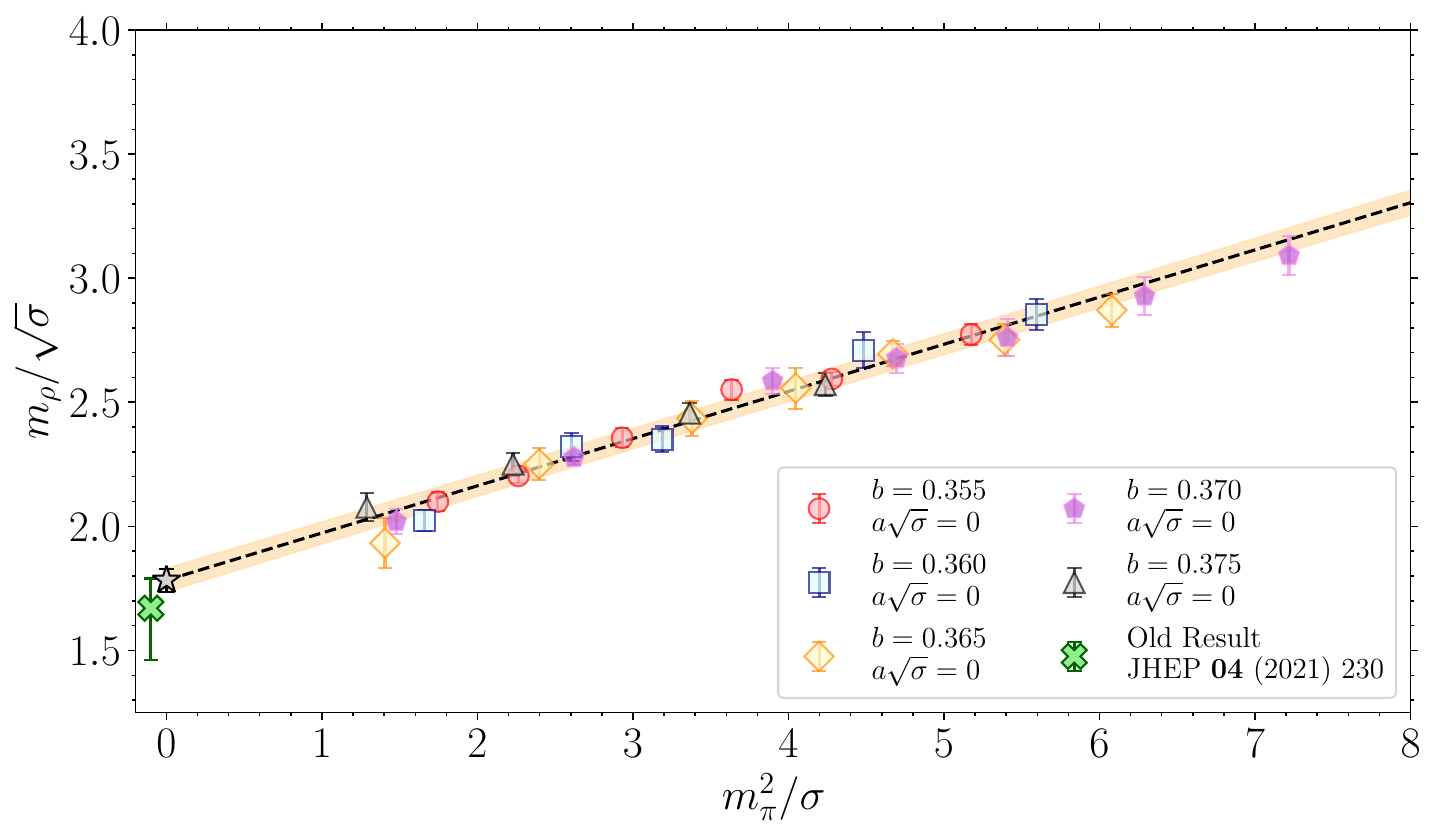}
\includegraphics[scale=0.29]{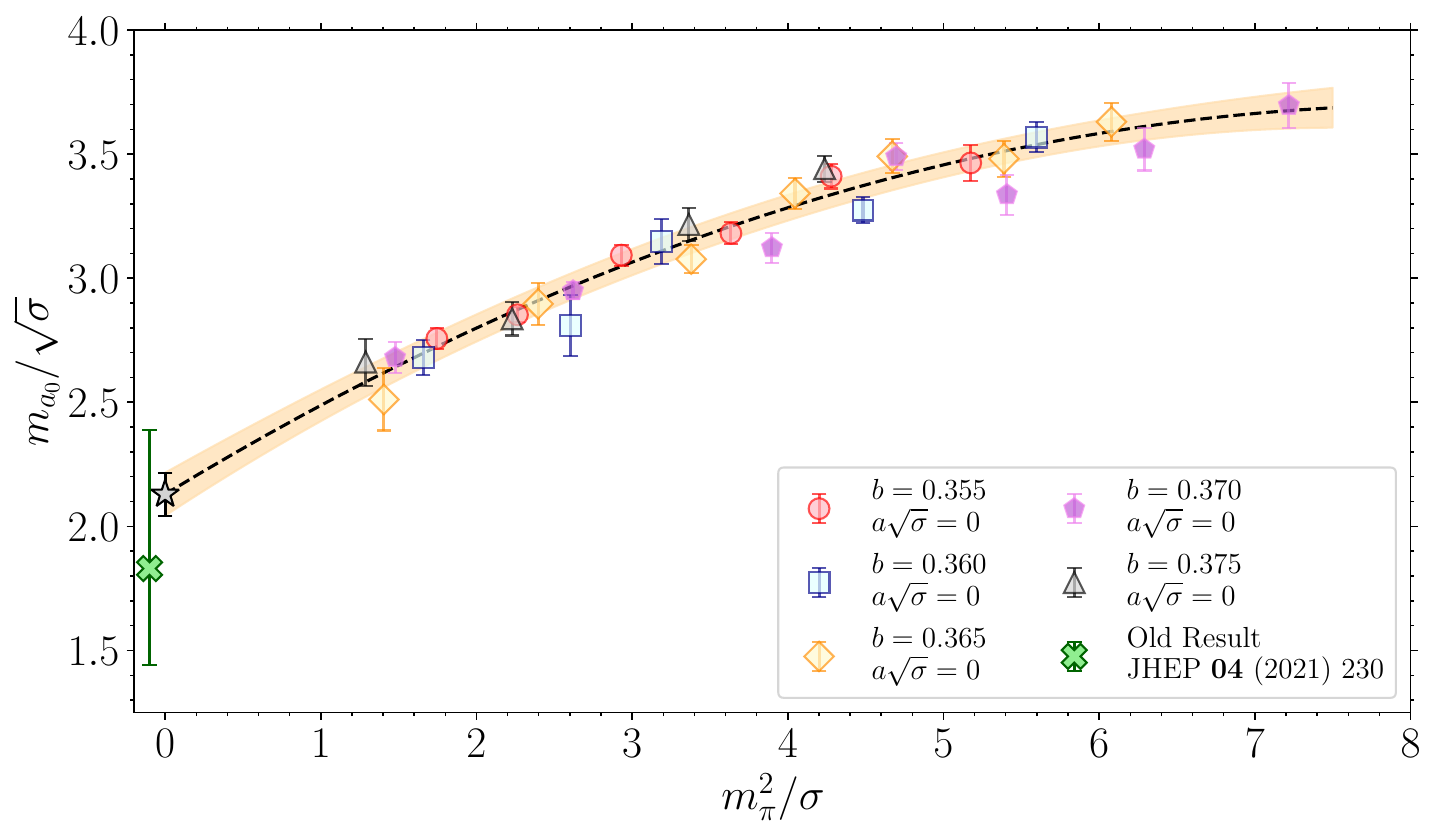}
\includegraphics[scale=0.29]{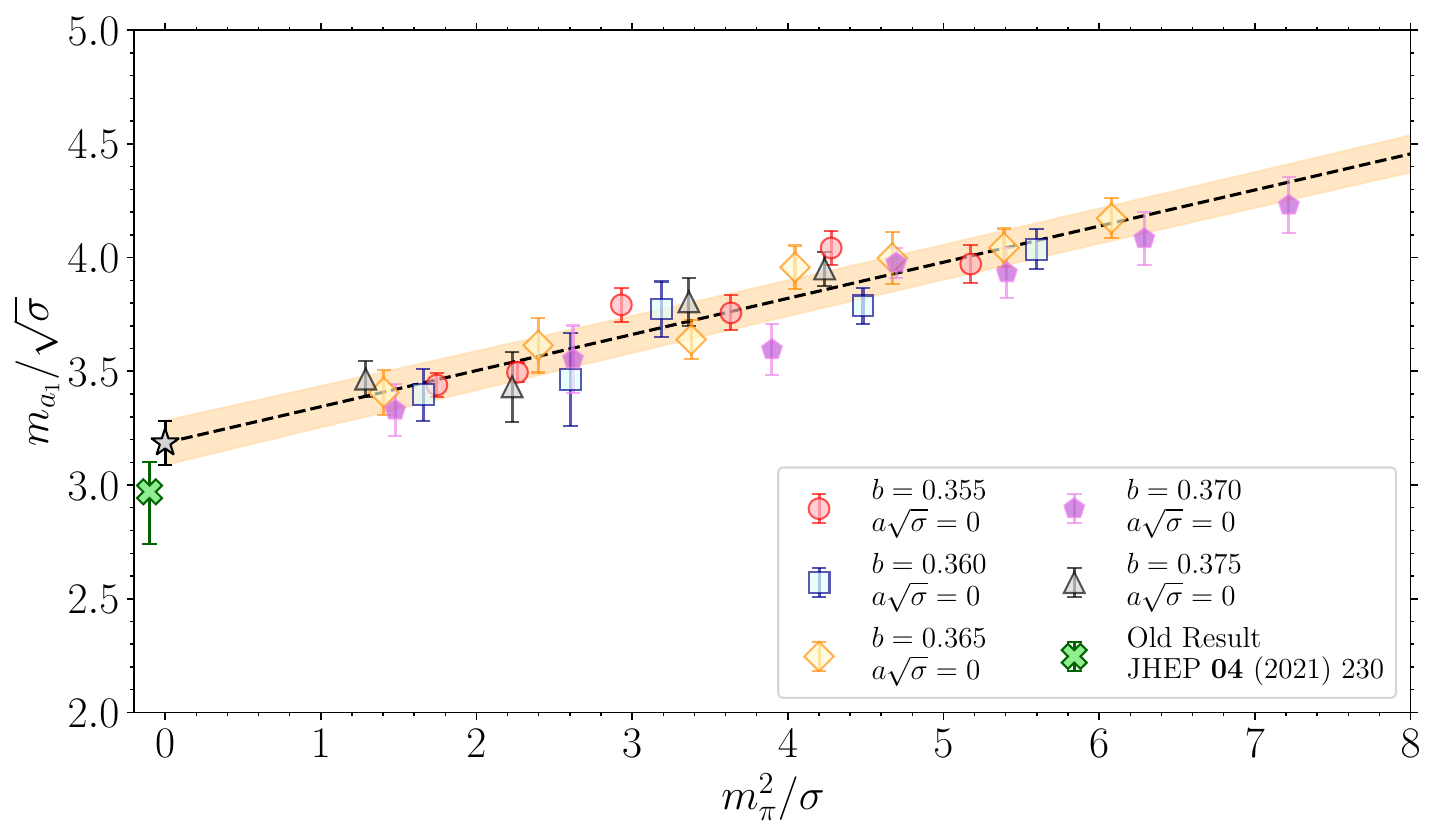}
\includegraphics[scale=0.29]{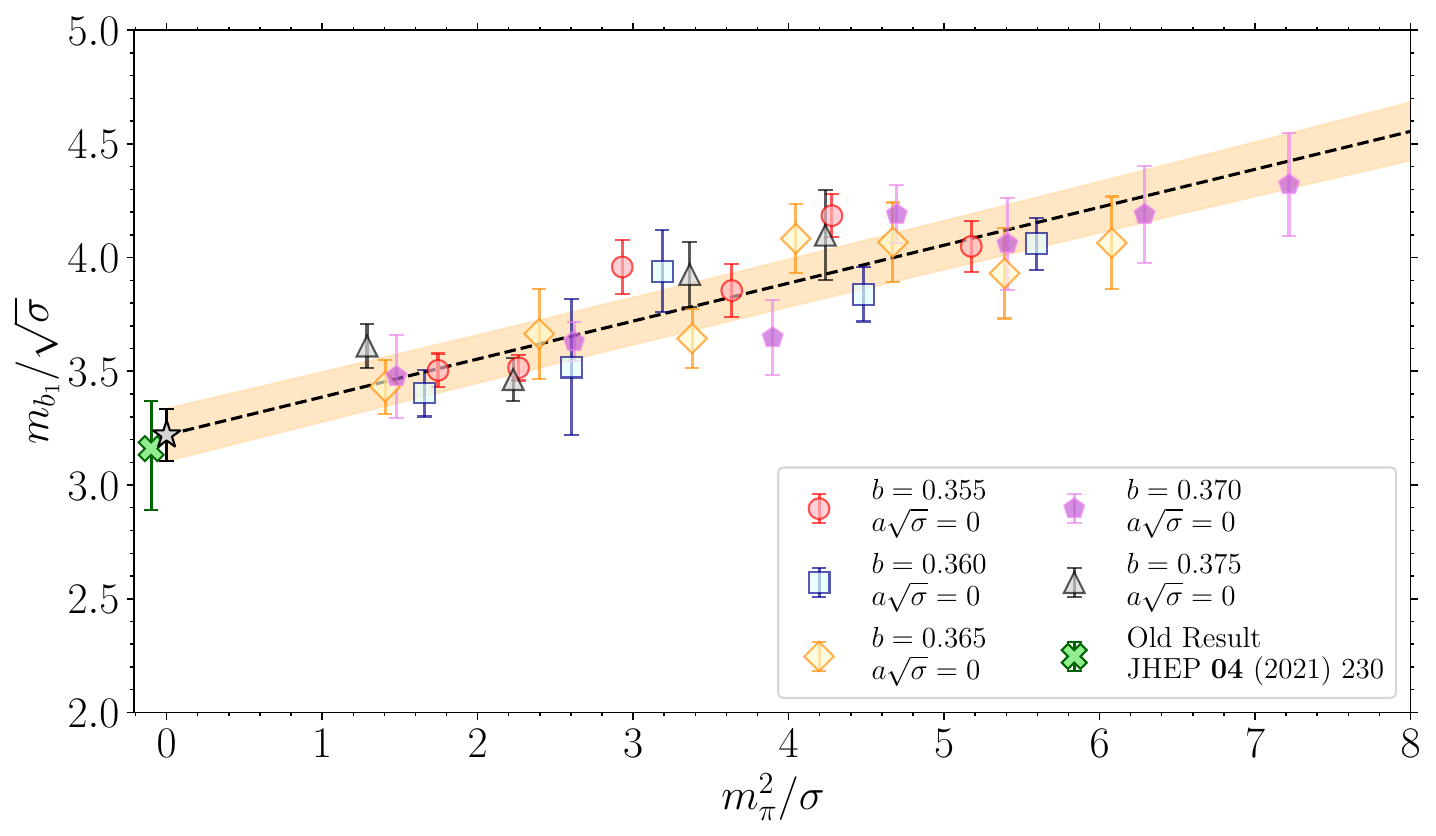}
\includegraphics[scale=0.29]{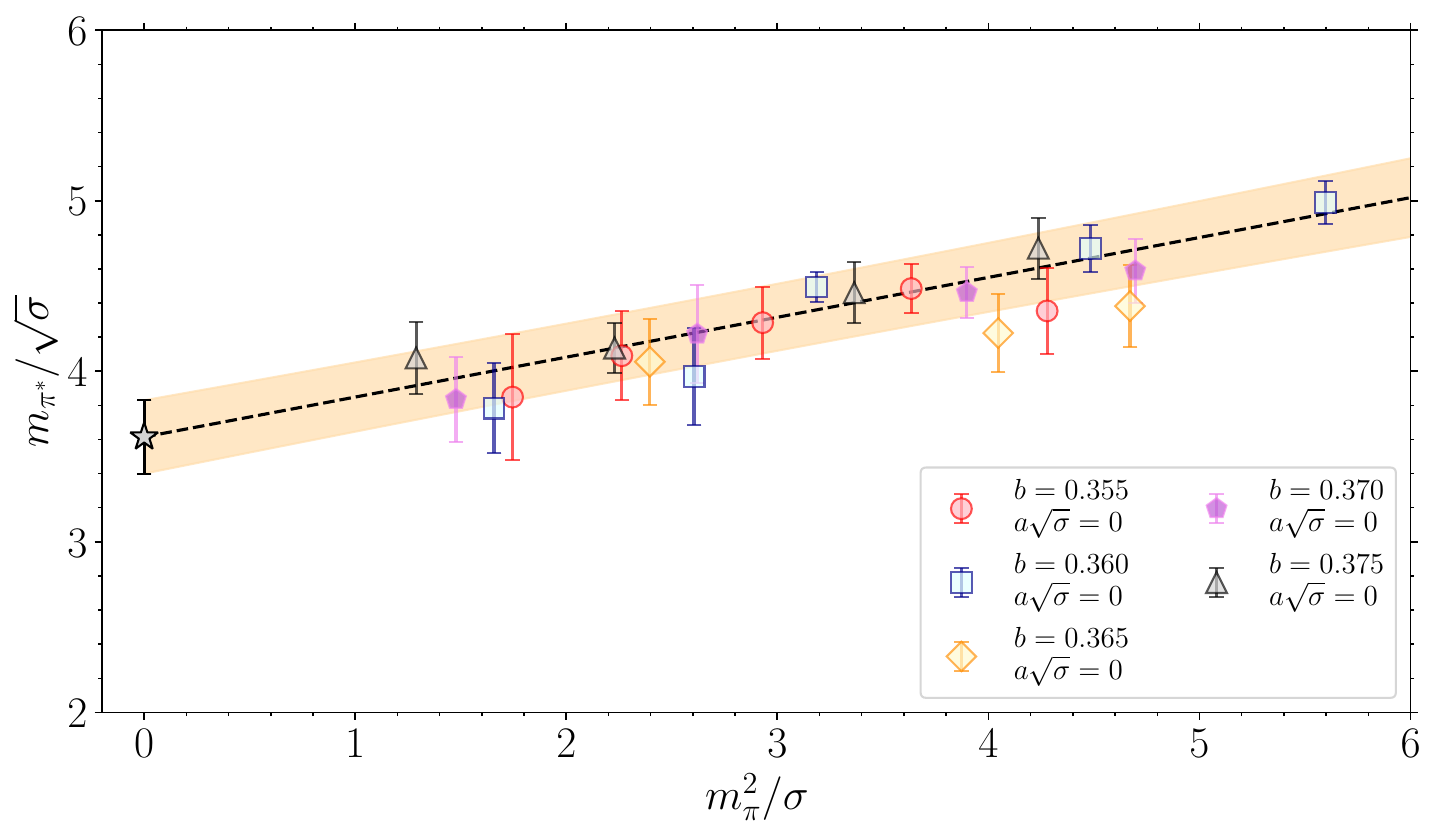}
\includegraphics[scale=0.29]{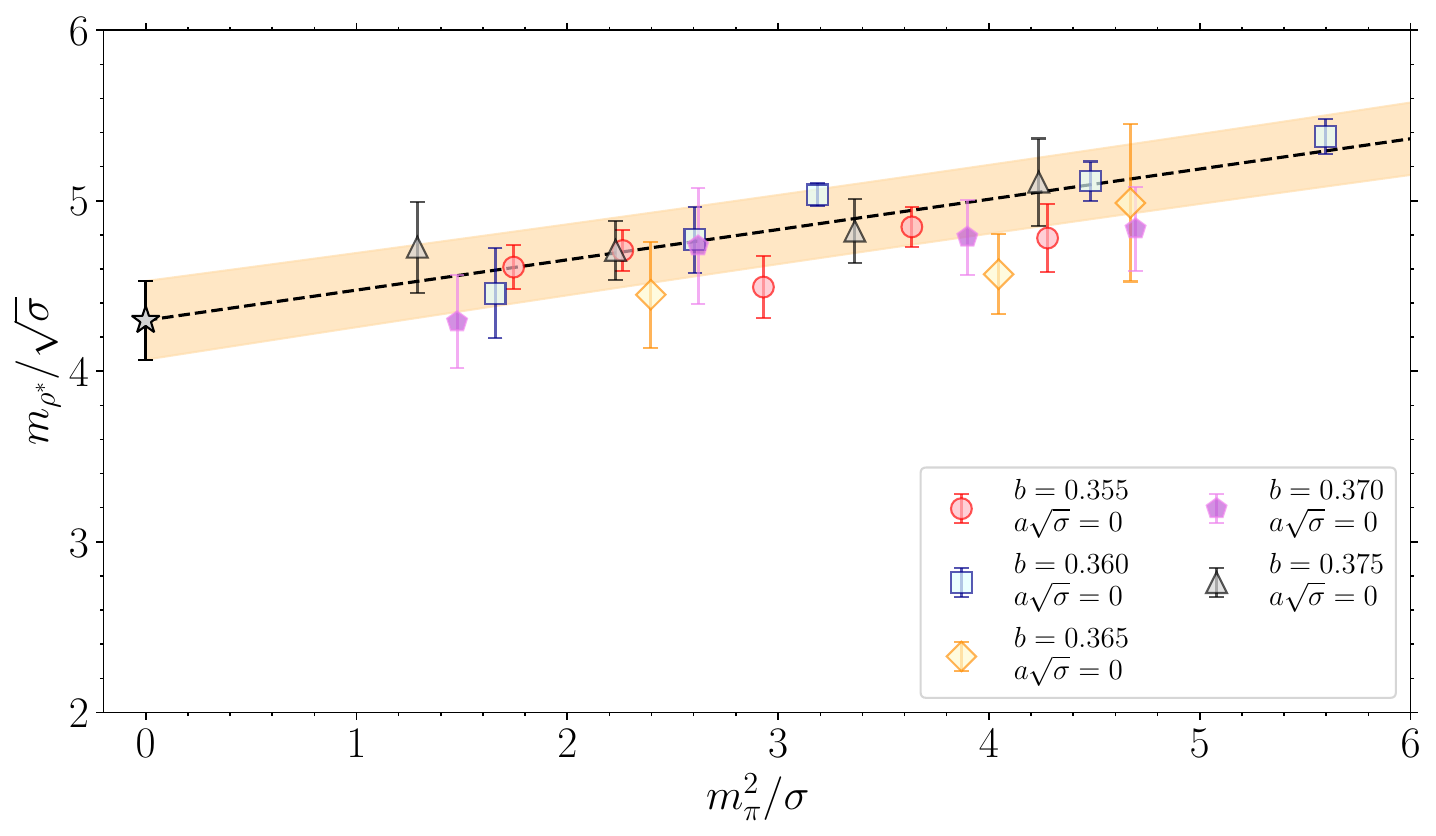}
\caption{Chiral-continuum extrapolations of the $\rho$, $a_0$, $a_1$, $b_1$, $\pi^{*}$ and $\rho^{*}$ masses.. In all cases it is sufficient to assume a linear dependence in $m_\pi^2/\sigma$ to describe the pion mass dependence of these masses, except for the $a_0$ meson, for which it is necessary to also include a further $m_\pi^4/\sigma^2$ correction. The displayed points represent the lattice determinations for each choice of $(b,\kappa)$ after the subtraction of the lattice artifact term $k_{\A} a\sqrt{\sigma}$. The dashed lines and shaded bands represent our continuum results for $m_\A(m_\pi)/\sqrt{\sigma}$. The starred points represent the chiral limit in the continuum, while the crossed points represent the previous large-$N$ TEK chiral determinations of Ref.~\cite{Perez:2020vbn}.}
\label{fig:meson_masses}
\end{figure}

In all cases we have verified that adding a further mixed term of the form $k_\A^{\prime} a\sqrt{\sigma} \times \frac{m_\pi^2}{\sigma}$ does not alter the obtained fit results within errors, and yields values of the coefficient $k_\A^{\prime}$ compatible with zero within errors. Concerning higher-order corrections in the pion mass, they turn out to be negligible in all cases except for the $a_0$ meson, where instead the inclusion of a $C_{\A}^{(2)} \frac{m_\pi^4}{\sigma^2}$ term is necessary to obtain a good fit with a reasonable reduced chi-squared. Finally, let us also stress that all our global best fits were cross-checked by performing separated chiral fits at fixed $b$ followed by a continuum extrapolation of these chiral-limit determinations.

The joint chiral-continuum best fits of our data are shown in Fig.~\ref{fig:meson_masses}. In the end, we quote these results in the continuum, depicted in Fig.~\ref{fig:meson_masses} as a shaded curve:
\beq
\label{eq:final_rho_vs_mpi}
\frac{1}{\sqrt{\sigma}}m_\rho(m_\pi)  &=& 1.783(47)+0.1902(65)\frac{m_\pi^2}{\sigma},\\
\label{eq:final_a0_vs_mpi}
\frac{1}{\sqrt{\sigma}}m_{a_0}(m_\pi) &=& 2.130(87)+0.381(40)\frac{m_\pi^2}{\sigma} -0.0231(52) \frac{m_\pi^4}{\sigma^2},\\
\label{eq:final_a1_vs_mpi}
\frac{1}{\sqrt{\sigma}}m_{a_1}(m_\pi) &=& 3.186(97)+0.159(11)\frac{m_\pi^2}{\sigma},\\
\label{eq:final_b1_vs_mpi}
\frac{1}{\sqrt{\sigma}}m_{b_1}(m_\pi) &=& 3.22(12)+0.167(16)\frac{m_\pi^2}{\sigma},\\
\label{eq:final_pi1exc_vs_mpi}
\frac{1}{\sqrt{\sigma}}m_{\pi^{*}}(m_\pi) &=& 3.62(21)+0.167(16)\frac{m_\pi^2}{\sigma},\\
\label{eq:final_rho1exc_vs_mpi}
\frac{1}{\sqrt{\sigma}}m_{\rho^{*}}(m_\pi) &=& 4.30(23)+0.177(30)\frac{m_\pi^2}{\sigma}.
\eeq

\begin{table}[!t]
\begin{center}
\begin{tabular}{|c|c|c|c|c|c|c|c|}
\cline{3-8}
\multicolumn{2}{c|}{} & $m^{\chir}_\rho/\sqrt{\sigma}$ & $m^{\chir}_{a_0}/\sqrt{\sigma}$ & $m^{\chir}_{a_1}/\sqrt{\sigma}$ & $m^{\chir}_{b_1}/\sqrt{\sigma}$ & $m^{\chir}_{\pi^{*}}/\sqrt{\sigma}$ & $m^{\chir}_{\rho^{*}}/\sqrt{\sigma}$\\
\hline
\multirow{3}{*}{\makecell{$\bullet$ Chiral limit\\$\bullet$ $b=0.36$}} & \makecell{This\\study} & 1.528(58) & 2.437(92) & 2.99(14) & 3.00(15) & 3.09(21) & 4.03(17) \\
\cline{2-8}
& \makecell{Ref.~\cite{Bali:2013kia}\\$\Nf=0$} & 1.5382(65) & 2.401(31) & 2.860(21) & 2.901(23) & 3.392(57) & 3.696(54)\\
\hline
\end{tabular}
\end{center}
\begin{center}
\begin{tabular}{|c|c|c|c|c|c|}
\cline{3-6}
\multicolumn{2}{c|}{} & $m^{\chir}_\rho/\sqrt{\sigma}$ & $m^{\chir}_{a_0}/\sqrt{\sigma}$ & $m^{\chir}_{a_1}/\sqrt{\sigma}$ & $m^{\chir}_{b_1}/\sqrt{\sigma}$ \\
\hline
\multirow{3}{*}{\makecell{$\bullet$ Chiral limit\\$\bullet$ Cont.~limit}} & \makecell{This\\study} & 1.783(47) & 2.130(87) & 3.186(97) & 3.22(12)\\
\cline{2-6}
& \makecell{Ref.~\cite{Castagnini:2015ejr}\\$\Nf=0$} & 1.687(24) & 1.81(17) & 2.93(11) & 2.97(13) \\
\hline
\end{tabular}
\end{center}
\begin{center}
\begin{tabular}{|c|c|c|c|c|}
\cline{3-5}
\multicolumn{2}{c|}{} & $m_\rho/\sqrt{\sigma}$ & $m_{a_0}/\sqrt{\sigma}$ & $m_{a_1}/\sqrt{\sigma}$ \\
\hline
\multirow{3}{*}{$m_\pi/\sqrt{\sigma}=1$} & \makecell{This study\\$a\sqrt{\sigma}=0$} & 1.973(44) & 2.489(64) & 3.344(91)\\
\cline{2-5}
& \makecell{Ref.~\cite{Baeza-Ballesteros:2025iee} ($\Nf=4$)\\$a/\sqrt{\sigma}\simeq 0.211$} &  2.2(3) & 2.1(6) & 3.4(3)\\
\hline
\end{tabular}
\end{center}
\caption{Comparison of our TEK findings with other available large-$N$ results obtained with standard methods ($N=\infty$ extrapolation from finite-$N$ results). See the text for more details.}
\label{tab:comparison_literature}
\end{table}

Our results are compatible with, and largely improve on, our previous TEK determinations in Ref.~\cite{Perez:2020vbn}, as it can be seen from the comparison shown in Fig.~\ref{fig:meson_masses} between the present (starred points in $m_\pi=0$) and the older (crossed points in $m_\pi=0$) chiral determinations. Our results are also in good agreement with previous determinations reported in the literature within at most $\sim 1.5$ standard deviations. This comparison is summarized in Tab.~\ref{tab:comparison_literature} and was performed as follows:
\vspace{-0.3\baselineskip}
\begin{itemize}
\item Comparison with Ref.~\cite{Bali:2013kia}.\\
This paper presents the masses of the $\rho$, $a_0$, $a_1$ and $b_1$ meson obtained from the large-$N$ extrapolation of quenched $\Nf=0$ QCD simulations with $2\le N \le 17$, at vanishing quark mass and for a fixed lattice spacing $a\sqrt{\sigma} \approx 0.2093$. This lattice spacing, for $N=\infty$, corresponds to $b\simeq 0.36$. Indeed, we find $a\sqrt{\sigma} = 0.2058(25)$ for this value of the bare coupling. Since the authors employ the same Wilson discretization also adopted in this study, we can directly compare the findings of Ref.~\cite{Bali:2013kia} with the results of a chiral extrapolation of our data at fixed $b=0.36$ assuming a linear behavior in $m_\pi^2$. In all cases, we find agreement with our results within less than two standard deviations at most.
\item Comparison with Ref.~\cite{Castagnini:2015ejr}.\\
In this thesis, the author considers further quenched ensembles in addition to the ones used in Ref.~\cite{Bali:2013kia}, in particular they consider 3 more values of the lattice spacing for just a few values of $N$ ($N=4,5,7$), and perform a global large-$N$-continuum best fit of all available data. The author quotes results in the chiral limit, which therefore can be compared with our chiral-continuum extrapolations. Also in this case, we always found agreement with our results within less than two standard deviations at most.
\item Comparison with Ref.~\cite{Baeza-Ballesteros:2025iee}.\\
This paper presents the masses of the $\rho$, $a_0$ and $a_1$ mesons obtained from the large-$N$ extrapolation of dynamical $\Nf=4$ QCD simulations with $3\le N\le6$. These results are obtained for a fixed value of the pion mass $m_\pi/\sqrt{\sigma}\approx 1$, and for a fixed value of the lattice spacing $(a/\sqrt{8t_0}) \approx 0.184$, which converted in string tension units using Eq.~\eqref{eq:conv_t0_sigma} yields $a\sqrt{\sigma}\approx 0.211$. This value of the lattice spacing roughly corresponds to the one of our $b=0.360$ ensemble, $a\sqrt{\sigma} = 0.2058(25)$. However, the authors employ an improved lattice discretization of both the gluonic and the fermionic actions, thus the comparison at fixed lattice spacing is not necessarily meaningful, as lattice artifacts affecting the two calculations will be in general completely different. Since improved actions are expected to lead to smaller lattice artifacts compared to unimproved ones, we will compare the results of Ref.~\cite{Baeza-Ballesteros:2025iee} with our continuum extrapolated ones at $m_\pi/\sqrt{\sigma} = 1$. Also in this case, we found agreement with our results within less than one standard deviation in all cases.
\end{itemize}
Let us also mention that our result for the large-$N$ $\rho$ mass in the chiral limit, $m_\rho^{\chir}/\sqrt{\sigma}=1.78(5)$, is also in good agreement with the phenomenological determination obtained in Ref.~\cite{Ledwig:2014cla} from the minimal hadronic resonance ansatz, $m_\rho^{\chir}/\sqrt{\sigma}\simeq 1.710(7)$.

\begin{figure}[!t]
\centering
\includegraphics[scale=0.28]{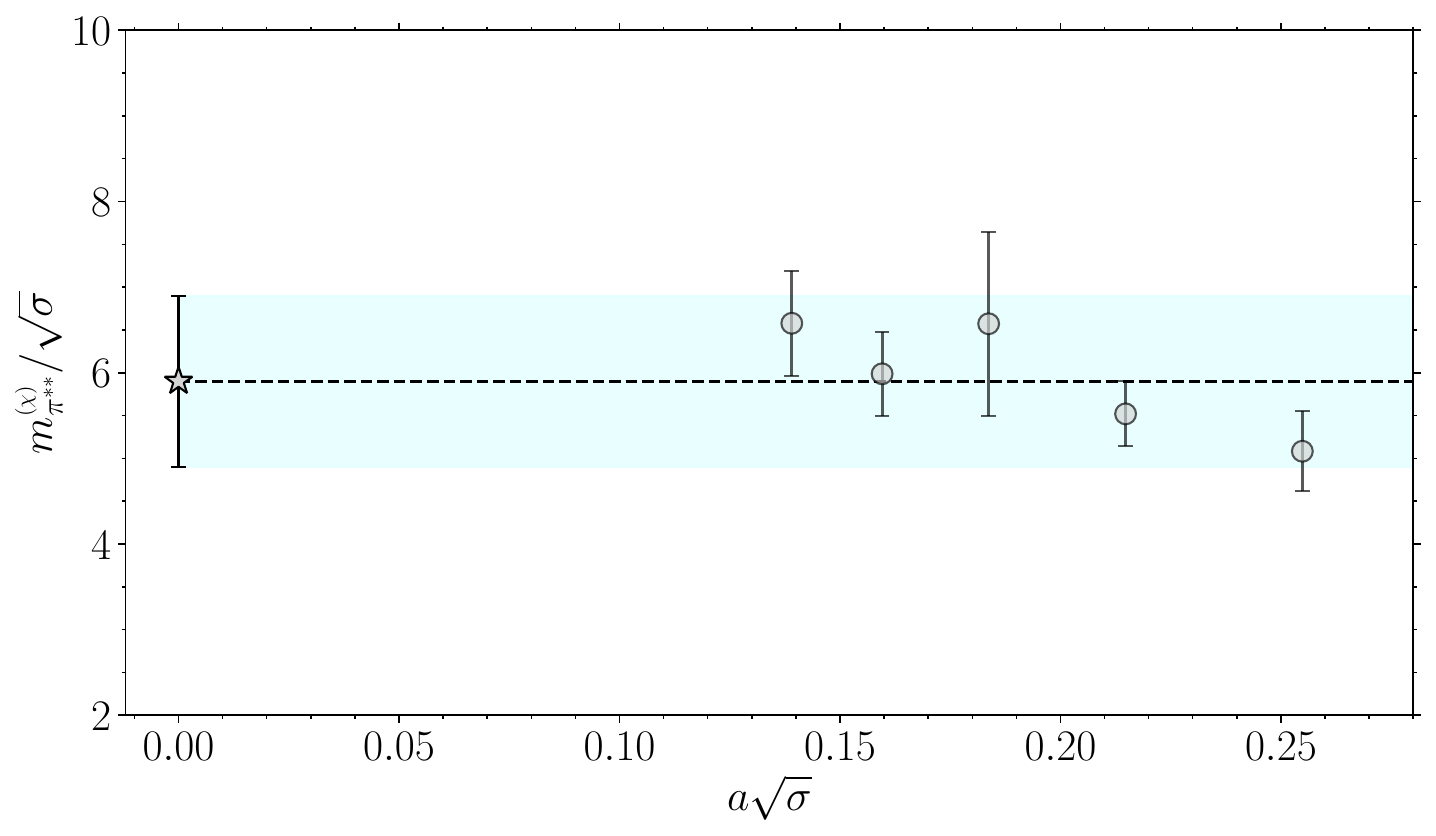}
\includegraphics[scale=0.28]{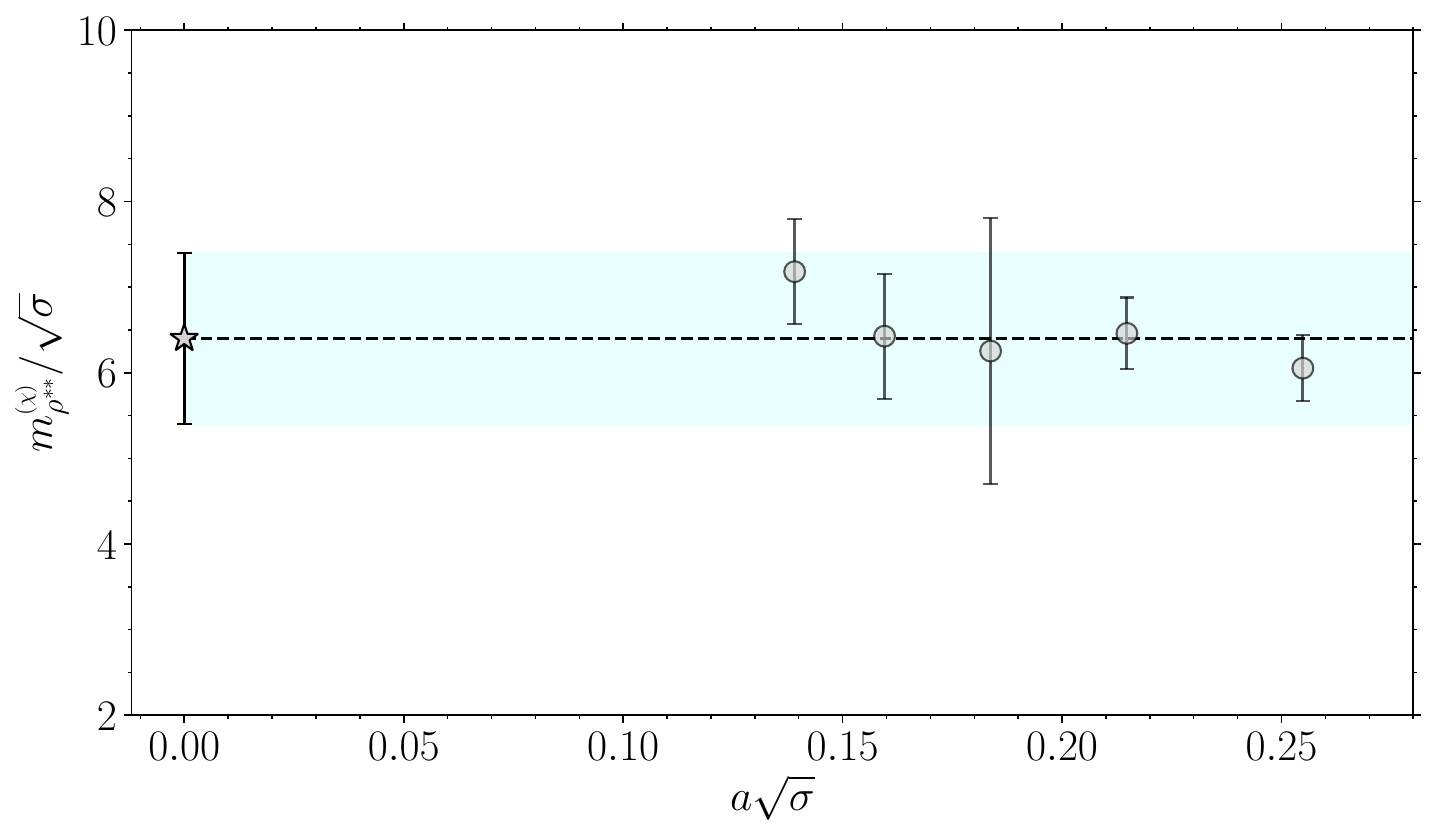}
\caption{In these figures we display our best determination of the final continuum results for the mass of the $\pi^{**}$ (left panel) and $\rho^{**}$ (right panel) mesons in the chiral limit, see the text for more details.}
\label{fig:exc_2_cont_lim}
\end{figure}

Finally, let us discuss our results for the $\pi^{**}$ and $\rho^{**}$ masses. In these two cases, we opted for a more conservative analysis of our data, as these masses have much larger error bars and are pretty large in lattice units (i.e., $am\sim 0.9-1.0$). Thus, we decided to give our conservative best estimate of the chiral limit of these masses as follows: first, we perform a chiral extrapolation of our data at fixed $b$; then, we take a conservative error band enveloping the chirally-extrapolated data as a function of the lattice spacing, see Fig.~\ref{fig:exc_2_cont_lim}. Since chiral extrapolations at fixed $b$ are practically flat within our statistical errors in all cases, the slopes as a function of $m_\pi^2/\sigma$ turn out always to be compatible with zero within errors for $\pi^{**}$ and $\rho^{**}$, thus we do not quote them here.\\
In the end, for the second excited states we report the following final  results for the masses in the chiral limit:
\beq
\frac{1}{\sqrt{\sigma}}m_{\pi^{**}}^{\chir} = 5.9(1.0), \qquad \qquad \frac{1}{\sqrt{\sigma}}m_{\rho^{**}}^{\chir} = 6.4(1.0).
\eeq

\FloatBarrier

\subsection{Meson spectrum tower and radial Regge trajectories}

\begin{figure}[!t]
\centering
\includegraphics[scale=0.48]{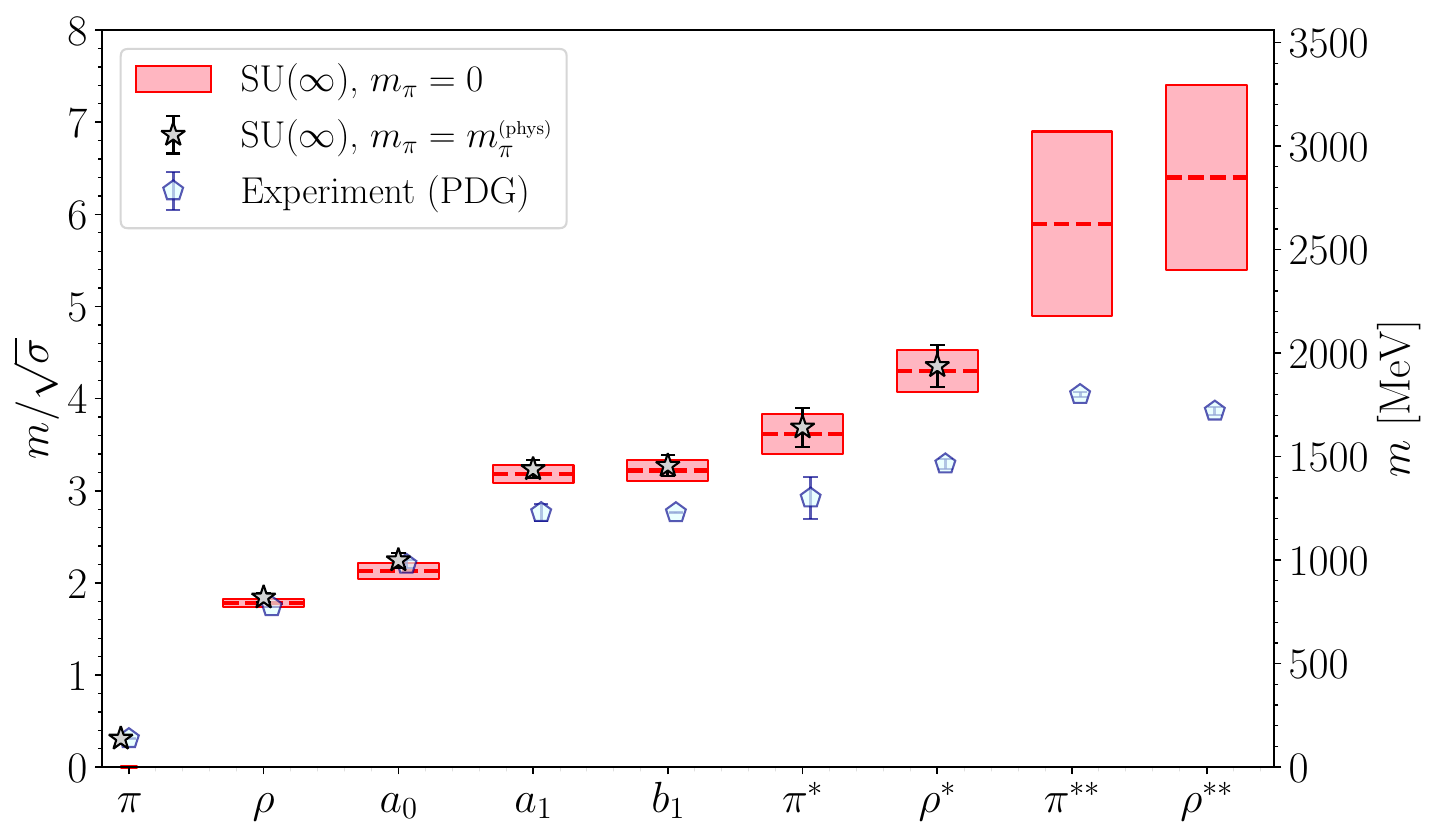}
\caption{Meson state tower at $N=\infty$, both in the chiral limit and at the physical point, compared to the experimental results taken from the PDG. In this plot we have assumed $\sqrt{\sigma}=445~\mathrm{MeV}$ and $m_\pi=138.04~\mathrm{MeV}$ at the physical point.}
\label{fig:MASTERFIG_meson}
\end{figure}

We conclude our discussion on the meson spectrum by presenting the large-$N$ meson state tower in Fig.~\ref{fig:MASTERFIG_meson}. Apart from the large-$N$ value of meson masses in the chiral limit $m_\pi=0$, we have also presented our results at the ``physical'' point, which we define as:
\beq\label{eq:physpoint_def}
\frac{m_\pi^{\scriptscriptstyle{(\rm phys)}}}{\sqrt{\sigma}} = 0.3102, \qquad (\text{physical point}).
\eeq
This definition stems from the combination of the physical value of the string tension~\cite{Bulava:2024jpj},
\beq\label{eq:string_tension_MeV}
\sqrt{\sigma} = 445~\mathrm{MeV}, \qquad (\text{physical point}),
\eeq
with the average of the masses of the charged and neutral pions~\cite{ParticleDataGroup:2024cfk}:
\beq
m_\pi^{\scriptscriptstyle{(\rm phys)}} = 138.04~\mathrm{MeV}, \qquad (\text{physical point}).
\eeq
The value of the string tension in~\eqref{eq:string_tension_MeV} is also used to present our meson masses results in MeV in Fig.~\ref{fig:MASTERFIG_meson} (see the right vertical axis). Clearly, while any dimensionless ratio extracted from lattice calculations is a rigorous non-perturbative theoretical prediction, any definition of the ``physical point'' for an unphysical theory such as large-$N$ QCD is arbitrary. Thus, this conversion should be regarded as just one possible prescription to convert our results into ``MeV'' units. This procedure is nonetheless interesting because it allows to compare our large-$N$ data with the experimental values of meson masses reported in the PDG~\cite{ParticleDataGroup:2024cfk}, which can give a rough idea of the magnitude of sub-leading $1/N$ corrections that should be expected in standard $N=3$ QCD. Experimental results for meson masses are also shown in Fig.~\ref{fig:MASTERFIG_meson}, both in physical MeV units and in string tension units, again using the value in Eq.~\eqref{eq:string_tension_MeV}. All data appearing in Fig.~\ref{fig:MASTERFIG_meson} can be found in Tab.~\ref{tab:comp_Ninf_EXP}.

\begin{table}[!t]
\begin{center}
\begin{tabular}{|c|c|c|c|c|}
\hline
Meson A &  $m_{\A}^{\chir}/\sqrt{\sigma}$ & $m_\A(m^{\scriptscriptstyle{(\rm phys)}}_\pi)/\sqrt{\sigma}$ & $m^{\scriptscriptstyle{(\rm exp)}}_{\A}/\sqrt{\sigma}$ & PDG meson name\\
\hline
$\pi$       & 0         & 0.3102     & 0.3102(63) & $\pi(140)$ \\
$\rho$      & 1.783(47) & 1.842(46)  & 1.7422(5)  & $\rho(770)$\\
$a_0$       & 2.130(87) & 2.246(78)  & 2.202(45)  & $a_0(980)$\\
$a_1$       & 3.186(97) & 3.235(95)  & 2.764(90)  & $a_1(1260)$\\
$b_1$       & 3.22(12)  & 3.27(11)   & 2.763(7)   & $b_1(1235)$\\
$\pi^{*}$   & 3.62(23)  & 3.69(21)   & 2.92(22)   & $\pi(1300)$\\
$\rho^{*}$  & 4.30(23)  & 4.35(23)   & 3.29(6)    & $\rho(1450)$\\
$\pi^{**}$  & 5.9(1.0)  &            & 4.04(2)    & $\pi(1800)$\\
$\rho^{**}$ & 6.4(1.0)  &            & 3.53(16)   & $\rho(1700)$\\
\hline
\end{tabular}
\end{center}
\caption{Collection of the $\pi, \pi^{*}, \pi^{**},\rho,\rho^{*}, \rho^{**}, a_0,a_1$ and $b_1$ large-$N$ masses both in the chiral limit and at the physical point, as defined in Eq.~\eqref{eq:physpoint_def}. We also report experimental measures of these meson masses as reported in the PDG~\cite{ParticleDataGroup:2024cfk}, which are expressed in units of the string tension using the result in Eq.~\eqref{eq:string_tension_MeV}.}
\label{tab:comp_Ninf_EXP}
\end{table}

\begin{figure}[!t]
\centering
\includegraphics[scale=0.4]{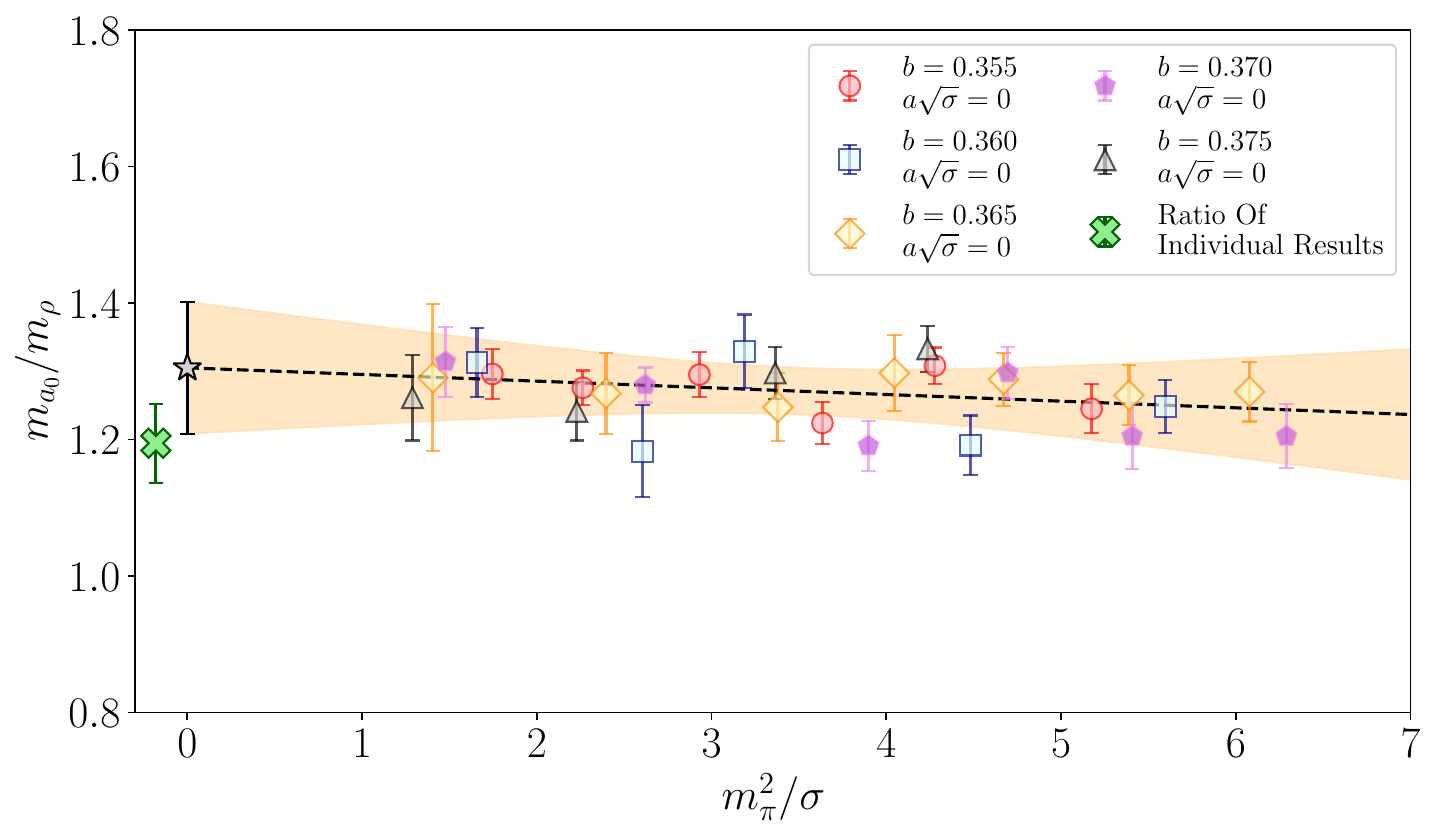}
\caption{Chiral-continuum extrapolation (starred point) of the $a_0$-$\rho$ mass ratio, compared to the determination obtained from the individual chiral-continuum extrapolation of $m_{a_0}$ and $m_{\rho}$ (crossed point). The direct determination yields $m_{a_0}^{\chir}/m_\rho^{\chir}=1.305(97)$ in the chiral limit, in agreement with the determination $m_{a_0}^{\chir}/m_\rho^{\chir} = 1.195(58)$ obtained from the individual chiral-continuum fits. Again, plotted points are lattice data to which we have subtracted lattice artifacts.}
\label{fig:rho_a0_massratio}
\end{figure}

Concerning ground state masses, we observe that the $a_1$ and $b_1$ mesons, which are found experimentally to be degenerate, remain degenerate even for $N=\infty$. Note that this is true not only in the chiral limit, but also for every value of the pion mass in the explored range, as the coefficients of the $m_\pi^2/\sigma$ term are compatible between these two states, cf.~Eqs.~\eqref{eq:final_a1_vs_mpi} and~\eqref{eq:final_b1_vs_mpi}. Moreover, we also observe that both the $a_1$ and the $b_1$ large-$N$ masses sensibly differ from their experimental values. On the other hand, the $\rho$ and $a_0$ large-$N$ masses sit much closer to the experimental data, albeit being both slightly heavier. This fact is interesting with respect to the possibility advocated in Refs.~\cite{Pelaez:2006nj,Nieves:2009ez,Nieves:2011gb} that the $a_0$ and $f_0(600)$ mesons (expected to become degenerate at $N=\infty$) could become degenerate with the $\rho$ too in the large-$N$ limit. Our data do not seem to support this possibility, as we find:
\beq
\frac{m_{a_0}^{\chir}}{m_{\rho}^{\chir}} = 1.195(58).
\eeq
Similar conclusions are drawn if, instead of performing separate chiral-continuum extrapolations of $m_\rho$ and $m_{a_0}$, one performs directly such extrapolation on the ratio $m_{a_0}/m_{\rho}$, which has the advantage of completely bypassing the necessity for scale setting. The outcome of this calculation is shown in Fig.~\ref{fig:rho_a0_massratio}, and yields the compatible result:
\beq
\frac{m_{a_0}^{\chir}}{m_{\rho}^{\chir}} = 1.305(97).
\eeq

Concerning the excited $\pi$ and $\rho$ masses, we observe that they also turn out to be somewhat larger than the experimental values, and that these differences grow as one climbs up the spectrum tower, thus confirming the indications already drawn from the analysis of ground state masses that $1/N$ corrections seem to be larger for heavier mesons. Curiously, the $\pi^{**}$ and $\rho^{**}$ seem to be approximately degenerate at large-$N$, similarly to what is observed experimentally for the $\pi(1800)$ and the $\rho(1700)$.\\
Given that in the $\pi$ and $\rho$ channels we were able to compute the mass of two excited states, it is interesting to check whether they can be described by radial Regge trajectories. For experimental masses, linear trajectories in the radial quantum number $n$, labeling the excited state number, are observed with parallel slopes~\cite{Anisovich:2000kxa}. Concerning large-$N$ QCD, there are several model calculations giving similar predictions, see, e.g., Refs.~\cite{Kaidalov:2001db,Afonin:2009xi,Afonin:2014nya,Afonin:2016wie}.\footnote{Recently, non-perturbative bootstrap results also indicate the existence of linear Regge trajectories in the total angular momentum $J$ in large-$N$ QCD, see Ref.~\cite{Albert:2023seb} (see also~\cite{Guerrieri:2024jkn} for an analytic S-matrix bootstrap study in $N=3$ QCD).}

\begin{figure}[!t]
\centering
\includegraphics[scale=0.4]{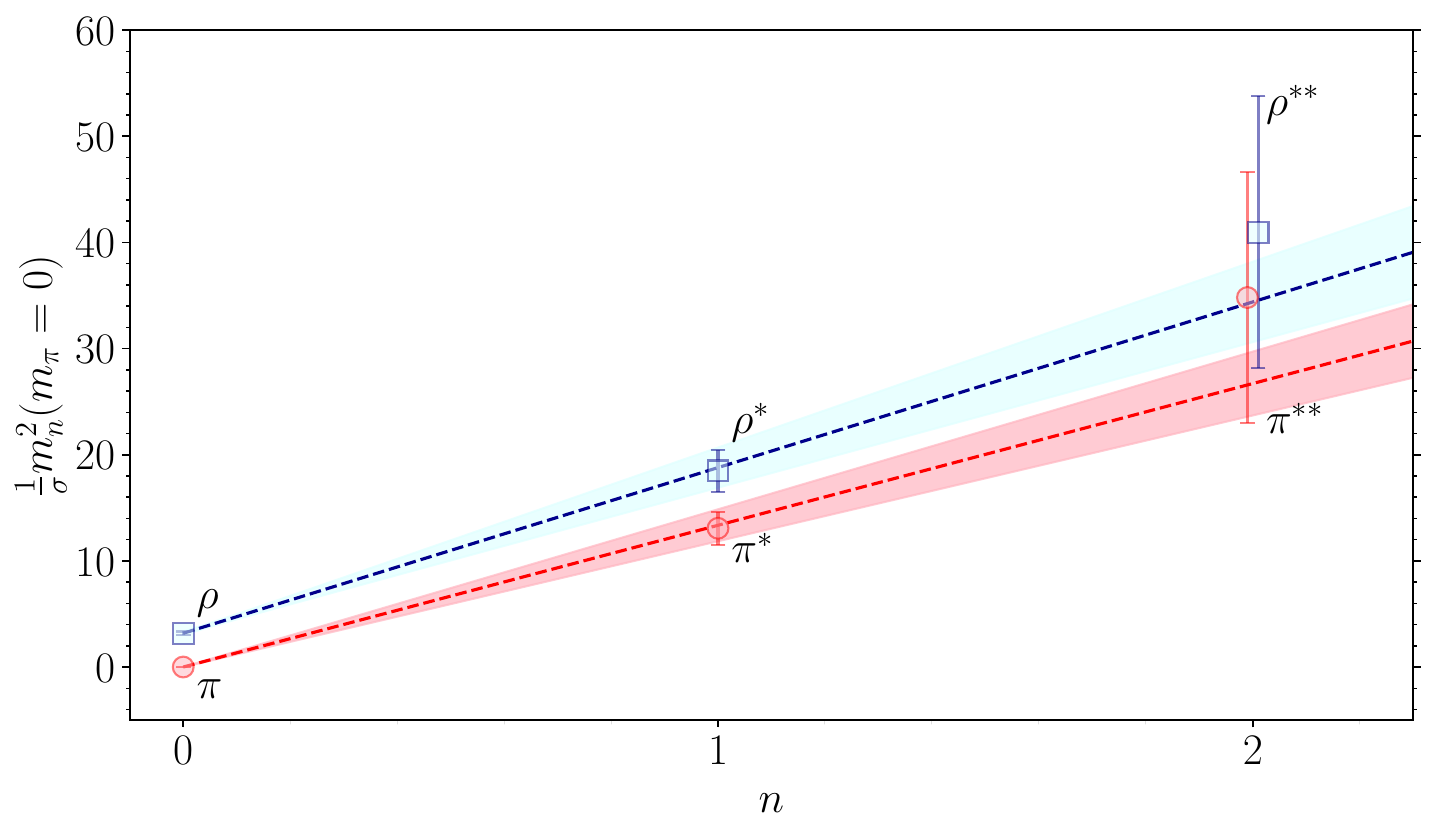}
\caption{Parallel radial linear Regge trajectories in the $\pi$ and $\rho$ channels in the chiral limit.}
\label{fig:Regge_traj}
\end{figure}

Plotting our data for the masses of the first three states found in the $\pi$ and $\rho$ channels in the chiral limit as a function of the radial quantum number $n$ ($n=0$ stands for the ground state, while $n=1,2$ for the first/second excited respectively), we find that a linear behavior in $n$:
\beq
\frac{1}{\sigma}m_n^2(m_\pi=0) = C + \frac{\mu^2_{\scriptscriptstyle{r}}}{\sigma} n,
\eeq
gives a very good description of our results. Strikingly, we observe parallel radial trajectory slopes $\mu_{\scriptscriptstyle{r}}$ in the $\pi$ and the $\rho$ channels, see Fig.~\ref{fig:Regge_traj}:
\beq
\frac{\mu_{\scriptscriptstyle{r}}}{\sqrt{\sigma}} = 3.65(21), \quad \qquad (\pi \text{ radial trajectory}),\\
\frac{\mu_{\scriptscriptstyle{r}}}{\sqrt{\sigma}} = 3.95(24), \quad \qquad (\rho \text{ radial trajectory}),
\eeq
thus supporting the hypothesized universal nature of Regge trajectories slope. These numbers can be compared with the slope found from a linear fit in $n$ to experimental $\pi$ and $\rho$ mass data: $\mu_{\scriptscriptstyle{r}}/\sqrt{\sigma} \approx 2.65$~\cite{Anisovich:2000kxa} (see also Ref.~\cite{Masjuan:2012gc}). Moreover, our results are in very good agreement with the large-$N$ model calculation of Ref.~\cite{Dubin:1994vn}, predicting the universal linear slope:
\beq
\frac{\mu_{\scriptscriptstyle{r}}}{\sqrt{\sigma}} = \sqrt{4\pi} \simeq 3.55.
\eeq
Clearly, given that we only have three points at our disposal, and that our data for the second excited states have larger error bars, also other functional forms could equally give good fits, thus more excited masses and smaller errors would be needed to check the possible presence of non-linear or non-universal features in large-$N$ radial Regge trajectories.

\subsection{Results for QCD low-energy constants}

This section is devoted to the discussion of our results for the QCD LECs, and is further subdivided into subsections, each one dedicated to, respectively, our calculations of $B=\Sigma/F_\pi^2$, $\Sigma$, $F_\pi$ and $\bar{\ell}_4$.

Since the LECs $B$ and $\Sigma$ are scheme- and scale-dependent, let us stress that in the following we will always quote our renormalized results $B_\R$ and $\Sigma_\R$ in the $\overline{\mathrm{MS}}$ scheme at the customary renormalization scale $\mu=2$ GeV. In order to compare our $N=\infty$ results for $B_\R$ and $\Sigma_\R$ with the finite-$N$ ones of Ref.~\cite{DeGrand:2023hzz} and compute their sub-leading $1/N$ corrections, we will adopt the same definition of GeV units used in that study to achieve the same renormalization scale. In particular, the authors of~\cite{DeGrand:2023hzz} convert their dimensionless ratios into physical units assuming
\beq
\sqrt{t_0} = 0.15~\mathrm{fm},
\eeq
with $t_0$ the scale in Eq.~\eqref{eq:t0_Ninf_Giusti}. This choice leads, for $\mu=2$ GeV, to:
\beq
\mu\sqrt{8t_0} = 4.30,
\eeq
\beq
\label{eq:2GeVdef_sigma}
\implies \frac{\mu}{\sqrt{\sigma}} = 3.75,
\eeq
where to obtain the last equation we used the result for $\sqrt{8t_0\sigma}$ in Eq.~\eqref{eq:conv_t0_sigma}.

\begin{table}[!t]
\begin{center}
\begin{tabular}{|c|c|c|}
\hline
$b$   & $\ZS$ & $B_\R/\sqrt{\sigma}$ \\
\hline
0.355 & 0.7287(24) & 3.860(80) \\
0.360 & 0.7226(67) & 4.22(12)  \\
0.365 & 0.732(11)  & 4.35(15)  \\
0.370 & 0.743(14)  & 4.52(13)  \\
0.375 & 0.755(18)  & 4.61(17)  \\
\hline
\end{tabular}
\end{center}
\caption{Non-perturbative large-$N$ determinations of $\ZS$ in the $\overline{\mathrm{MS}}$ scheme at $\mu=2~\mathrm{GeV}$ used in this study. These determination come from Ref.~\cite{Castagnini:2015ejr}, and have been suitably extrapolated towards $N=\infty$, interpolated at our values of $(a\sqrt{\sigma})(b)$, and run from the lattice scale to the conventional scale so as to match the definition of $\mu=2~\mathrm{GeV}$ in Eq.~\eqref{eq:2GeVdef_sigma}. These data are used to obtain the renormalized QCD LEC $B_\R = \Sigma_\R/F_\pi^2$ from the bare data in Tab.~\ref{tab:kappac_res}.}
\label{tab:ZS_and_B_res}
\end{table}

\subsubsection{The pion mass slope \texorpdfstring{$B=\Sigma/F_\pi^2$}{B=Sigma/Fpi2}}

The bare pion mass slope $B=B_\R/\ZS$ has been already extracted from the chiral fits of $m_\pi$ studied in Sec.~\ref{sec:kappac_determination} to define the chiral point, see Tab.~\ref{tab:kappac_res}. These bare values need to be renormalized using the renormalization constant $\ZS$. A non-perturbative determination of $\ZS$ from the TEK model would require a substantial computational effort, which is beyond the scope of this paper. As an alternative, in this study we will use the non-perturbative large-$N$ determinations of Ref.~\cite{Castagnini:2015ejr}, which have been extrapolated towards $N=\infty$ and interpolated at our values of the lattice spacing, following the strategy used for our previous determination of the chiral condensate in Ref.~\cite{Bonanno:2023ypf}. Moreover, since in Ref.~\cite{Castagnini:2015ejr} the author defines $\mu=2$ GeV as $\mu/\sqrt{\sigma}=4.59$, we have also re-performed the running of $\ZS(a\mu=1)$ up to $\mu=2$ GeV as defined in Eq.~\eqref{eq:2GeVdef_sigma}. These results for $\ZS$ are reported in Tab.~\ref{tab:ZS_and_B_res}; for the sake of completeness, further details on their determination can be found in App.~\ref{app:ZS}.

\begin{figure}[!t]
\centering
\includegraphics[scale=0.45]{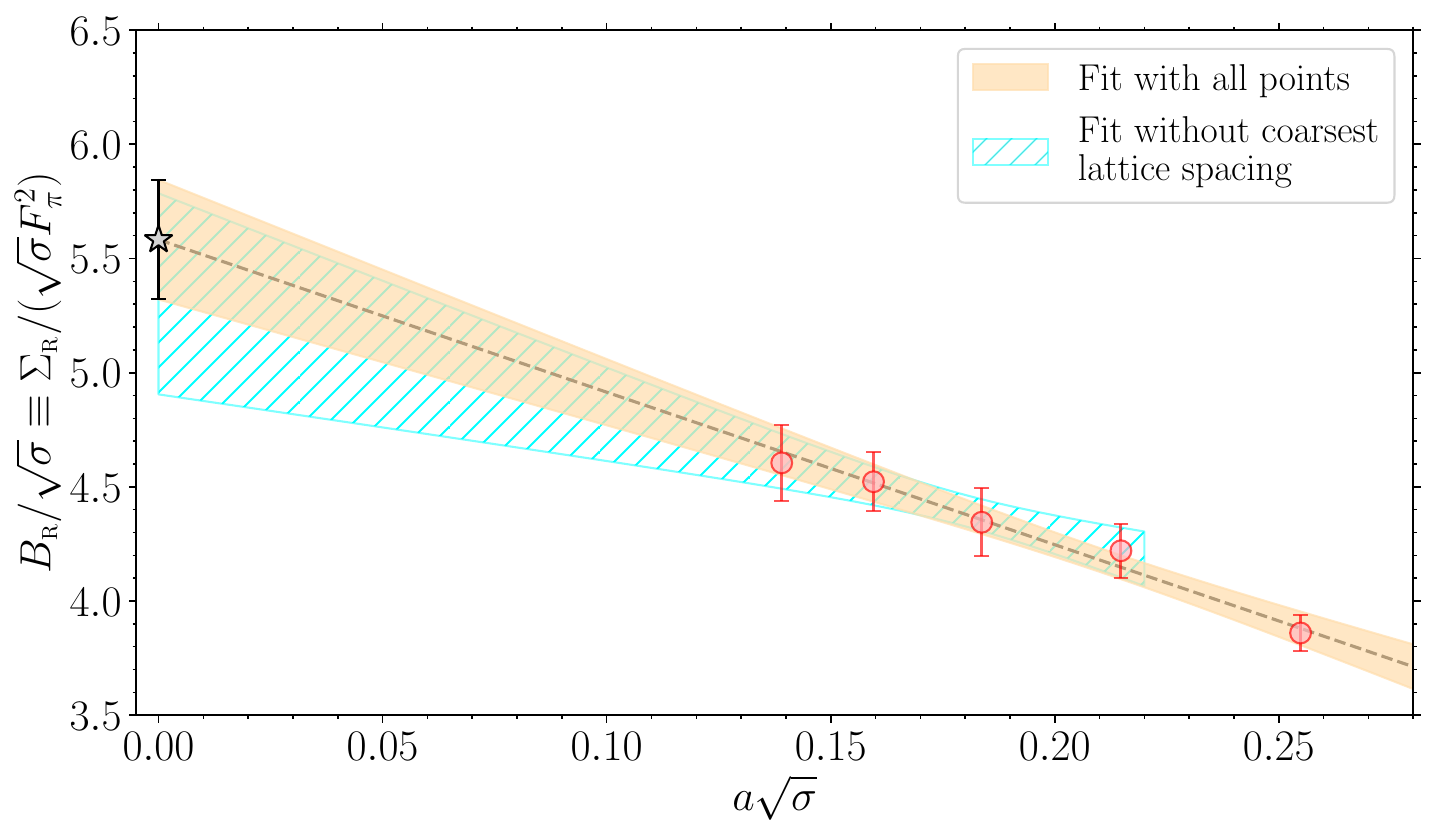}
\caption{Continuum extrapolation of the QCD renormalized LEC $B_\R = \Sigma_\R/F_\pi^2$ in the $\overline{\mathrm{MS}}$ scheme at $\mu=2~\mathrm{GeV}$ in units of $\sqrt{\sigma}$.}
\label{fig:B_contlim}
\end{figure}

Once our bare determinations of the LEC $B$ are renormalized, we are able to extrapolate our results towards the continuum limit assuming $\mathcal{O}(a)$ lattice artifacts. As already done for meson masses, we express our results in units of $\sqrt{\sigma}$ passing through our determinations of the $\sqrt{8t_1}$ scale in order to obtain smaller errors and more constrained fits, cf.~Eq.~\eqref{eq:conv_sigma_through_t1}. We find that $B_\R$ is affected by significant lattice artifacts, cf.~Tab.~\ref{tab:ZS_and_B_res} and Fig.~\ref{fig:B_contlim}, thus taking the continuum limit is crucial to obtain a reliable final result. Despite of this, we observe perfect compatibility between the result obtained by extrapolating all data points, and the one obtained using all data points but the coarsest lattice spacing one. Thus, we use the former to quote our final continuum result:
\beq
\frac{B_\R}{\sqrt{\sigma}} = \frac{\Sigma_\R}{F_\pi^2\sqrt{\sigma}} = 5.58(26).
\eeq
Excluding the coarsest lattice spacing, one instead would find $B_\R/\sqrt{\sigma}=5.35(44)$.

\subsubsection{The chiral condensate \texorpdfstring{$\Sigma$}{Sigma}}

The chiral condensate alone is extracted from Dirac eigenvalues via the mode number \emph{à la} Giusti--L\"uscher~\cite{Giusti:2008vb}. The eigenvectors obtained from the resolution of the same eigenproblems also allows to compute the RGI ratio of renormalization constants $\ZP/\ZS$~\cite{Giusti:2008vb}, which is needed to renormalize $\Sigma$, as explained in Sec.~\ref{sec:LEC_lattice_setup}.

For the calculation of Dirac spectra we considered values of $N$ satisfying $\ell\sqrt{\sigma} = a L \sqrt{\sigma} = a \sqrt{N} \sqrt{\sigma} \gtrsim 4$ (corresponding to $\ell\sim$ 1.8 -- 2 fm assuming $\sqrt{\sigma}=$ 445 MeV~\cite{Bulava:2024jpj}), which is enough to contain finite-volume effects, as we will show. Concerning the values of the pion mass employed for the chiral extrapolations, our choices of $\kappa$ typically translate into $m_\pi\ell \gtrsim 4$. Note that, based on the general theoretical arguments of~\cite{Giusti:2008vb}, the chiral condensate extracted from the mode number is expected to suffer for exponentially small finite-size effects which are even more suppressed compared to those affecting, e.g., the pion mass extraction. Thus, we could push the calculation of Dirac spectra towards slightly smaller values of $m_\pi$ compared to the mass spectra one. The parameters employed for Dirac spectra calculations are found in Tab.~\ref{tab:Diracspectra_params} in Appendix~\ref{app:rawdata}.

\begin{figure}[!t]
\centering
\includegraphics[scale=0.295]{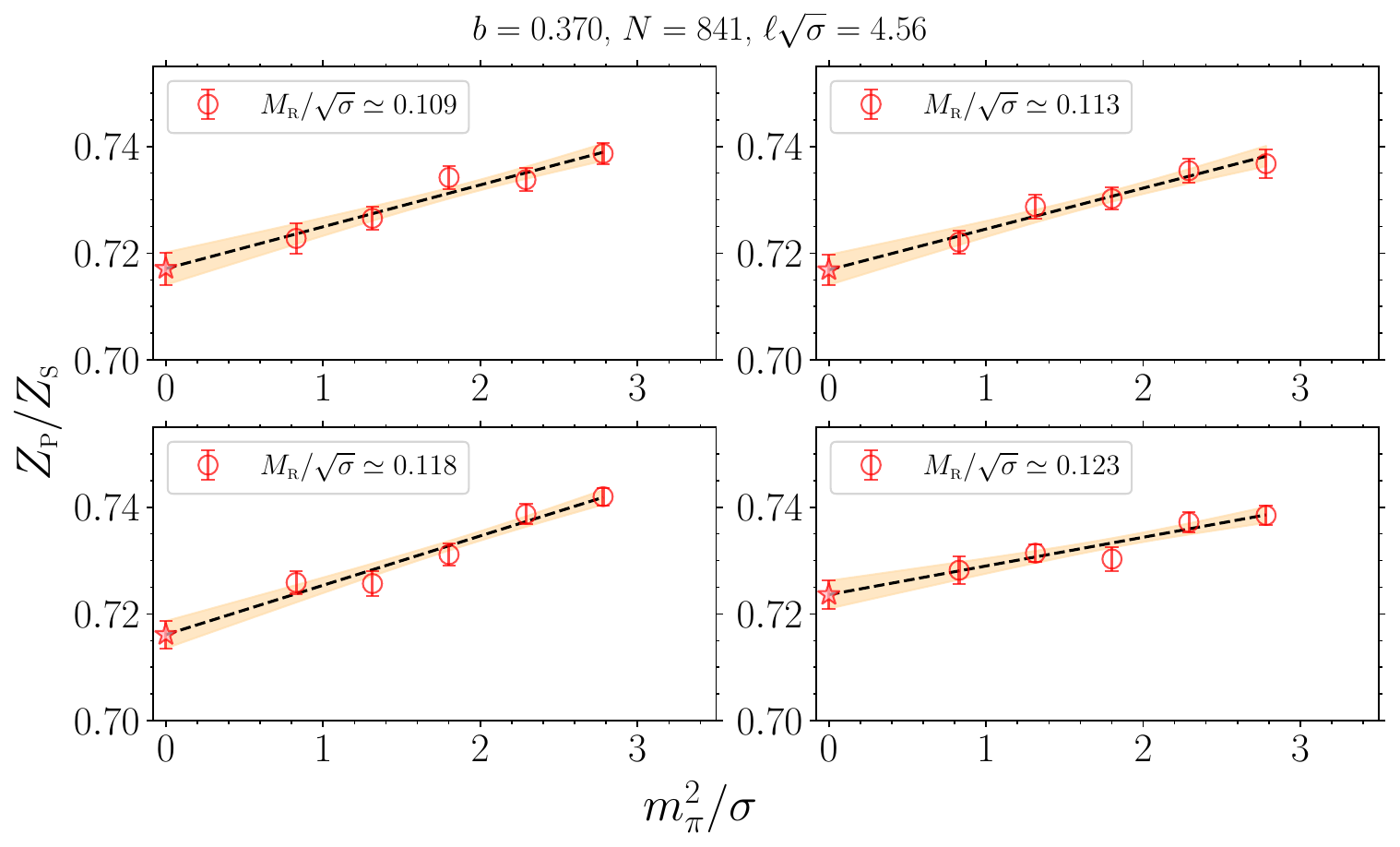}
\includegraphics[scale=0.29]{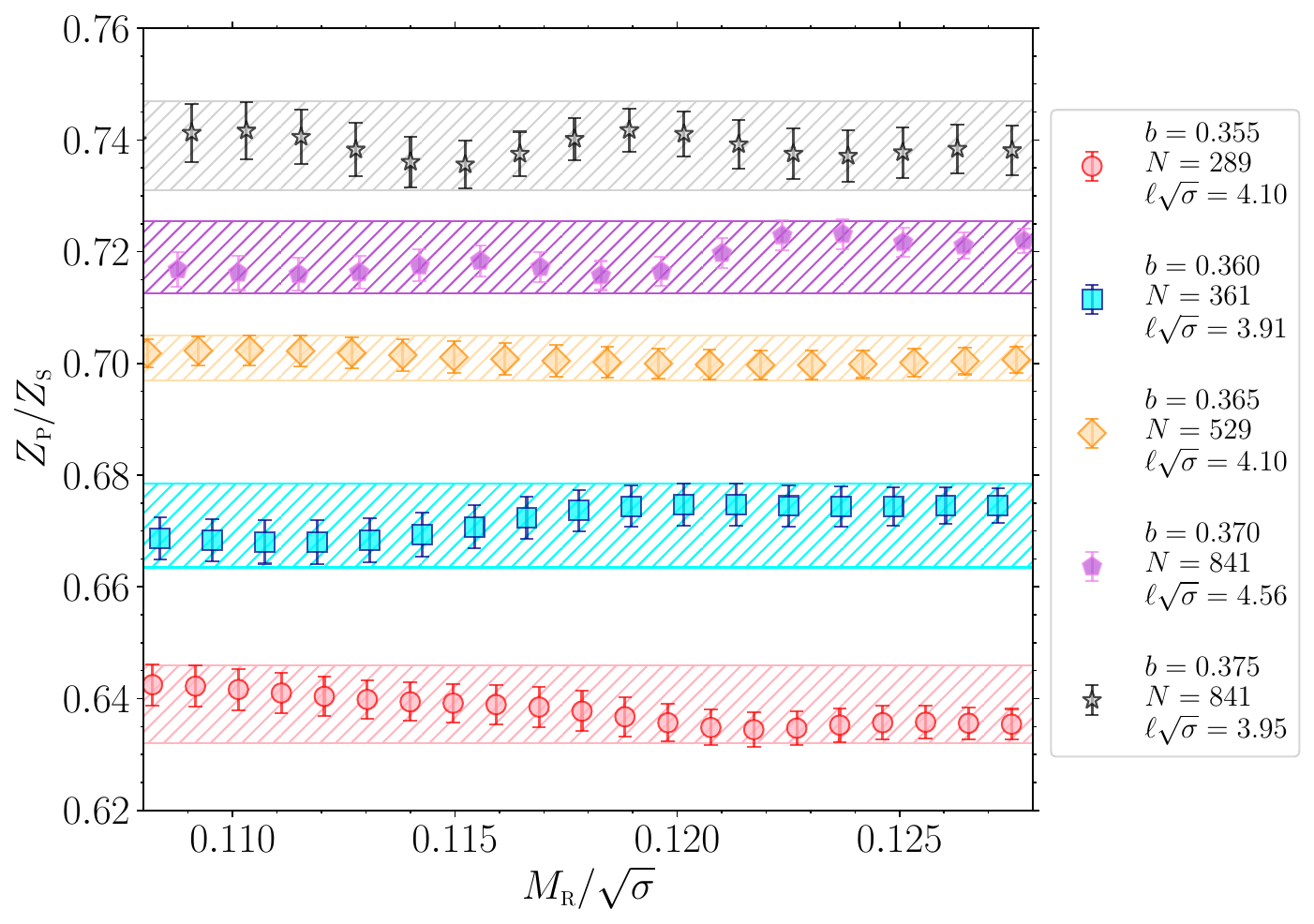}
\caption{Chiral-extrapolation of the spectral determination of $\ZP/\ZS$ for a few choices of the renormalized spectral threshold $M_\R$ (left panels). Plateaus of the chiral-extrapolated determinations of $\ZP/\ZS$ as a function of the spectral threshold (right panel). Shaded areas represent our final determinations, reported in Tab.~\ref{tab:ZPZS_determinations}.}
\label{fig:ZPZS}
\end{figure}

\begin{table}[!t]
\small
\begin{center}
\begin{tabular}{|c|c|c|c|c|c|}
\hline
$b$ & $\ZP/\ZS$ & $\ZP$ & $\ZA$ \\
\hline
0.355 & 0.6390(70) & 0.4656(53) & 0.759(10)\\
0.360 & 0.6710(75) & 0.4848(70) & 0.778(13)\\
0.365 & 0.7010(40) & 0.5128(81) & 0.831(12)\\
0.370 & 0.7190(65) & 0.535(11)  & 0.831(10)\\
0.375 & 0.7390(80) & 0.558(14)  & 0.824(15)\\
\hline
\end{tabular}
\end{center}
\caption{Final spectral determinations of the RGI ratio of renormalization constants $\ZP/\ZS$. For the reader's benefit, we also report the values of the renormalization constants $\ZP=\ZS \times (\ZP/\ZS)$ and $\ZA = (\ZS\ZA/\ZP)\times(\ZP/\ZS)$ obtained combining our spectral determinations of $\ZP/\ZS$ with other renormalization constants earlier reported in Tab.~\ref{tab:kappac_res} and Tab.~~\ref{tab:ZS_and_B_res}. Recall that $\ZP$, just like $\ZS$, is expressed in the $\overline{\mathrm{MS}}$ renormalization scheme at a renormalization scale $\mu=2~\mathrm{GeV}$.}
\label{tab:ZPZS_determinations}
\end{table}

Let us start from the calculation of $\ZP/\ZS$. The renormalization constants are obtained from a ratio of spectral sums, extrapolated to the chiral limit, according to Eq.~\eqref{eq:specsum_ZPZS}. When adopting this particular discretization, it is very important to check that the final result does not depend significantly on the chosen threshold $M$ to cut the spectral sums. In Fig.~\ref{fig:ZPZS} we show, for $N=841$ and $b=0.370$, the chiral extrapolations of $\ZP/\ZS$ obtained from spectral methods for a few values of the renormalized cut-off $M_\R/\sqrt{\sigma}$. In that same figure we also show the chiral-extrapolated values of $\ZP/\ZS$ as a function of $M_\R/\sqrt{\sigma}$. Remarkably, we observe not only that our results for $\ZP/\ZS$ exhibit a plateau as a function of $M_\R/\sqrt{\sigma}$ once extrapolated to vanishing pion mass, but also that for all values of $b$ these plateaus are found in the same range of values of $M_\R/\sqrt{\sigma}$. The final values for $\ZP/\ZS$, reported in Tab.~\ref{tab:ZPZS_determinations}, are determined by taking into account the observed fluctuations around the plateau, and are practically chosen so as to encompass all points in the plateau range, cf.~the shaded bands in the right panel of Fig.~\ref{fig:ZPZS}. Note that $\ZP/\ZS$ is smaller than 1 and grows towards 1 as the continuum limit is approached. This is the expected behavior due to the explicit chiral symmetry breaking of Wilson fermions vanishing in the continuum limit, where $\ZP/\ZS\underset{a \,\to\, 0}{\longrightarrow} 1$.

\begin{figure}[!t]
\centering
\includegraphics[scale=0.42]{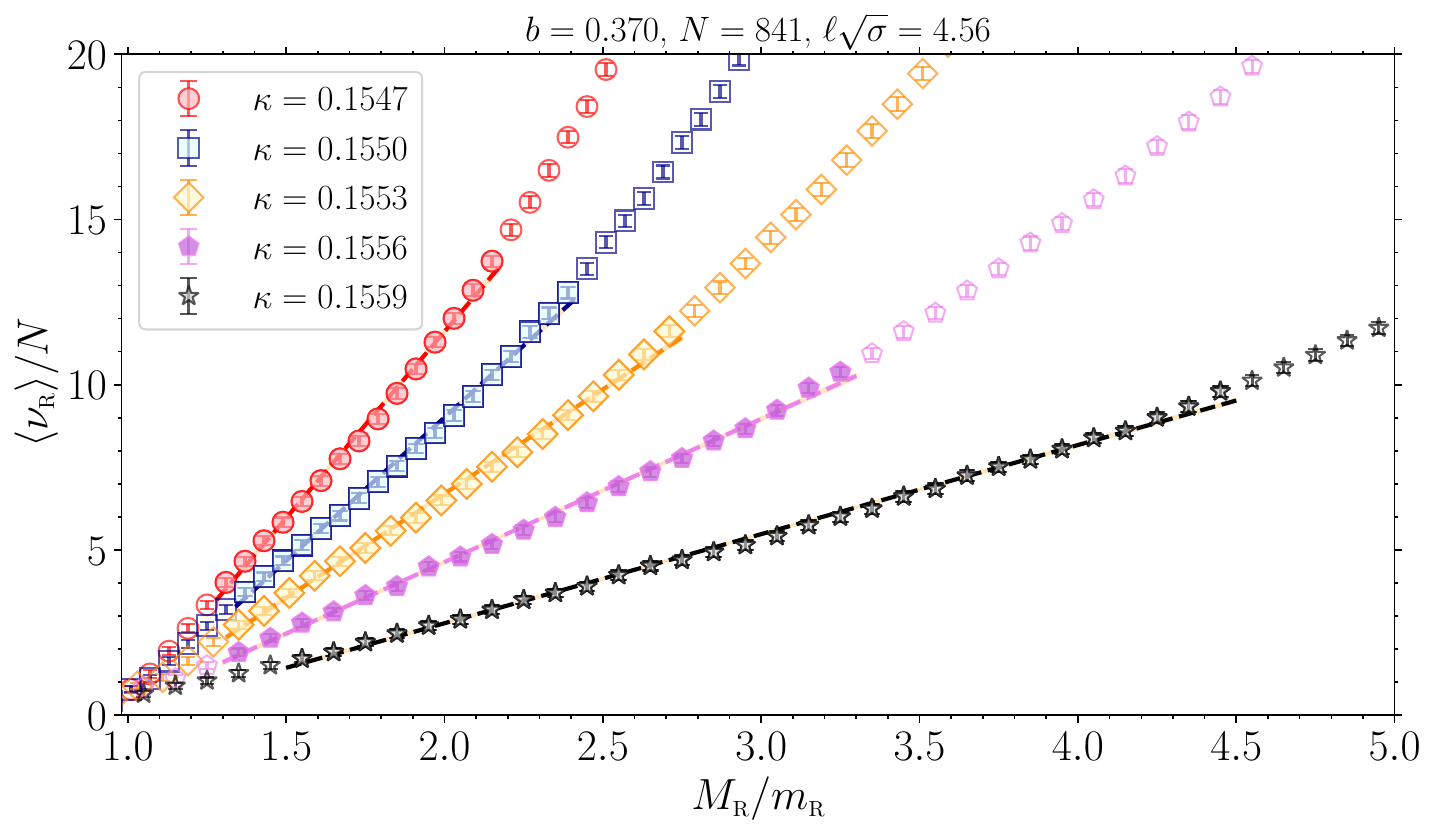}\\
\includegraphics[scale=0.42]{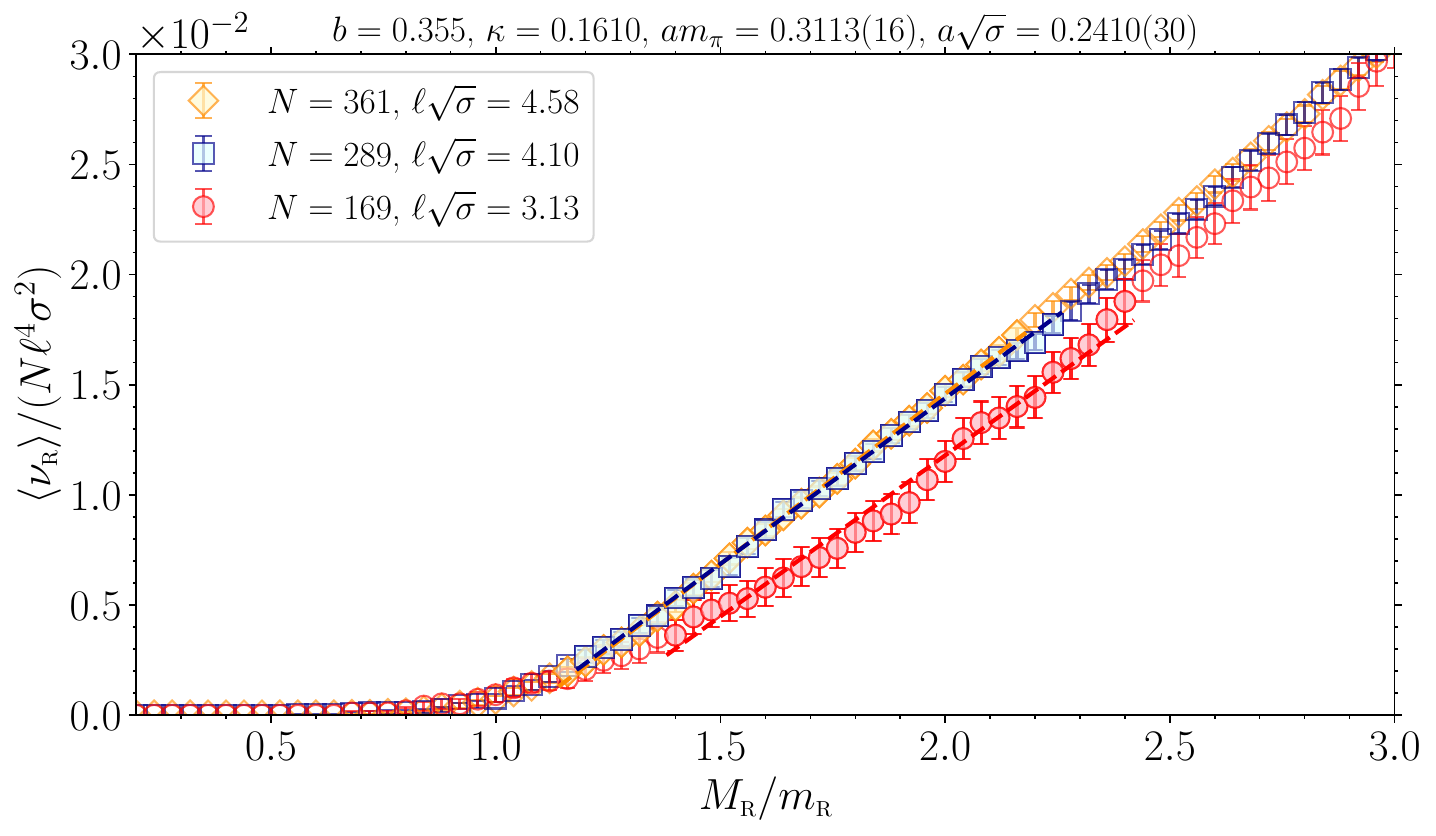}
\caption{Top panel: examples of linear fits to the mode number for the ensemble with $N=841$ and $b=0.370$ as a function of the Wilson hopping parameter $\kappa$. Bottom panel: finite-effective-volume study of the mode number slope at fixed $b=0.355$ and $\kappa=0.1610$ for 3 values of $N$.}
\label{fig:fit_mode_number_ex}
\end{figure}

Let us now move on to the determination of the effective chiral condensate $\Sigma_\R^{\effsup}/N$ from the mode number slope. In Fig.~\ref{fig:fit_mode_number_ex}, we exemplify the calculation by showing the linear fits to the mode number to extract its slope for $N=841$ and $b=0.370$, and for all the five values of the pion mass explored. The horizontal axis is expressed in terms of the renormalized ratio:
\beq
\frac{M_\R}{m_\R} = \frac{M}{\ZP m_\R}, \qquad \ZP m_\R = \left(\frac{\ZS\ZA}{\ZP}\right)\times\left(\frac{\ZP}{\ZS}\right)\times m_\PCAC,
\eeq
where the first factor giving $\ZP m_\R$ comes from Tab.~\ref{tab:kappac_res}, while the second from Tab.~\ref{tab:ZPZS_determinations}.

In that same figure, we also show linear fits to the mode number performed at fixed values of $b$ and $\kappa$ for a few values of $N$. As it can be seen, while finite-volume effects for the smaller value of $N$ cause a shift of the mode number towards larger values, the slope of the mode number close to $M_\R/m_\R$ is essentially unaffected, and turns out to be parallel to the one extracted from larger effective volumes, where instead finite-volume effects are negligible, as mode numbers (once divided for the effective volume) fall on top of each other. Thus, these calculations show that choosing $\ell\sqrt{\sigma}\gtrsim 4$ is a safe threshold to contain finite-volume effects in our computation of the mode number slope. All determinations of the bare effective condensate $\Sigma^{\effsup}_\R/\ZP$ from the mode number, done according to Eq.~\eqref{eq:effective_cond_def}, are collected in Tab.~\ref{tab:Diracspectra_params} in Appendix~\ref{app:rawdata}.

In order to give a final result for the chiral condensate, we perform a combined chiral-continuum limit assuming $\mathcal{O}(a)$ and $\mathcal{O}(m_\pi^2)$ corrections (as expected from general theoretical arguments~\cite{Giusti:2008vb}), in the same fashion as the meson masses analysis. This is the fit function that we employed:
\beq\label{eq:fitfunction_Sigmaeff_chircont}
\frac{\Sigma_\R^{\effsup}(a,m_\pi)}{N \sqrt{\sigma^3}} = \frac{\Sigma_\R}{N \sqrt{\sigma^3}} + C_0 \frac{m_\pi^2}{\sigma} + k \, a \sqrt{\sigma}.
\eeq
As already outlined in Sec.~\ref{sec:LEC_lattice_setup}, the bare effective chiral condensate extracted from the mode number is renormalized through the combination $\ZP = (\ZP/\ZS) \times \ZS$, achieved combining results from Tab.~\ref{tab:ZS_and_B_res} and Tab.~\ref{tab:ZPZS_determinations}, cf.~also Eq.~\eqref{eq:renormalization_Sigmaeff}. It is also important to stress that the chiral condensate in units of the string tension is obtained as:
\beq
\frac{\Sigma_\R^{\effsup}}{N\sqrt{\sigma^3}} = \left(\frac{\sqrt{8t_1}}{a}\right)^3 \times \frac{a^3\Sigma_\R}{N} \times \left(\frac{1}{\sqrt{8t_1\sigma}}\right)^3,
\eeq
with $\sqrt{8t_1\sigma}$ the continuum-extrapolated result of Eq.~\eqref{eq:conv_t1_sigma}. Since the relative error on the scale gets amplified by a factor of 3 when putting $\Sigma$ in physical units, this scale setting procedure is crucial to attain a good precision for the chiral condensate, as results for $\sqrt{8 t_1}$ are much more precise than those for $\sqrt{\sigma}$.

\begin{figure}[!t]
\centering
\includegraphics[scale=0.5]{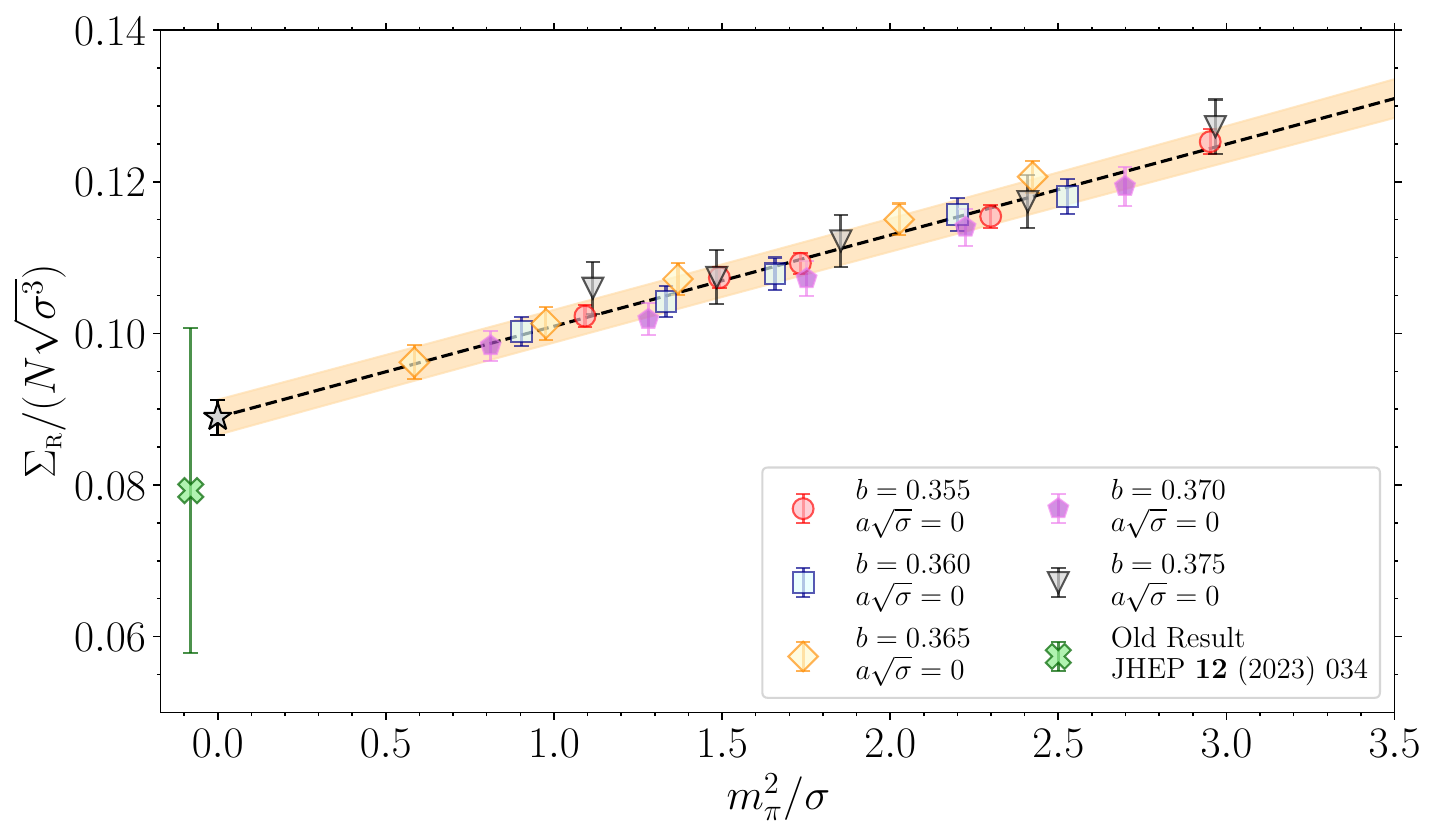}
\caption{Joint chiral-continuum extrapolation of the effective chiral condensate obtained from the mode number slope. The starred point represents our final value for the chiral condensate, while the crossed point represents our previous TEK determination presented in Ref.~\cite{Bonanno:2023ypf}. The shaded band represents the result reported in Eq.~\eqref{eq:mpi_dep_Sigmaeff}, while the points at various lattice spacings and pion masses represent the actual data points to which the lattice artifact term $k \, a \sqrt{\sigma}$ has been subtracted.}
\label{fig:chircont_lim_Sigma}
\end{figure}

The best fit according to Eq.~\eqref{eq:fitfunction_Sigmaeff_chircont} is displayed in Fig.~\ref{fig:chircont_lim_Sigma}, and, as it can be seen, this functional form provides an excellent description of our data. Also in this case we have verified that the addition of a mixed $k^\prime \, a \sqrt{\sigma} \times  m_\pi^2 / \sigma$ term in the fit function does not alter the result and yields a value of $k^\prime$ which is compatible with zero within errors. Moreover, the analysis was cross-checked by performing a chiral extrapolation at fixed $b$ for each lattice spacing followed by a continuum limit of the chiral-extrapolated results, which yielded perfectly compatible results.

The joint fit depicted in Fig.~\ref{fig:chircont_lim_Sigma} yields the following final result for the chiral condensate:
\beq
\frac{\Sigma_\R}{N\sqrt{\sigma^3}} = 0.0889(23).
\eeq
We also quote this result for the pion mass dependence of the effective condensate in the continuum limit:
\beq\label{eq:mpi_dep_Sigmaeff}
\frac{\Sigma_\R^{\effsup}(m_\pi)}{N\sqrt{\sigma^3}} = 0.0889(23) + 0.1201(64)\frac{m_\pi^2}{\sigma}.
\eeq
As it can be seen, our final result for $\Sigma_\R$ largely improves on our previous determination in Ref.~\cite{Bonanno:2023ypf}, reducing the uncertainty by about a factor of 7. This achievement was possible thanks to the combination of several factors: the more precise scale setting procedure adopted involving $t_1$; the employment of larger values of $N$ up to 841 at fixed statistics (whereas in our earlier study we only used $N=289$ ensembles), which allowed to improve the statistical accuracy of determinations at finer lattice spacings thanks to volume self-averaging by up to factors of $(841/289)^2\sim 8$; the addition of one finer lattice spacing and of two more values of $\kappa$ for each $b$; and the joint chiral-continuum extrapolation fitting strategy.

\subsubsection{The pion decay constant \texorpdfstring{$F_\pi$}{Fpi} and the NLO coupling \texorpdfstring{$\bar{\ell}_4$}{l4}}

Combining the results of the previous subsections for the pion mass slope obtained from the GMOR relation, $B_\R = \Sigma_\R/F_\pi^2$, and for the chiral condensate $\Sigma_\R$ obtained from the BC formula, we obtain the following determination of the pion decay constant:
\beq\label{eq:Fpi_BC_GMOR}
\frac{F_\pi}{\sqrt{\sigma}\sqrt{N}} = 0.1262(34), \qquad (\text{BC $+$ GMOR}).
\eeq
The goal of this section is to cross-check this result from a direct determination of $F_\pi$, using the discretization discussed in Sec.~\ref{sec:LEC_lattice_setup}, based on its definition in the continuum theory. This calculation will also allow to assess its pion mass dependence, from which the NLO LEC $\bar{\ell}_4/N\sim\mathcal{O}(N^0)$ will be extracted.

\begin{figure}[!t]
\centering
\includegraphics[scale=0.5]{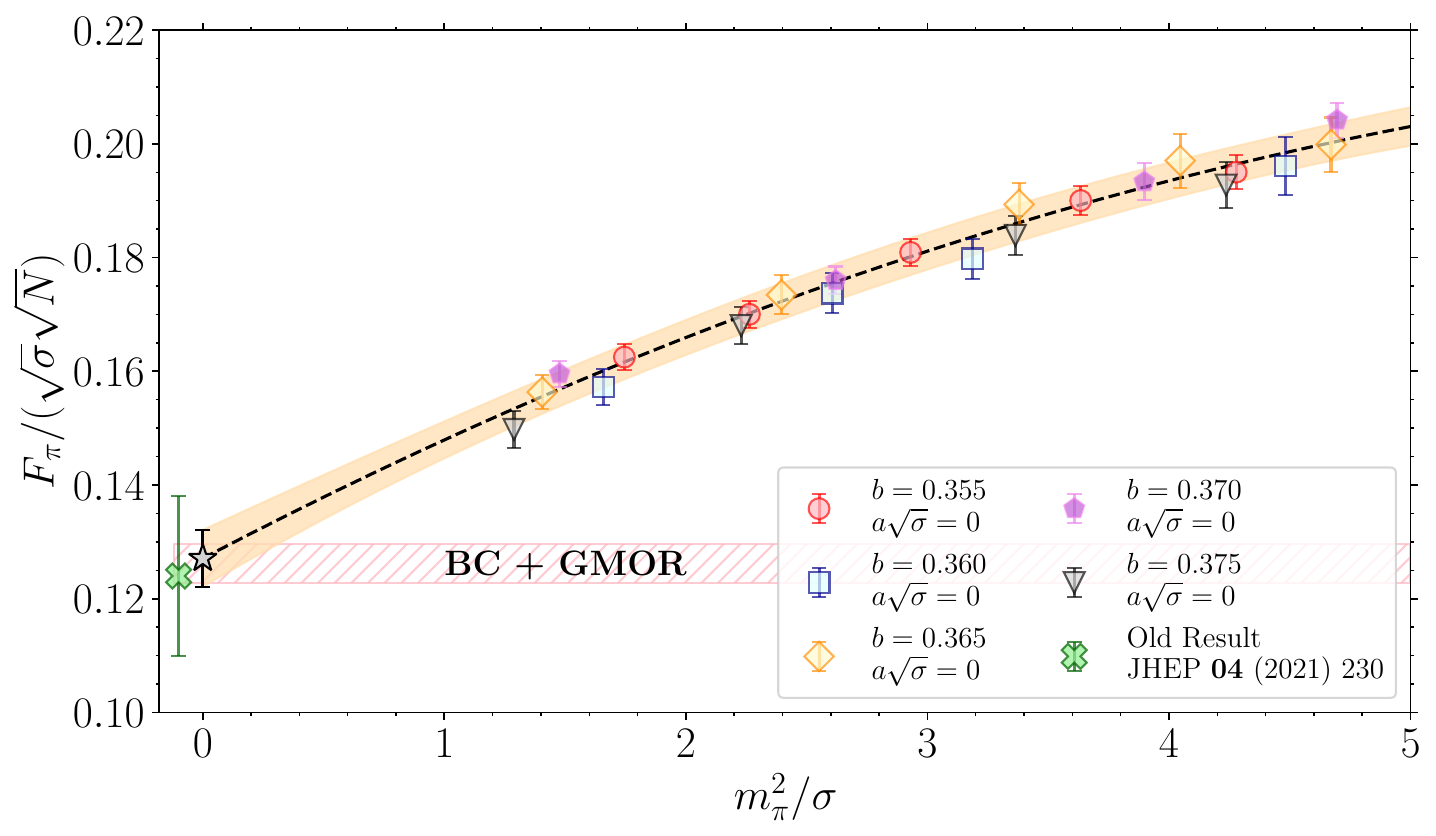}
\caption{Joint chiral-continuum extrapolation of our direct TEK determinations of $F_\pi/(\sqrt{\sigma}\sqrt{N})$. The chiral limit (starred point) is compared to our previous determination~\cite{Perez:2020vbn} (crossed point) and with the determination obtained combining the Banks--Casher (BC) and Gell-Mann--Oakes--Renner (GMOR) equations (dashed shaded area). The continuous shaded area represents the pion mass dependence of the pion decay constant in the continuum limit, cf.~Eq.~\eqref{eq:pionmass_dep_Fpi_continuum}, while the points at various lattice spacings and pion masses represent the actual data points to which the lattice artifact term $k \, a \sqrt{\sigma}$ has been subtracted.}
\label{fig:Fpi_chircont}
\end{figure}

As already explained in Sec.~\ref{sec:LEC_lattice_setup}, our bare determinations of $F_\pi/(\ZA\sqrt{N})$, collected in Tab.~\ref{tab:Fpi_rawdata} in the Appendix~\ref{app:rawdata}, are renormalized using the combination $\ZA=(\ZA\ZS/\ZP)\times(\ZP/\ZS)$, cf.~Tab.~\ref{tab:kappac_res} and Tab.~\ref{tab:ZPZS_determinations}. Scale setting is again performed in terms of $\sqrt{\sigma}$, but passing through $\sqrt{8t_1}$, cf.~Eq.~\eqref{eq:conv_sigma_through_t1}. It should also be noted that, concerning $F_\pi$, we only considered the ensembles used for meson mass computations that satisfied $\ell\sqrt{\sigma}> 3.9$, as this threshold was shown in Ref.~\cite{Perez:2020vbn} to be sufficient to contain finite-volume effects within the typical achieved statistical error on this quantity. After renormalization, we perform again a combined chiral-continuum extrapolation of $F_\pi/(\sqrt{\sigma}\sqrt{N})$ assuming the following fit function:
\beq
\frac{F_\pi(a,m_\pi)}{\sqrt{\sigma}\sqrt{N}} = \frac{F_\pi}{\sqrt{\sigma}\sqrt{N}} + C_0 \frac{m_\pi^2}{\sigma} + C_2 \frac{m_\pi^4}{\sigma^2}+ k\, a \sqrt{\sigma}.
\eeq
The best fit procedure is displayed in Fig.~\ref{fig:Fpi_chircont} and yields the following result in the continuum limit:
\beq\label{eq:pionmass_dep_Fpi_continuum}
\frac{F_\pi}{\sqrt{\sigma}\sqrt{N}} = 0.1271(50) + 0.0222(34) \frac{m_\pi^2}{\sigma} -0.00141(54) \frac{m_\pi^4}{\sigma^2}.
\eeq
This result was also cross-checked by performing individual chiral extrapolations at fixed $b$ followed by a continuum limit extrapolation, and, as before, we verified that mixed terms of the form $a \times m_\pi^2$ do not alter the final result and have coefficients which are compatible with zero within errors.

As it can be seen from Fig.~\ref{fig:Fpi_chircont}, the final result in the continuum and chiral limit (starred point) is in perfect agreement both with the previous TEK determination in~\cite{Perez:2020vbn} (crossed point), and with the determination in Eq.~\eqref{eq:Fpi_BC_GMOR} obtained from the BC and GMOR formulas (dashed shaded area). This cross-check thus fully validates our results for $B_\R$ and $\Sigma_\R$ too. In light of this result, we quote Eq.~\eqref{eq:Fpi_BC_GMOR} as our final result for $F_\pi$, because it is slightly more precise than the one in Eq.~\eqref{eq:pionmass_dep_Fpi_continuum}. Moreover, our determination is also in very good agreement with previous ones that can be retrieved in the literature. The comparison is detailed in the following.

\begin{itemize}
\item Comparison with Ref.~\cite{Bali:2013kia}.\\
The authors compute the large-$N$ limit of $F_\pi/(\sqrt{\sigma}\sqrt{N})$ in the chiral limit from quenched simulations with $2\le N \le17$ and for a fixed value of the lattice spacing, corresponding to $b\simeq 0.36$ (cf.~the discussion of meson mass results from the same study in Sec.~\ref{sec:mesonmass_res}). Since they employ the Wilson discretization, this result can be compared with a chiral extrapolation of our data at fixed $b=0.36$. One finds:
\beq
\frac{F_\pi}{\sqrt{\sigma}\sqrt{N}} &=& 0.1255(17), \qquad\,\left(\text{Ref.~\cite{Bali:2013kia}, chiral limit at fixed $b\simeq 0.36$}\right),\\
\frac{F_\pi}{\sqrt{\sigma}\sqrt{N}} &=& 0.1228(45), \qquad\,\left(\text{this study, chiral limit at fixed $b=0.36$}\right).
\eeq
\item Comparison with Ref.~\cite{Castagnini:2015ejr}.\\
The author quotes a result for the large-$N$ limit of $F_\pi$, which is also extrapolated towards the chiral and continuum limit (cf.~the discussion of meson mass results from the same study in Sec.~\ref{sec:mesonmass_res}). After removing an extra factor of $\sqrt{3}$ to match the definition of Ref.~\cite{Castagnini:2015ejr} with the one employed here, we find:
\beq
\frac{F_\pi}{\sqrt{\sigma}\sqrt{N}} = 0.114(12).
\eeq
\item Comparison with Ref.~\cite{Hernandez:2019qed}.\\
The authors obtain a large-$N$ determination of $F_\pi/\sqrt{N}$ extrapolating $\Nf=4$ results with $3\le N \le 6$ at fixed lattice spacing $a\sqrt{\sigma} \approx 0.211$ (see the discussion in Sec.~\ref{sec:mesonmass_res} relative to the meson mass determinations of Ref.~\cite{Baeza-Ballesteros:2025iee}, which uses the same lattice spacing and lattice setup). After converting their result from $\sqrt{8 t_0}$ to $\sqrt{\sigma}$ as done in Sec.~\ref{sec:mesonmass_res} (cf.~again the discussion related to Ref.~\cite{Baeza-Ballesteros:2025iee}), we find for the result of Ref.~\cite{Hernandez:2019qed}:
\beq
\frac{F_\pi}{\sqrt{\sigma}\sqrt{N}} = 0.1212(55),
\eeq
which, despite not being extrapolated towards the continuum limit, agrees with our continuum determination.
\end{itemize}

We conclude this section by extracting the dimensionless RGI NLO LEC $\bar{\ell}_4/N$~\cite{Gasser:1983yg} from the coefficient $C_0$ in Eq.~\eqref{eq:pionmass_dep_Fpi_continuum}, according to the following formula descending from Eq.~\eqref{eq:ell4_def}:
\beq
C_0 = \frac{\bar{\ell}_4}{N}\frac{\sqrt{\sigma}\sqrt{N}}{16\pi^2F_\pi}.
\eeq
From this equation we find:
\beq
\frac{\bar{\ell}_4}{N} = 16 \pi^2 \, \frac{F_\pi}{\sqrt{\sigma}\sqrt{N}} \, C_0 = 0.446(55).
\eeq

\subsection{The \texorpdfstring{$1/N$}{1/N} expansion of the QCD low-energy constants}\label{sec:1_on_N}

\begin{table}[!t]
\vspace*{-\baselineskip}
\small
\begin{center}
\begin{tabular}{|c|c|c|c|c|c|c|c|}
\hline
Study & $N$ & $\Nf$ & $(8t_0)^{3/2}\Sigma_\R/N$ & $\sqrt{8t_0}B_\R$ & $\sqrt{8t_0}F_\pi/\sqrt{N}$ & \makecell{$\bar{\ell}_4/N$\\ $[$SU(2) $\chi$PT$]$} & \makecell{$\bar{\ell}_4/N$\\ $[$U(2) $\chi$PT$]$} \\
\hline
Ref.~\cite{Wennekers:2005wa} & 3 & 0 & 0.1128(84) & \multicolumn{4}{c}{}\\
\cline{1-7}
\multirow{3}{*}{Ref.~\cite{DeGrand:2023hzz}} & 3 & 2 & 0.0530(68) & 5.37(28) & 0.0993(58) & 1.40(23) & \multicolumn{1}{c}{} \\
\cline{8-8}
& 4 & 2 & 0.0690(68) & 5.60(23) & 0.1110(50) & 1.23(15) & 0.22(6) \\
& 5 & 2 & 0.0773(54) & 5.15(20) & 0.1225(36) & 1.04(8)  & 0.28(6) \\
\hline
This study & $\infty$ & 0 & 0.1347(70) & 6.41(31) & 0.1449(45) & \multicolumn{2}{c|}{0.446(55)} \\
\hline
\end{tabular}
\end{center}
\caption{Summary of continuum determinations of the QCD LECs $B_\R$, $\Sigma_\R/N$, $F_\pi/\sqrt{N}$ and $\bar{\ell}_4/N$ obtained for various number of colors $N$ and number of flavors $\Nf$. All dimensionful quantities are expressed in terms of the gradient flow scale $\sqrt{8t_0}$ defined in Eq.~\eqref{eq:t0_Ninf_Giusti}.}
\label{tab:Ndep_COMP}
\end{table}

As explained in the Introduction, the $N$-dependence of TEK results amounts essentially to finite-effective-volume effects affecting the $N=\infty$ results. In order to grasp the physical $N$-dependence of any physical observable and determine the coefficients of the $1/N$ series, one has to combine finite-$N$ and TEK results.

In this section we aim exactly at studying the $N$-dependence of the LECs computed in this study---$\Sigma_\R/N$, $B_\R$, $F_\pi/\sqrt{N}$ and $\bar{\ell}_4/N$---by combining our TEK determinations with the $N=3,4,5$ determinations of these quantities of Ref.~\cite{DeGrand:2023hzz}, where both the chiral and continuum limits (despite from coarser lattice spacings compared to those adopted here) have been taken in the $\Nf=2$ theory. Concerning the chiral condensate $\Sigma_\R$, we will also include in our study the $\Nf=0$ determination (extrapolated towards the chiral and continuum limits) obtained in Ref.~\cite{Wennekers:2005wa} for the SU(3) Yang--Mills theory using overlap fermions. The data that will be analyzed in this section are collected in Tab.~\ref{tab:Ndep_COMP}.

Before proceeding with the actual data analysis, a few comments are in order about the numbers reported in Tab.~\ref{tab:Ndep_COMP}. First, since the authors of Ref.~\cite{DeGrand:2023hzz} express their findings in units of $\sqrt{8 t_0}$, we will also express our results in these units (following the same conversion procedure already explained in the previous sections). Moreover, the authors of Ref.~\cite{DeGrand:2023hzz} utilize a different convention for $F_\pi$ which carries an extra factor of $\sqrt{2}$ with respect to ours that will be removed to match the definition employed here. Secondly, we are reporting two different determinations of $\bar{\ell}_4$ from Ref.~\cite{DeGrand:2023hzz}, obtained respectively assuming $\SU(2)$ or $\mathrm{U}(2)$ $\chi$PT fit strategies. These two determinations clearly differ at finite values of $N$, but are expected to come together in the large-$N$ limit, when also the $\eta^\prime$ meson becomes a Nambu--Goldstone boson in the chiral limit. Thirdly, the quenched $N=3$ result for $\Sigma_\R$ of Ref.~\cite{Wennekers:2005wa} is expressed in terms of the Sommer scale $r_0$, thus we will express it in units of $\sqrt{8t_0}$ using the continuum value of $\sqrt{8t_0}/r_0 = 0.941(7)$ determined in Ref.~\cite{Giusti:2018cmp} in the SU(3) pure-gauge theory from Master Field simulations. Finally, let us recall that, on general theoretical grounds, one expects for the color and flavor dependence of a generic observable with a finite large-$N$ limit:
\beq\label{eq:1N_exp}
\braket{\mathcal{O}}(N,\Nf) = A_0 + A_1 \frac{\Nf}{N} + (A_2+A_2^\prime \Nf^2) \frac{1}{N^2} + \mathcal{O}\left(\frac{1}{N^3}\right),
\eeq
with the large-$N$ limit $A_0$ being independent of $\Nf$.

\begin{figure}[!t]
\centering
\includegraphics[scale=0.44]{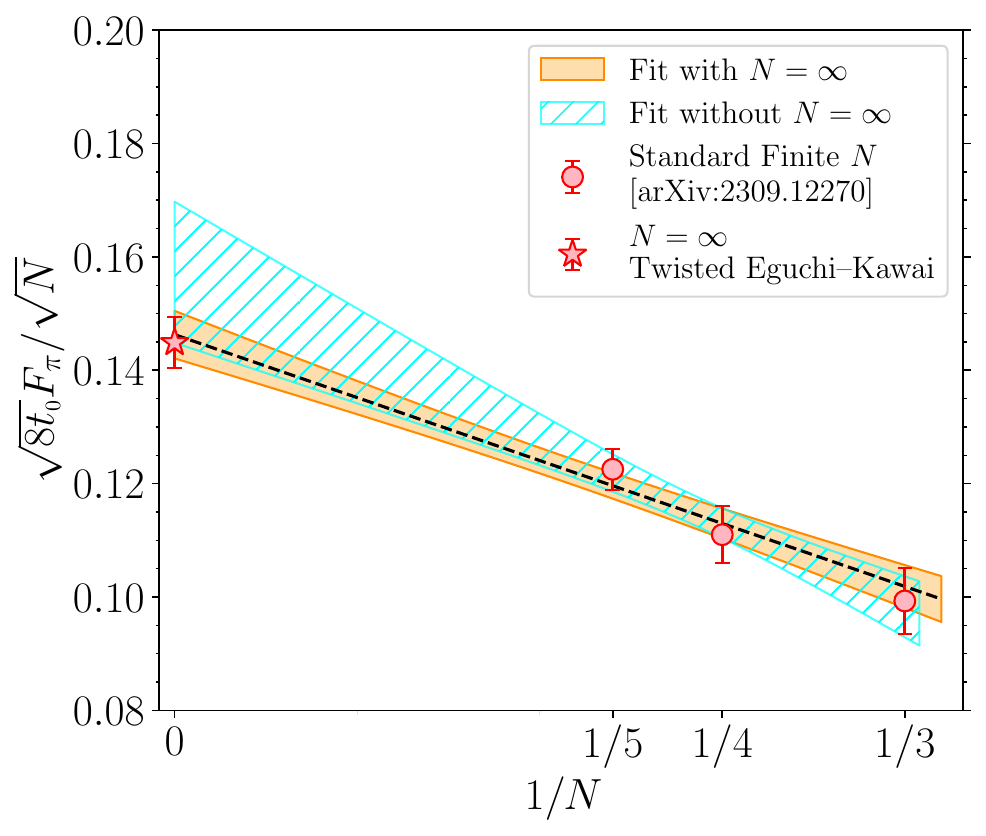}
\includegraphics[scale=0.44]{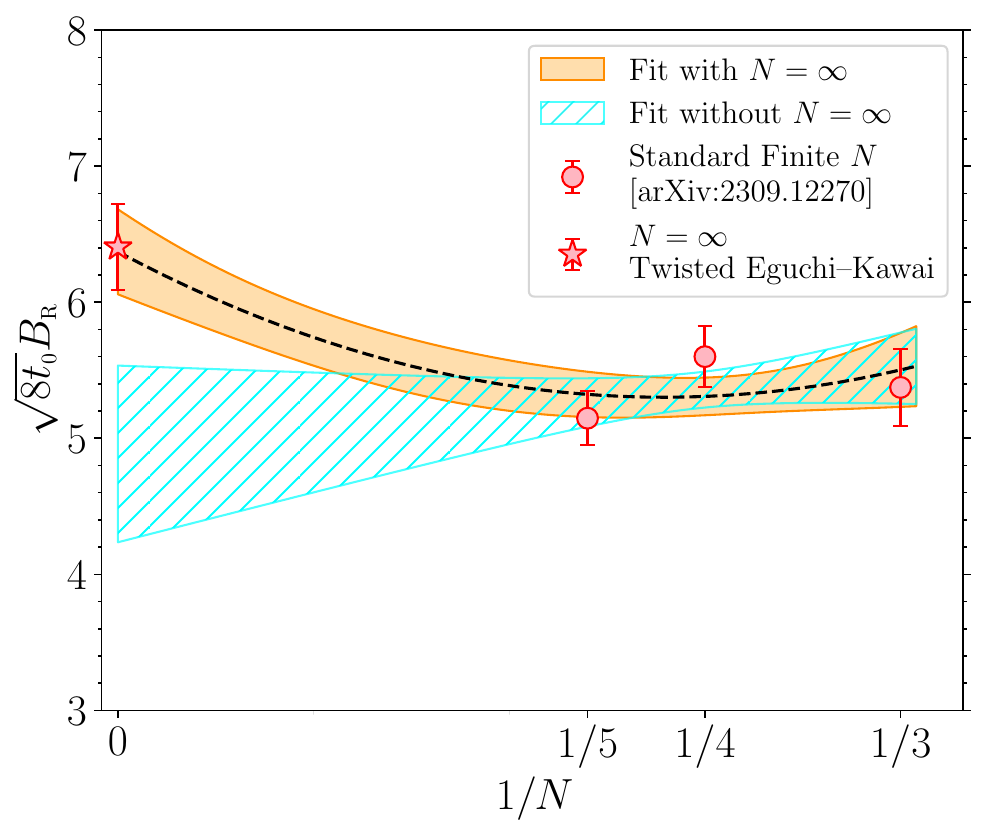}
\caption{The $1/N$ dependence of the LO QCD LECs $B_\R$ and $F_\pi/\sqrt{N}$, both expressed in units of the gradient flow scale $\sqrt{8 t_0}$ in Eq.~\eqref{eq:t0_Ninf_Giusti}. The shaded areas represent best fits of the data according to Eq.~\eqref{eq:1N_exp} with or without our $N=\infty$ TEK determination. The finite-$N$ determinations come from Ref.~\cite{DeGrand:2023hzz}.}
\label{fig:Ndep_LEC_B_Fpi}
\end{figure}

Let us start by discussing the LECs $F_\pi$ and $B_\R$. First of all, it should be noted that Ref.~\cite{DeGrand:2023hzz}'s results have $\Nf/N$ which is $2/5 = 0.4$ at best, which is not excessively small, thus one can in principle expect sizable corrections with respect to the $N=\infty$ results. This is indeed confirmed by our $N=\infty$ results for both $F_\pi$ and $B_\R$, which are significantly different (in particular, larger) compared to the $N=3,4,5$, $\Nf=2$ determinations. By performing a joint fit according to Eq.~\eqref{eq:1N_exp} of these determinations along with our $N=\infty$ ones, we are able to assess the magnitude of sub-leading terms in $1/N$. We obtain:
\beq
\label{eq_Ndep_F}
\frac{\sqrt{8t_0}F_\pi}{\sqrt{N}} &=& 0.1463(42) - 0.067(10)\frac{\Nf}{N},\\
\label{eq_Ndep_B}
\sqrt{8t_0}B_\R &=& 6.37(31) - 4.6(1.8)\frac{\Nf}{N} + 19.8(10.5)\frac{1}{N^2}.
\eeq
The best fits are depicted in Fig.~\ref{fig:Ndep_LEC_B_Fpi}. We observe that a $1/N^2$ term is not necessary to obtain a good fit quality for $F_\pi$ data: indeed, its coefficient turns out to be compatible with zero, $-0.17(20)$, and its inclusion leaves the one of the $\Nf/N$ term unchanged within errors. On the other hand, including a $1/N^2$ term in the fit function turns out to be necessary to describe the $N$-dependence of $B_\R$, as a simple linear fit in $1/N$ would have a poor reduced chi-squared, with a small $p$-value.

Interestingly, we observe an important difference between these two observables. On the one hand, despite the presence of sizable $\Nf/N$ corrections, we observe that the naive extrapolation that would be obtained by only taking into account finite-$N$ data for $F_\pi/\sqrt{N}$ of Ref.~\cite{DeGrand:2023hzz} is in agreement with our $N=\infty$ result. On the other hand, this is not at all true for $B_\R$, where a naive extrapolation of finite-$N$ data would give a completely different $N=\infty$ result, and a much different sub-leading $\Nf/N$ correction. Indeed, the $N=3,4,5$ data are almost flat and exhibit a slightly descendant trend, which would give a small and positive $\mathcal{O}(\Nf/N)$ term, while the indication of the best fit obtained including our $N=\infty$ result is that this term has actually a large and negative coefficient, and that the almost flat behavior is due to an accidental cancellation between the $\mathcal{O}(\Nf/N)$ and $\mathcal{O}(1/N^2)$ terms, which have opposite signs. This fact thus shows the importance of precisely pinpointing the value of the $N=\infty$ limit, a challenge that can be efficiently tackled via our TEK approach. Finally, let us also add that the large-$N$ $\Nf=4$ analysis of Ref.~\cite{Hernandez:2019qed}, which also uses $N=6$ data, also indicates that both $F_\pi$ and $B_\R$ grow as a function of $N$ towards the large-$N$ limit, thus being in agreement with our conclusions for both observables.

\begin{figure}[!t]
\centering
\includegraphics[scale=0.44]{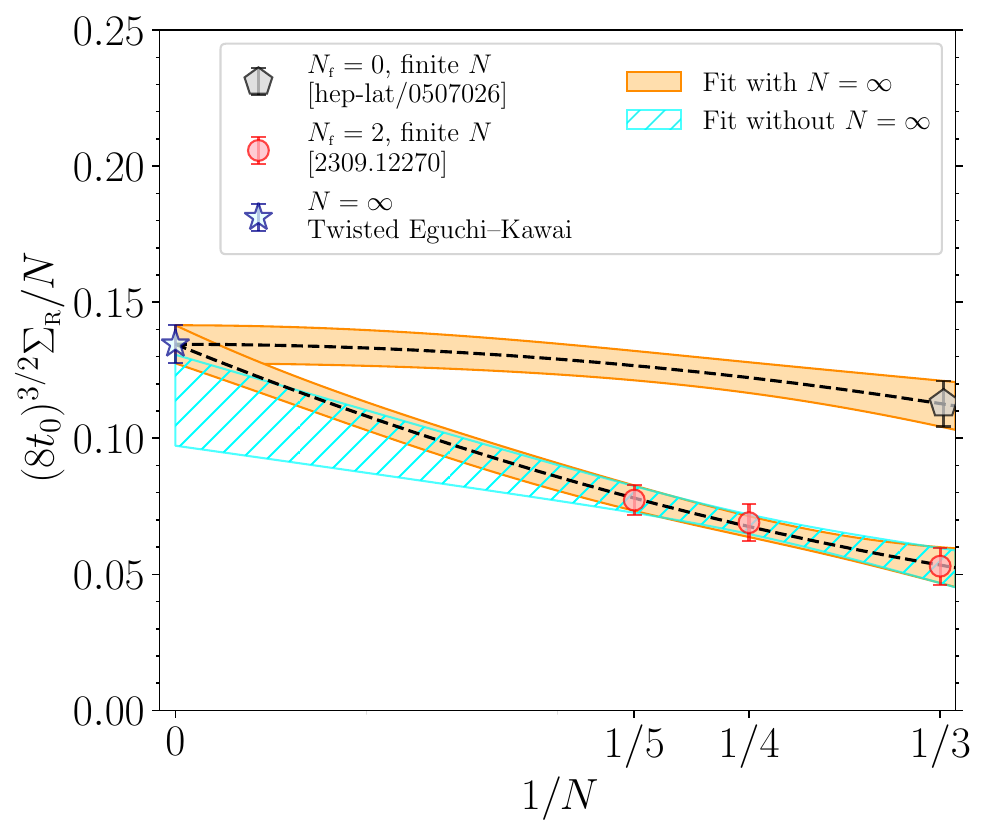}
\includegraphics[scale=0.44]{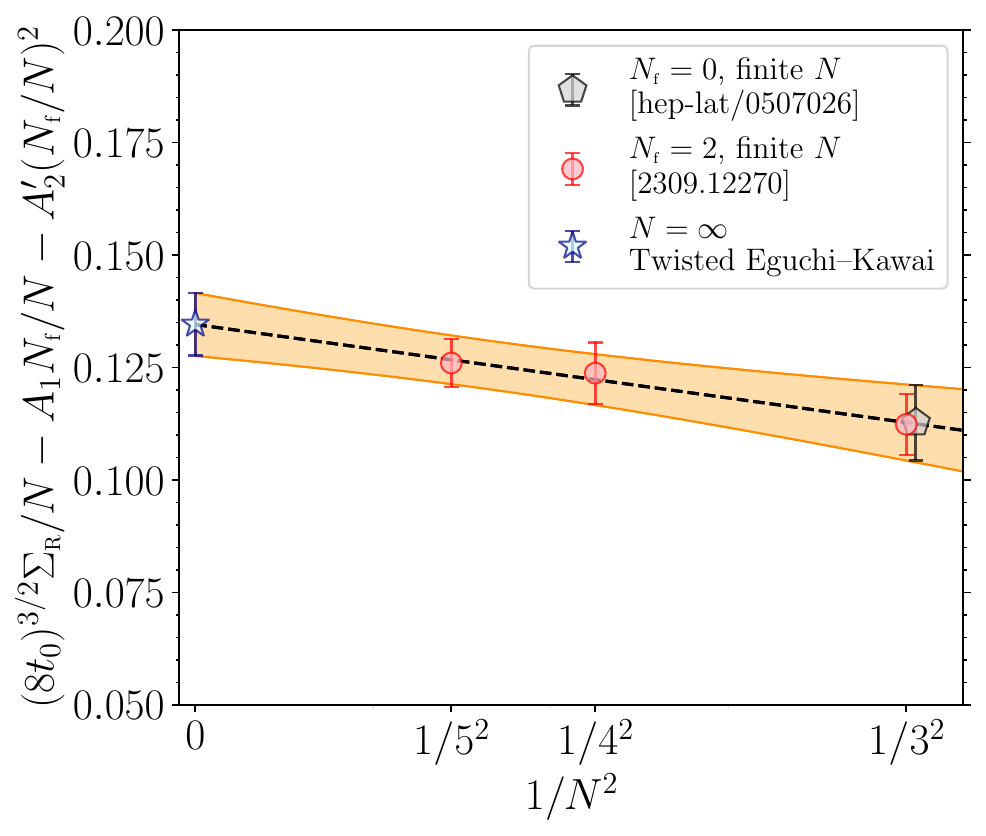}
\caption{The $1/N$ dependence of the LO QCD LEC $\Sigma_\R/N$ expressed in units of the gradient flow scale $\sqrt{8 t_0}$ in Eq.~\eqref{eq:t0_Ninf_Giusti}. Left panel: the shaded areas represent best fits of the data according to Eq.~\eqref{eq:1N_exp} with or without our $N=\infty$ TEK determination. Right panel: same data as in the left panel, but after the subtraction of the $\mathcal{O}(\Nf/N)$ and $\mathcal{O}[(\Nf/N)^2]$ corrections affecting the $\Nf=2$ finite-$N$ results. In this case the shaded area just represents the surviving $\mathcal{O}(1/N^2)$ term. The finite-$N$ determinations come from Ref.~\cite{DeGrand:2023hzz} (unquenched results) and Ref.~\cite{Wennekers:2005wa} (quenched result).}
\label{fig:Ndep_LEC_Sigma}
\end{figure}

In order to further investigate this issue related to the magnitude of sub-leading $\mathcal{O}(\Nf/N)$ corrections, it is interesting to look directly at the chiral condensate $\Sigma_\R/N$. Indeed, in this case, we also have at our disposal the quenched $\Nf=0$ determination of Ref.~\cite{Wennekers:2005wa}, which is unaffected by the $\mathcal{O}(\Nf/N)$ term. All available data for $\sqrt{8t_0}\Sigma_\R(N,\Nf)/N$ are plotted in Fig.~\ref{fig:Ndep_LEC_Sigma}. Interestingly, by looking at the left panel, we observe that the quenched $\Nf=0$ result is much larger (a factor of 2) compared to the $\Nf=2$ one at the same value of $N$, which is already indicative of a sizable negative $\mathcal{O}(\Nf/N)$ term by itself. This is further confirmed by the fact that the $\Nf=0$ quenched determination is just $\sim 17\%$ smaller than our $N=\infty$ result, which is also insensitive to the $\mathcal{O}(\Nf/N)$ term. Overall, thus, these data support the scenario in which unquenched results receive sizable negative $\mathcal{O}(\Nf/N)$ corrections, as argued from the analysis of $F_\pi$ and $B_\R$. This is further pointed out by the fact that the naive large-$N$ extrapolation of finite-$N$ data of Ref.~\cite{DeGrand:2023hzz} is only marginally compatible with our direct $N=\infty$ determination.


Performing a best fit to all available determinations of $\Sigma_\R$, we obtain the following $1/N$ expansion of $\Sigma_\R/N$:
\beq
\label{eq:Ndep_Sigma}
\frac{(8t_0)^{3/2}\Sigma_\R}{N} = 0.1346(70) -0.171(44)\frac{\Nf}{N} +[-0.196(99)+0.123(76)\Nf^2]\frac{1}{N^2}.
\eeq
A nice consequence of the inclusion of the quenched result among the fitted data is that it allows to assess the $\Nf$-dependence of the $1/N^2$ term. By subtracting the $\mathcal{O}(\Nf/N)$ and the $\mathcal{O}[(\Nf/N)^2]$ terms in Eq.~\eqref{eq:Ndep_Sigma} from the unquenched finite-$N$ results, we observe a collapse of all data for the chiral condensate on the remaining curve, parameterized by $A_0 + A_2/N^2$, see Fig.~\ref{fig:Ndep_LEC_Sigma}, right panel. Clearly, it would be interesting to further investigate the sign and magnitude of sub-leading $1/N$ corrections with more finite-$N$ data, especially from quenched $\Nf=0$ simulations, which would help improving the precision of the coefficient of the $1/N^2$ term.

\begin{figure}[!t]
\centering
\includegraphics[scale=0.48]{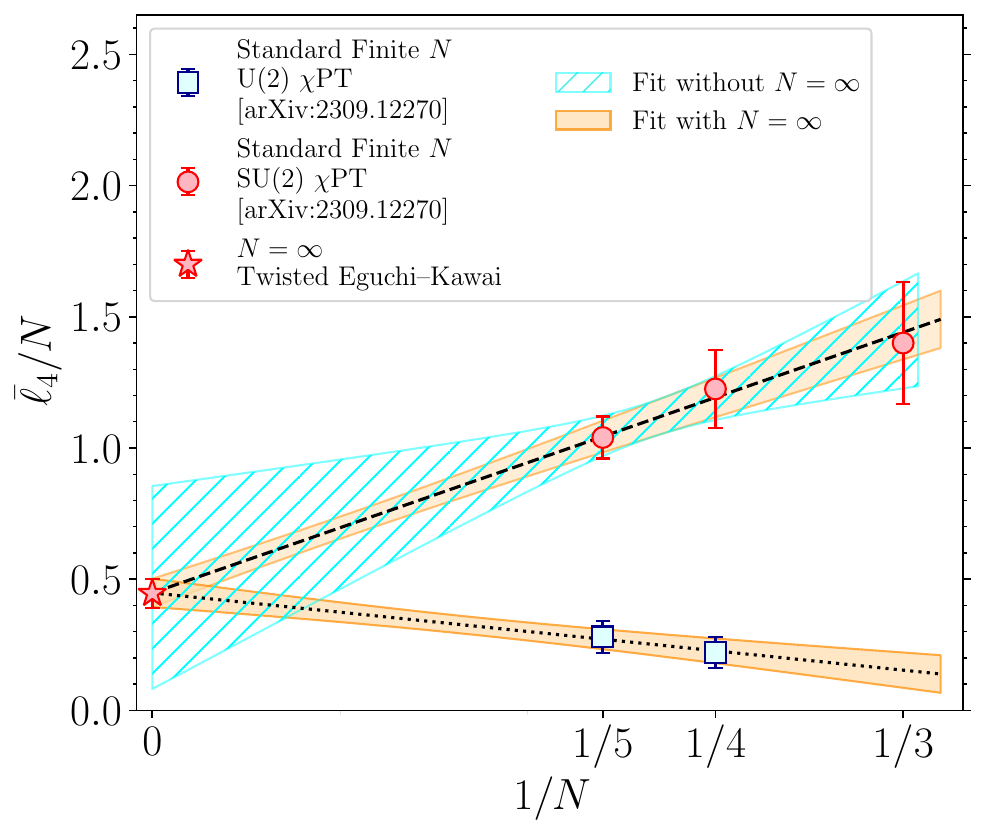}
\caption{The $1/N$ dependence of the dimensionless NLO QCD LEC $\bar{\ell}_4/N$. The continuous shaded areas represent the combined best fits of the SU(2) and U(2) data sets imposing a common large-$N$ limit, while the dashed shaded area represents the best fit of the SU(2) data alone without the TEK $N=\infty$ result. The finite-$N$ determinations come from Ref.~\cite{DeGrand:2023hzz}.}
\label{fig:Ndep_LEC_L4}
\end{figure}

We conclude this section by analyzing the available results for the NLO LEC $\bar{\ell}_4/N$. Thanks to our TEK $N=\infty$ determination, we are able to show that the SU(2) and U(2) data sets obtained at finite-$N$ indeed converge towards the same large-$N$ limit, as expected on general theoretical arguments based on the Witten--Veneziano solution to the U(1)$_{\A}$ puzzle. Moreover, as expected, the results obtained from the U(2) $\chi$PT fitting strategies suffer for much smaller $1/N$ corrections, again in agreement with the expectation that the $\eta^\prime$ becomes light in the large-$N$ limit. Notice indeed that removing our $N=\infty$ result it would be impossible to obtain a good fit quality combining the SU(2) and U(2) data sets obtained at finite values of $N$, while a linear best fit to the SU(2) data set alone would give a large-$N$ extrapolation with $\sim 80\%$ error: $\bar{\ell}_4/N=0.47(39)$. Thus, this is yet another remarkable example where it is necessary to combine standard finite-$N$ and TEK large-$N$ determinations to extract meaningful physics from lattice data.

In the end, we quote the following final results, obtained from a joint fit of the SU(2) and U(2) data sets (including the $N=\infty$ TEK result) imposing a common large-$N$ limit:
\beq
\frac{\bar{\ell}_4}{N} &=& 0.448(54) + 1.49(19)\frac{\Nf}{N}, \qquad [\mathrm{SU}(2)~\chi\mathrm{PT}],\\
\frac{\bar{\ell}_4}{N} &=& 0.448(54) - 0.44(15)\frac{\Nf}{N}, \qquad [\mathrm{U}(2)~\chi\mathrm{PT}].
\eeq
These best fit results are also depicted in Fig.~\ref{fig:Ndep_LEC_L4} as shaded bands.

\section{Conclusions}\label{sec:conclu}

In this paper we have presented an extended study of meson masses and low-energy constants (LECs) in large-$N$ QCD utilizing the TEK approach, which leverages twisted boundary conditions to achieve large-$N$ volume reduction on a one-point box. Our results, reaching values up to $N=841$, and employing 5 lattice spacings and several values of the quark mass, allowed us to provide fully non-perturbative first-principles determinations of several interesting observables with negligible finite-effective-volume effects, controlled chiral-continuum extrapolations, and non-perturbative renormalization. When possible, we compared our results with previous available results in the literature, finding always good agreement.

First, we provided large-$N$ determinations of the ground state masses of the $\rho,a_0,a_1$ and $b_1$ mesons and of the first and second $\pi,\rho$ excited ones, including their pion mass dependence down to the chiral limit. These results allowed us to show the linearity and the universality of Regge trajectories in the $\pi$ and $\rho$ channels at large $N$. Then, we discussed our determinations of the QCD LECs $\Sigma_\R/N$, $B_\R$, $F_\pi/\sqrt{N}$ and $\bar{\ell}_4/N$. Combining our $N=\infty$ results for these QCD LECs with previous finite-$N$ determinations, we were also able to compute the coefficients of the first few sub-leading terms in the $1/N$ expansion. A comprehensive summary of the main results presented in this paper is reported in Tab.~\ref{tab:final_summary}.

\begin{table}[!htb]
\small
\begin{center}
\begin{tabular}{|c|}
\hline
\textbf{Scale Setting At $N=\infty$}\\
\hline\\[-1em]
$\sqrt{8t_0\sigma} = 1.148(17)$, $\qquad$ $\frac{1}{N}t^2\braket{E(t)}\bigg\vert_{t\,=\,t_0}=0.1125$\\[-1em]
\multicolumn{1}{|c|}{}\\\hline
\hline
\textbf{Meson Masses At $N=\infty$}\\
\hline
\\[-1em]
\\
$\dfrac{1}{\sqrt{\sigma}}m_\rho(m_\pi)  = 1.783(47)+0.1902(65)\dfrac{m_\pi^2}{\sigma}$\\
$\dfrac{1}{\sqrt{\sigma}}m_{a_0}(m_\pi) = 2.130(87)+0.381(40)\dfrac{m_\pi^2}{\sigma} -0.0231(52) \dfrac{m_\pi^4}{\sigma^2}$\\
$\dfrac{1}{\sqrt{\sigma}}m_{a_1}(m_\pi) = 3.186(97)+0.159(11)\dfrac{m_\pi^2}{\sigma}$ $\qquad$
$\dfrac{1}{\sqrt{\sigma}}m_{b_1}(m_\pi) = 3.22(12)+0.167(16)\dfrac{m_\pi^2}{\sigma}$\\
$\dfrac{1}{\sqrt{\sigma}}m_{\pi^{*}}(m_\pi) = 3.62(21)+0.167(16)\dfrac{m_\pi^2}{\sigma}$ $\quad$
$\dfrac{1}{\sqrt{\sigma}}m_{\rho^{*}}(m_\pi) = 4.30(23)+0.177(30)\dfrac{m_\pi^2}{\sigma}$\\
\\[-1em]
$\dfrac{1}{\sqrt{\sigma}}m_{\pi^{**}} = 5.9(1.0)+\mathcal{O}\left(\dfrac{m_\pi^2}{\sigma}\right)$ $\qquad$
$\dfrac{1}{\sqrt{\sigma}}m_{\rho^{**}}(m_\pi) =6.4(1.0)+\mathcal{O}\left(\dfrac{m_\pi^2}{\sigma}\right)$\\
\\[-1em]
$\dfrac{1}{\sqrt{\sigma}}\mu_{\scriptscriptstyle{r}}=3.65(21)$, $\qquad$ (linear radial Regge traj.~slope for $\pi$)\\
\\[-1em]
$\dfrac{1}{\sqrt{\sigma}}\mu_{\scriptscriptstyle{r}}=3.95(24)$ $\qquad$ (linear radial Regge traj.~slope for $\rho$)\\
\\[-1em]
\multicolumn{1}{|c|}{}\\\hline
\hline
\textbf{QCD Low-Energy Constants At $N=\infty$}\\
\hline
\\[-1em]
\\
$\dfrac{B_\R}{\sqrt{\sigma}} \equiv \dfrac{\Sigma_\R}{F_\pi^2\sqrt{\sigma}} = 5.58(26)$ $\qquad$ [Gell-Mann--Oakes--Renner (GMOR)]\\
\\[-1em]
\\[-1em]
$\dfrac{\Sigma_\R(m_\pi)}{N\sqrt{\sigma^3}} = 0.0889(23) + 0.1201(64)\dfrac{m_\pi^2}{\sigma}$ $\qquad$ [Banks--Casher (BC)]\\
\\[-1em]
$\implies$ $\dfrac{F_\pi}{\sqrt{\sigma}\sqrt{N}} = 0.1262(34)$ $\qquad$ [BC + GMOR]\\
\\[-1em]
\hline
\\[-1em]
$\dfrac{F_\pi(m_\pi)}{\sqrt{\sigma}\sqrt{N}} = 0.1271(50) + 0.0222(34) \dfrac{m_\pi^2}{\sigma} -0.00141(54) \dfrac{m_\pi^4}{\sigma^2}$ $\qquad$ [Direct Determination]\\
$\implies \bar{\ell}_4/N=0.446(55)$\\[-1em]
\multicolumn{1}{|c|}{}\\\hline
\hline
\\[-1em]
\textbf{$\dfrac{1}{N}$ Expansion Of The QCD Low-Energy Constants}\\
\\[-1em]
\hline
\\[-1em]
\\
$\dfrac{\sqrt{8t_0}F_\pi}{\sqrt{N}} = 0.1463(42) - 0.067(10)\dfrac{\Nf}{N}$\\
\\[-1em]
$\sqrt{8t_0}B_\R = 6.37(31) - 4.6(1.8)\dfrac{\Nf}{N} + 19.8(10.5)\dfrac{1}{N^2}$\\
\\[-1em]
\\[-1em]
$\dfrac{(8t_0)^{3/2}\Sigma_\R}{N} = 0.1346(70) -0.171(44)\dfrac{\Nf}{N} +[-0.196(99)+0.123(76)\Nf^2]\dfrac{1}{N^2}$\\
\\[-1em]
\\[-1em]
$\dfrac{\bar{\ell}_4}{N} = 0.448(54) + 1.49(19)\dfrac{\Nf}{N}, \qquad [\mathrm{SU}(2)~\chi\mathrm{PT}]$\\
\\[-1em]
$\dfrac{\bar{\ell}_4}{N} = 0.448(54) - 0.44(15)\dfrac{\Nf}{N}, \qquad [\mathrm{U}(2)~\chi\mathrm{PT}]$\\
\multicolumn{1}{|c|}{}\\\hline
\end{tabular}
\end{center}
\vspace*{-0.5\baselineskip}
\caption{Summary of the main results of the present study. All renormalized quantities are expressed in the $\overline{\mathrm{MS}}$ scheme at a renormalization scale $\mu=2~\mathrm{GeV}$.}
\label{tab:final_summary}
\end{table}

What aspects of the present study could we look forward to refining in the next future? We have presented non-perturbative TEK determinations of all the necessary renormalization constants except for the scalar one, $\ZS$, entering $\Sigma_\R$ and $B_\R$. In the absence of TEK results, we have extrapolated the finite-$N$ determinations of~\cite{Castagnini:2015ejr}, whose reliability is discussed in App.~\ref{app:ZS}. Clearly, it would be interesting to perform a direct TEK calculation of $\ZS$. This requires a non-trivial effort, especially in the study of systematic errors, and a dedicated project. This is something we plan to do, being $\ZS$ important in other gauge theories too, such as in Super-Yang--Mills to renormalize the gluino condensate~\cite{Bonanno:2024bqg}.

Apart from this point, this study shows the overall robustness and solidity of the TEK approach as a non-perturbative method to efficiently tackle the study of large-$N$ gauge theories, and how TEK results can be conveniently combined with standard finite-$N$ determinations to accurately compute sub-leading terms in the $1/N$ expansion. A natural future outlook of the present work would be to further extend our investigation of large-$N$ QCD meson dynamics by looking at meson scattering amplitudes, a topic which has been recently investigated with standard lattice techniques in full QCD with varying number of colors~\cite{Baeza-Ballesteros:2022azb,Baeza-Ballesteros:2025iee}, and which has intriguing phenomenological implications.

\FloatBarrier

\section*{Acknowledgements}
This work is partially supported by the Spanish Research Agency (Agencia Estatal de Investigaci\'on) through the grant IFT Centro de Excelencia Severo Ochoa CEX2020-001007-S and, partially, by the grant PID2021-127526NB-I00, both funded by MCIN/AEI/ \\10.13039/501100011033. It is also partially funded by the European Commission – NextGenerationEU, through Momentum
CSIC Programme: Develop Your Digital Talent. K.-I.~I.~is supported in part by MEXT as ``Feasibility studies for the next-generation computing infrastructure''. This research was supported in part by grant NSF PHY-2309135 to the Kavli Institute for Theoretical Physics (KITP). Numerical calculations have been performed on the \texttt{Finisterrae~III} cluster at CESGA (Centro de Supercomputaci\'on de Galicia), on the \texttt{Drago} cluster at CSIC (Consejo Superior de Investigaciones Científicas) and on the Hydra cluster at IFT. We acknowledge HPC support by Emilio Ambite, staff hired under the Generation D initiative, promoted by Red.es, an organisation attached to the Ministry for Digital Transformation and the Civil Service, for the attraction and retention of talent through grants and training contracts, financed by the Recovery, Transformation and Resilience Plan
through the European Union's Next Generation funds. We have also used computational resource of Oakbridge-CX, at the University of Tokyo through the HPCI System Research Project (Project ID: hp230021 and hp220011), of Cygnus at Center for Computational Sciences, University of Tsukuba, of SQUID at D3 Center of Osaka university through the RCNP joint use program, and of the Genkai system at the Research Institute for Information Technology, Kyushu University, under the category of trial use projects.

\FloatBarrier
\newpage
\clearpage

\appendix

\section*{Appendix}
\section{Raw data}\label{app:rawdata}

This appendix collects the raw data for meson masses and low energy constants, whose analysis is presented in the main text. More precisely:
\begin{itemize}
\item Tab.~\ref{tab:raw_mpi_mPCAC} reports the simulation parameters employed to compute all meson masses, as well as the results for the pion and the PCAC mass in lattice units;
\item Tab.~\ref{tab:raw_other_mesons} reports the results for the $\rho,a_0,a_1,b_1,\pi^{*},\rho^{*},\pi^{**},\rho^{**}$ masses in lattice units;
\item Tab.~\ref{tab:Diracspectra_params} reports the simulation parameters employed to compute Dirac spectra, as well as the results for the bare effective chiral condensate in lattice units extracted from the mode number;
\item Tab.~\ref{tab:Fpi_rawdata} reports the results for the bare pion decay constant in lattice units computed from Eq.~\eqref{eq:Fpi_def}.
\end{itemize}

\begin{table*}[!htb]
\tiny
\begin{center}
\begin{tabular}{|c|c|c|c|c|c|c|}
\hline
$b$ & $\kappa$ & $N$ & $\ell\sqrt{\sigma}$ & $m_\pi \ell$ & $am_\pi$ & $am_\PCAC$ \\
\hline
\hline
\multirow{6}{*}{0.355} & 0.1565 & 289 & 4.10 & 9.86 & 0.5798(64) & 0.10781(45) \\
\cline{2-7}
 & 0.1575 & \multirow{5}{*}{529} & \multirow{5}{*}{5.54} & 12.12 & 0.5272(39) & 0.09030(51) \\
 & 0.1585 &  & & 11.18 & 0.4859(30) & 0.07513(50) \\
 & 0.1592 &  & & 10.03 & 0.4363(33) & 0.06221(43) \\
 & 0.1600 &  & & 8.82 & 0.3834(31) & 0.04888(37) \\
 & 0.1607 &  & & 7.74 & 0.3367(34) & 0.03760(43) \\
\hline
\hline
\multirow{5}{*}{0.360} & 0.1550 & \multirow{5}{*}{529} & \multirow{5}{*}{4.73} & 11.69 & 0.5081(68) & 0.09336(98) \\
 & 0.1560 & & & 10.46 & 0.4547(71) & 0.07475(88) \\
 & 0.1570 & & & 8.82 & 0.3834(44) & 0.05626(41) \\
 & 0.1577 & & & 7.97 & 0.3467(51) & 0.04500(52) \\
 & 0.1585 & & & 6.36 & 0.2765(20) & 0.03099(27) \\
\hline
\hline
\multirow{7}{*}{0.365} & 0.1535 & \multirow{2}{*}{289} & \multirow{2}{*}{3.03} & 7.70 & 0.4532(70) & 0.0814(11) \\
 & 0.1540 & & & 7.25 & 0.4267(65) & 0.0720(10) \\
\cline{2-7}
 & 0.1545 & \multirow{5}{*}{529} & \multirow{5}{*}{4.10} & 9.14 & 0.3973(54) & 0.06544(90) \\
 & 0.1550 & & & 8.50 & 0.3698(63) & 0.05648(59) \\
 & 0.1555 & & & 7.77 & 0.3379(65) & 0.04793(60) \\
 & 0.1562 & & & 6.54 & 0.2845(37) & 0.03487(55) \\
 & 0.1570 & & & 5.01 & 0.2179(78) & 0.02093(45) \\
\hline
\hline
\multirow{7}{*}{0.370} & 0.1520 & \multirow{3}{*}{289} & \multirow{3}{*}{2.67} & 7.29 & 0.4287(88) & 0.0821(13) \\
 & 0.1525 & & & 6.80 & 0.4002(86) & 0.0722(12) \\
 & 0.1530 & & & 6.31 & 0.3710(86) & 0.0624(11) \\
\cline{2-7}
 & 0.1535 & \multirow{4}{*}{841} & \multirow{4}{*}{4.56} & 10.03 & 0.3458(29) & 0.05552(40) \\
 & 0.1540 & & & 9.14 & 0.3150(41) & 0.04569(42) \\
 & 0.1547 & & & 7.49 & 0.2583(21) & 0.03274(22) \\
 & 0.1555 & & & 5.62 & 0.1939(26) & 0.01858(17) \\
\hline
\hline
\multirow{4}{*}{0.375} & 0.1527 & \multirow{4}{*}{841} & \multirow{4}{*}{3.95} & 8.30 & 0.2862(27) & 0.04348(43) \\
 & 0.1532 & & & 7.40 & 0.2551(21) & 0.03478(31) \\
 & 0.1538 & & & 6.02 & 0.2076(19) & 0.02315(23) \\
 & 0.1543 & & & 4.58 & 0.1579(31) & 0.01336(30) \\
\hline
\end{tabular}
\end{center}
\caption{Summary of our TEK determinations of $m_\pi$ and $m_\PCAC$ in lattice units as a function of the bare coupling $b$ and the hopping parameter $\kappa$. For completeness we also report the effective lattice size $\ell=aL=a\sqrt{N}$ in units of the string tension and of the pion mass.}
\label{tab:raw_mpi_mPCAC}
\end{table*}

\begin{table*}[!t]
\tiny
\begin{center}
\begin{tabular}{|c|c|c|c|c|c|c|}
\hline
$b$ & $\kappa$ & $N$ & $a m_{\rho}$ & $a m_{a_0}$ & $a m_{a_1}$ & $a m_{b_1}$ \\
\hline
\hline		
\multirow{6}{*}{0.355} & 0.1565 & 289 & 0.657(11) &	0.917(18) &	0.962(21) &	0.980(28) \\
\cline{2-7}
 & 0.1575 & \multirow{5}{*}{529} & 0.6116(70) &	0.904(12) &	0.980(19) &	1.014(24) \\
 & 0.1585 & & 0.601(10)  &	0.845(11) &	0.907(20) &	0.930(30) \\
 & 0.1592 & & 0.5512(97) &	0.823(11) &	0.916(19) &	0.956(31) \\
 & 0.1600 & & 0.5120(70) &	0.7615(75) & 0.840(11) & 0.844(14) \\
 & 0.1607 & & 0.4855(95) &	0.737(10) &	0.826(13) &	0.841(19) \\
\hline
\hline		
\multirow{5}{*}{0.360} & 0.1550 & \multirow{5}{*}{529} & 0.578(13) &	0.791(13) &	0.831(19) &	0.835(25) \\
 & 0.1560 & & 0.547(16) &	0.728(11) &	0.777(17) &	0.787(26) \\
 & 0.1570 & & 0.470(11) &	0.701(19) &	0.775(26) &	0.809(39) \\
 & 0.1577 & & 0.463(12) &	0.628(26) &	0.708(44) &	0.718(64) \\
 & 0.1585 & & 0.3994(87) &	0.600(15) &	0.694(25) &	0.694(22) \\
\hline
\hline		
\multirow{7}{*}{0.365} & 0.1535 & \multirow{2}{*}{289} & 0.502(12) &	0.685(14) &	0.740(16) &	0.719(37) \\
 & 0.1540 &  & 0.480(12) &	0.658(13) &	0.717(15) &	0.695(36) \\
\cline{2-7}
 & 0.1545 & \multirow{5}{*}{529} & 0.4694(92) &	0.659(12) &	0.708(21) &	0.720(32) \\
 & 0.1550 & & 0.444(15) &	0.632(11) &	0.701(17) &	0.723(28) \\
 & 0.1555 & & 0.422(13) &	0.583(10) &	0.642(16) &	0.642(24) \\
 & 0.1562 & & 0.388(12) &	0.550(15) &	0.638(22) &	0.646(36) \\
 & 0.1570 & & 0.329(19) &	0.479(11) &	0.600(18) &	0.603(22) \\
\hline
\hline		
\multirow{7}{*}{0.370} & 0.1520 & \multirow{3}{*}{289} & 0.474(13) &	0.603(14) &	0.655(20) &	0.669(36) \\
 & 0.1525 &  & 0.448(12) &	0.575(14) &	0.632(18) &	0.648(34) \\
 & 0.1530 &  & 0.421(12) &	0.546(13) &	0.608(17) &	0.627(32) \\
\cline{2-7}
 & 0.1535 & \multirow{4}{*}{841} & 0.4078(93) &	0.5703(84) &	0.615(10) &	0.648(20) \\
 & 0.1540 &  & 0.3932(83) &	0.5119(96) &	0.554(18) &	0.562(26) \\
 & 0.1547 &  & 0.3439(48) &	0.4842(53) &	0.547(24) &	0.559(13) \\
 & 0.1555 &  & 0.3029(81) &	0.4413(99) &	0.511(18) &	0.534(29) \\
\hline
\hline
\multirow{4}{*}{0.375} & 0.1527 & \multirow{4}{*}{841} & 0.3429(62) &	0.4889(72) &	0.534(11) &	0.554(27) \\
 & 0.1532 &  & 0.3268(55) &	0.4576(93) &	0.514(14) &	0.530(20) \\
 & 0.1538 &  & 0.2982(63) &	0.4047(93) &	0.462(21) &	0.466(13) \\
 & 0.1543 &  & 0.2741(79) &	0.380(13) &	0.467(11) &	0.487(13) \\
\hline
\end{tabular}
\end{center}

\begin{center}
\begin{tabular}{|c|c|c|c|c|c|c|}
\hline
$b$ & $\kappa$ & $N$ & $a m_{\pi^{*}}$ & $a m_{\pi^{**}}$ & $a m_{\rho^{*}}$ & $a m_{\rho^{**}}$ \\
\hline
\multirow{5}{*}{0.355} & 0.1575 & \multirow{5}{*}{529} & 0.953(32) &      1.289(28) &     1.091(26) &     1.526(41) \\
 & 0.1585 &  & 0.986(19) &      1.403(29) &     1.107(15) &     1.614(30) \\
 & 0.1592 &  & 0.935(27) &      1.380(27) &     1.018(23) &     1.463(31) \\
 & 0.1600 &  & 0.886(33) &      1.291(45) &     1.072(15) &     1.593(25) \\
 & 0.1607 &  & 0.824(47) &      1.234(46) &     1.047(16) &     1.527(33) \\
\hline
\multirow{5}{*}{0.360} & 0.1550 & \multirow{5}{*}{529} & 0.960(13) &      1.364(22) &     1.064(11) &     1.572(21) \\
 & 0.1560 &  & 0.902(15) &      1.350(24) &     1.008(12) &    1.542(25) \\
 & 0.1570 &  & 0.8536(91)&      1.305(14) &     0.9904(70)&    1.524(20) \\
 & 0.1577 &  & 0.741(31) &      1.187(40) &     0.934(21) &    1.408(35) \\
 & 0.1585 &  & 0.701(28) &      0.71(16)  &               &      \\
\hline
\multirow{3}{*}{0.365} & 0.1545 & \multirow{3}{*}{529} & 0.723(22) &      0.976(61) &     0.850(42) &     1.110(74) \\
 & 0.1550 &  & 0.694(21) &      1.046(34) &     0.773(21) &     1.249(56) \\
 & 0.1562 &  & 0.664(23) &      1.099(40) &     0.751(28) &     1.150(60) \\
\hline
\multirow{4}{*}{0.370} & 0.1535 & \multirow{4}{*}{841} & 0.671(15) &      1.020(17) &     0.721(19) &     1.063(24) \\
 & 0.1540 &  & 0.650(12) &      1.047(45) &     0.714(17) &     1.113(62) \\
 & 0.1547 &  & 0.611(23) &      0.993(27) &     0.705(27) &     1.048(35) \\
 & 0.1555 &  & 0.550(20) &      0.973(32) &     0.634(22) &     1.032(52) \\
\hline
\multirow{4}{*}{0.375} & 0.1527 & \multirow{4}{*}{841} & 0.610(12) &      1.043(23) &     0.673(18) &     1.089(27) \\
 & 0.1532 &  & 0.574(12) &    0.965(16) &     0.633(13) &     1.017(17) \\
 & 0.1538 &  & 0.528(10) &    0.968(23) &     0.617(12) &     1.012(28) \\
 & 0.1543 &  & 0.520(15) &    0.989(44) &     0.619(18) &     1.053(35) \\
\hline
\end{tabular}
\end{center}

\caption{Summary of our TEK determinations of ground state meson masses $m_\A$ in lattice units ($\mathrm{A}=\rho,a_0,a_1,b_1$) and of excited meson masses $m_{\A^{*}}$ and $m_{\A^{**}}$ in lattice units ($\mathrm{A}=\pi, \rho$) as a function of the bare coupling $b$ and the hopping parameter $\kappa$.}
\label{tab:raw_other_mesons}
\end{table*}

\begin{table}[!t]
\tiny
\begin{center}
\begin{tabular}{|c|c|c|c|c|c|c|c|}
\hline
$N$ & $b$ & $\kappa$ & $am_\pi$ & $am_{\PCAC}$ & $m_\pi\ell$ & $\ell\sqrt{\sigma}$ & $10^3 \times a^3\Sigma^{\effsup}_\R/(N\ZP)$ \\
\hline
\hline
\multirow{5}{*}{289} & \multirow{5}{*}{0.355} & 0.1592 & 0.43786(71) & 0.06230(18) & 7.44 & \multirow{5}{*}{4.10} & 3.319(42) \\
 & & 0.1600 & 0.38642(86) & 0.04909(23) & 6.57 & & 2.969(41) \\
 & & 0.1607 & 0.3355(13) & 0.03763(28)  & 5.70 & & 2.748(38) \\
 & & 0.1610 & 0.3113(16) & 0.03275(31)  & 5.29 & & 2.681(36) \\
 & & 0.1615 & 0.2664(23) & 0.02466(37)  & 4.53 & & 2.501(42) \\
\hline
\hline
\multirow{5}{*}{361} & \multirow{5}{*}{0.360} & 0.1577 & 0.34147(86) & 0.04470(20) & 6.49 & \multirow{5}{*}{3.91} & 1.862(37) \\
 & & 0.1580 & 0.31858(80) & 0.03951(20) & 6.05 & & 1.814(33) \\
 & & 0.1585 & 0.27649(98) & 0.03090(25) & 5.25 & & 1.654(36) \\
 & & 0.1588 & 0.2480(13)  & 0.02576(29) & 4.71 & & 1.579(34) \\
 & & 0.1592 & 0.2041(21)  & 0.01894(36) & 3.88 & & 1.498(33) \\
\hline
\hline
\multirow{5}{*}{529} & \multirow{5}{*}{0.365} & 0.1562 & 0.2860(14) & 0.03501(27) & 6.58 & \multirow{5}{*}{4.10} & 1.181(15) \\
 & & 0.1565 & 0.2616(17) & 0.02984(30) & 6.02 & & 1.113(16) \\
 & & 0.1570 & 0.2149(26) & 0.02125(38) & 4.94 & & 1.018(18) \\
 & & 0.1573 & 0.1815(36) & 0.01612(43) & 4.17 & & 0.946(20) \\
 & & 0.1576 & 0.1405(52) & 0.01102(49) & 3.23 & & 0.885(22) \\
\hline
\hline
\multirow{5}{*}{841} & \multirow{5}{*}{0.370} & 0.1547 & 0.26210(66) & 0.03292(12) & 7.60 & \multirow{5}{*}{4.56} & 0.755(10) \\
 & & 0.1550 & 0.23795(71) & 0.02751(12) & 6.90 & & 0.7143(94) \\
 & & 0.1553 & 0.21116(91) & 0.02212(14) & 6.12 & & 0.6627(87) \\
 & & 0.1556 & 0.1806(13)  & 0.01675(17) & 5.24 & & 0.6219(78) \\
 & & 0.1559 & 0.1438(20)  & 0.01141(20) & 4.17 & & 0.5954(65) \\
\hline
\hline
\multirow{5}{*}{841} & \multirow{5}{*}{0.375} & 0.1534 & 0.23956(60) & 0.03064(17) & 6.95 & \multirow{5}{*}{3.95} & 0.5294(96) \\
 & & 0.1537 & 0.21586(59) & 0.02493(15) & 6.26 & & 0.482(10) \\
 & & 0.1540 & 0.18934(78) & 0.01925(18) & 5.49 & & 0.457(11) \\
 & & 0.1542 & 0.1694(11)  & 0.01548(22) & 4.91 & & 0.434(13) \\
 & & 0.1544 & 0.1469(15)  & 0.01171(26) & 4.26 & & 0.427(12) \\
\hline
\end{tabular}
\end{center}
\caption{Simulation parameters for the Dirac spectra calculations, along with our TEK determinations of the bare effective condensate $\Sigma_\R^{\effsup}/\ZP$ from the mode number slope in lattice units. The values of $am_\pi$ and $am_\PCAC$ for the Dirac spectrum study were obtained from chiral interpolation/extrapolation of the data in Tab.~\ref{tab:raw_mpi_mPCAC} via their linear best fit in $1/(2\kappa)$ presented in Sec.~\ref{sec:kappac_determination}.}
\label{tab:Diracspectra_params}
\end{table}

\begin{table}[!t]
\tiny
\begin{center}
\begin{tabular}{|c|c|c|c|c|}
\hline
$b$ & $\kappa$ & $N$ & $\ell\sqrt{\sigma}$ & $a F_\pi/(\ZA\sqrt{N})$ \\
\hline
\hline
\multirow{5}{*}{0.355} & 0.1575 & \multirow{5}{*}{529} & \multirow{5}{*}{5.54} & 0.05894(61) \\
 & 0.1585 & & & 0.05727(38) \\
 & 0.1592 & & & 0.05421(35) \\
 & 0.1600 & & & 0.05055(41) \\
 & 0.1607 & & & 0.04802(42) \\
\hline
\hline
\multirow{5}{*}{0.360} & 0.1550 & \multirow{5}{*}{529} & \multirow{5}{*}{4.73} & 0.0521(11) \\
 & 0.1560 & & & 0.0496(12) \\
 & 0.1570 & & & 0.04510(61) \\
 & 0.1577 & & & 0.04344(65) \\
 & 0.1585 & & & 0.03888(60) \\
\hline
\hline
\multirow{5}{*}{0.365} & 0.1545 & \multirow{5}{*}{529} & \multirow{5}{*}{4.10} & 0.04112(86) \\
 & 0.1550 & & & 0.04049(87) \\
 & 0.1555 & & & 0.03879(59) \\
 & 0.1562 & & & 0.03527(56) \\
 & 0.1570 & & & 0.03149(49) \\
\hline
\hline
\multirow{4}{*}{0.370} & 0.1535 & \multirow{4}{*}{841} & \multirow{4}{*}{4.56} & 0.03685(37) \\
 & 0.1540 & & & 0.03478(43) \\
 & 0.1547 & & & 0.03144(25) \\
 & 0.1555 & & & 0.02829(24) \\
\hline
\hline
\multirow{4}{*}{0.375} & 0.1527 & \multirow{4}{*}{841} & \multirow{4}{*}{3.95} & 0.03074(36) \\
 & 0.1532 & & & 0.02925(16) \\
 & 0.1538 & & & 0.02657(25) \\
 & 0.1543 & & & 0.02349(31) \\
\hline
\end{tabular}
\end{center}
\caption{Summary of our TEK determinations of the bare pion decay constant $F_\pi/(\ZA\sqrt{N})$ in lattice units.}
\label{tab:Fpi_rawdata}
\end{table}

\clearpage
\newpage

\section{The scalar renormalization constant \texorpdfstring{$\ZS$}{ZS}}\label{app:ZS}

In absence of non-perturbative determinations of the renormalization constant $\ZS$ within the TEK model, we relied on the large-$N$ determinations of Ref.~\cite{Castagnini:2015ejr}, obtained with the same action (Wilson plaquette gauge action plus Wilson fermions) adopted in this study. Given that Eguchi--Kawai large-$N$ reduction holds for lattice Yang--Mills theories, in the large-$N$ limit the standard Wilson discretization of QCD is expected to coincide with our 1-site twisted reduced model, lattice artifacts included. 

\begin{figure}[!t]
\centering
\includegraphics[scale=0.5]{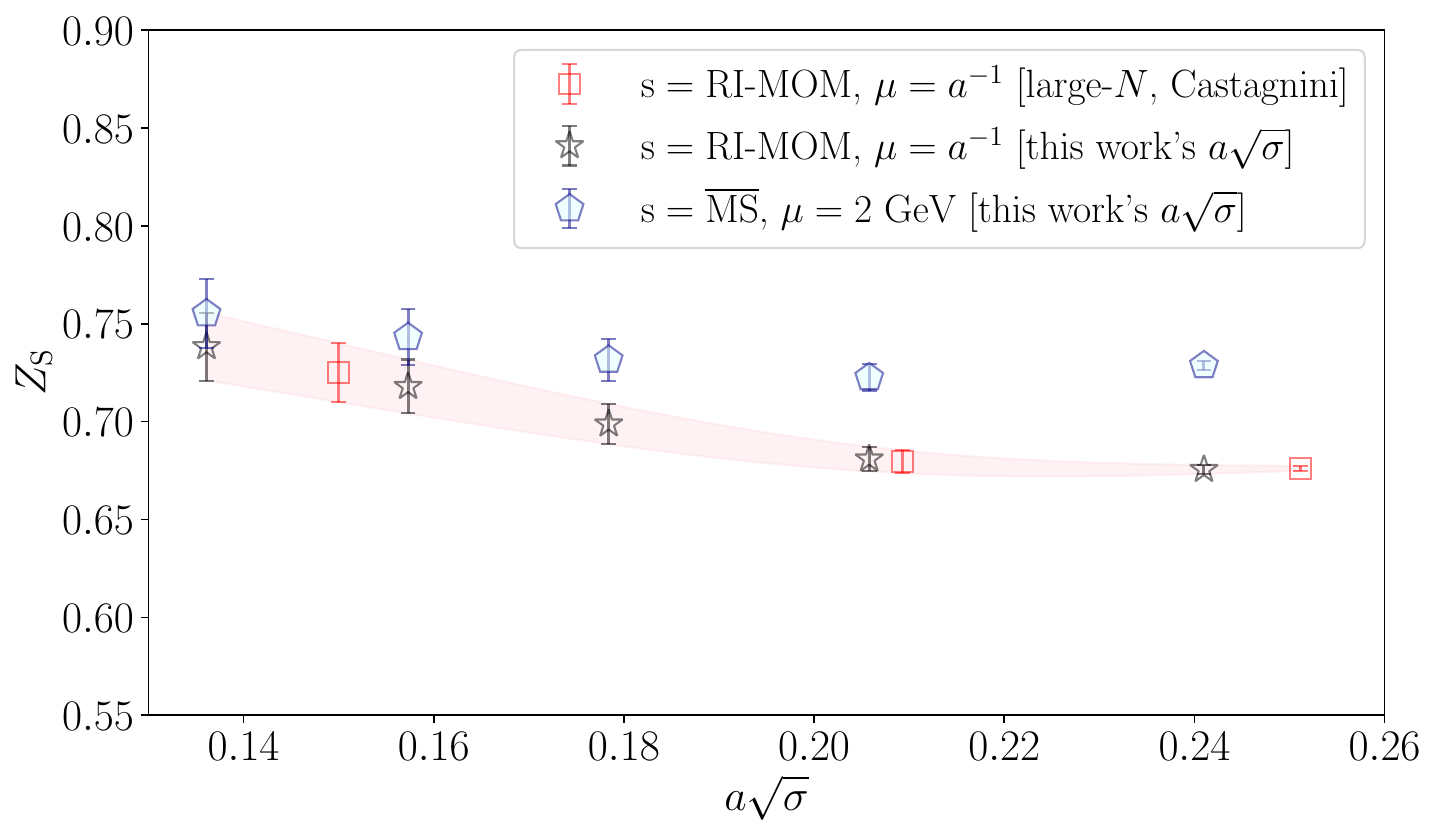}
\caption{Determination of $\ZS$ for the values of the lattice spacing employed in this study starting from those of Ref.~\cite{Castagnini:2015ejr}. Pentagon points are the final results employed in this study, reported in Tab.~\ref{tab:ZS_and_B_res}.}
\label{fig:ZS_computation_app}
\end{figure}

The procedure to determine the values of $\ZS$ in Tab.~\ref{tab:ZS_and_B_res} from Ref.~\cite{Castagnini:2015ejr} is summarized below, and illustrated in Fig.~\ref{fig:ZS_computation_app}.
\begin{itemize}
\item We have extrapolated Ref.~\cite{Castagnini:2015ejr} results for $\ZS^{\scriptscriptstyle{(\mathrm{RI-MOM})}}(a\mu=1)$ obtained for $N=4$--$7$ towards $N=\infty$ assuming $\mathcal{O}(1/N^2)$ leading corrections to the large-$N$ limit. Such extrapolations have been performed at fixed lattice spacing for the three finest values of $a\sqrt{\sigma}$ reported in~\cite{Castagnini:2015ejr}.
\item We have interpolated $N=\infty$ results for $\ZS^{\scriptscriptstyle{(\mathrm{RI-MOM})}}(a\mu=1)$ at our values of the lattice spacing for $b=0.355,0.360,0.365,0.370$. For $b=0.375$ we had instead to perform a slight extrapolation, as our finest lattice spacing was $\sim 13\%$ smaller than the finest one explored in Ref.~\cite{Castagnini:2015ejr}. However, due to the flat behavior exhibited by $\ZS^{\scriptscriptstyle{(\mathrm{RI-MOM})}}(a\mu=1)$, cf.~Fig.~\ref{fig:ZS_computation_app}, this extrapolation is pretty safe. Moreover, conservative error bars have been assigned in this case.
\item Finally, we have converted $\ZS^{\scriptscriptstyle{(\mathrm{RI-MOM})}}(a\mu=1)$ from the RI-MOM to the $\MSbar$ scheme, and we have run the scale from $\mu/\sqrt{\sigma} = 1/(a\sqrt{\sigma})$ to $\mu/\sqrt{\sigma}=3.75$ (corresponding to $\mu=2$ GeV). Details on the running of $\mu$ and on the conversion from RI-MOM to $\MSbar$ can be found in~\cite{Castagnini:2015ejr} and in the appendix of~\cite{Bonanno:2023ypf}.
\end{itemize}

In Ref.~\cite{Castagnini:2015ejr} we could also retrieve determinations of $\ZA$ and $\ZP/\ZS$ for $a\sqrt{\sigma}\simeq 0.2093$, which is very close to the value we found for $b=0.360$. Thus, for this coupling, we could obtain $N=\infty$ determinations of these renormalization constants too. These are not bound to agree with the ones obtained from the TEK model, as they are obtained with different methods, and are thus affected by different lattice artifacts.

\begin{figure}[!t]
\centering
\includegraphics[scale=0.5]{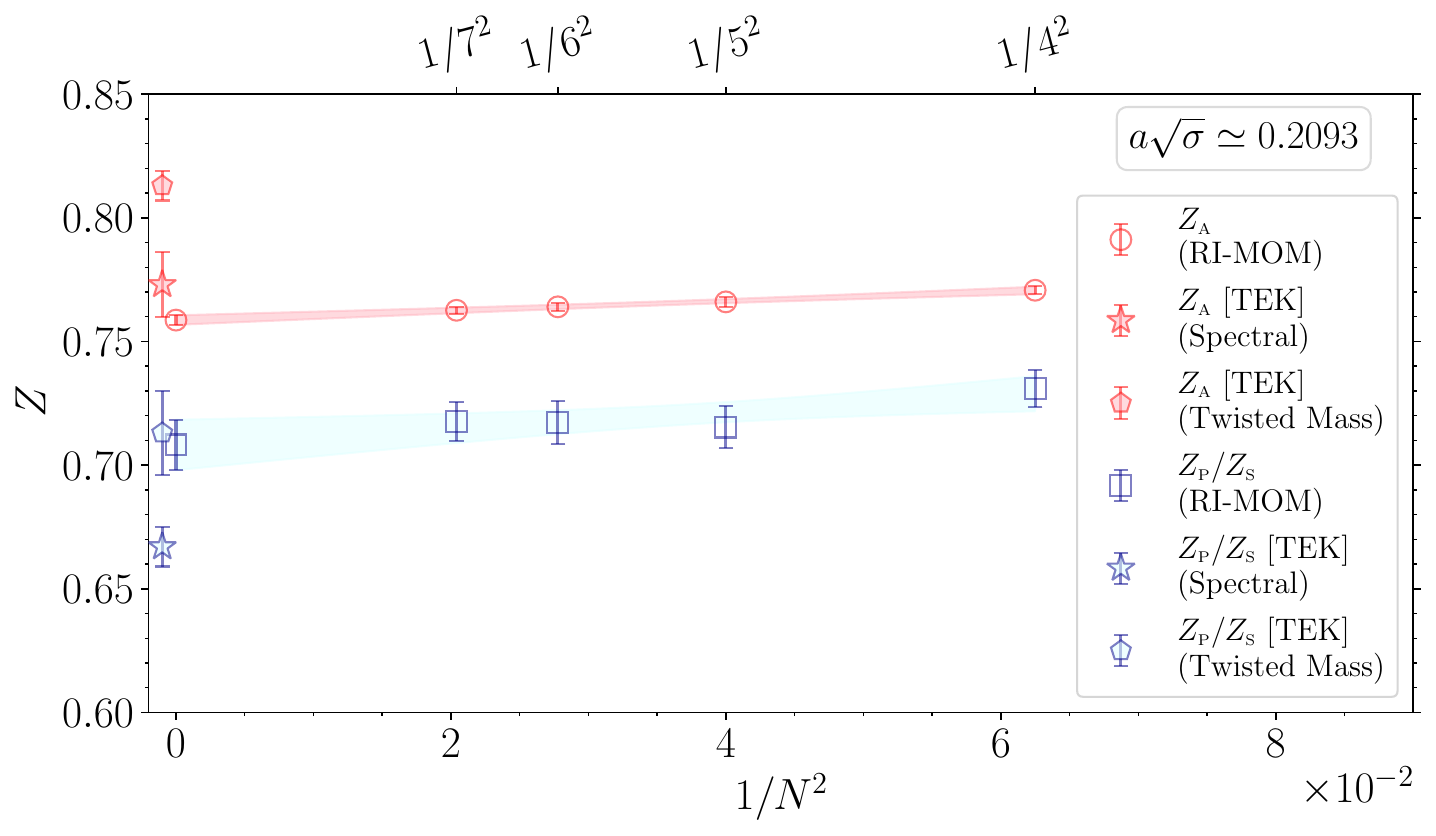}
\caption{Comparison of different large-$N$ determinations of $\ZP/\ZS$ and $\ZA$ for a lattice spacing $a\sqrt{\sigma}\simeq 0.2093$ obtained from the Rome--Southampton method (RI-MOM)~\cite{Castagnini:2015ejr}, the spectral method (this work) and via twisted mass Wilson fermions~\cite{Perez:2020vbn}.}
\label{fig:Z_COMP_app}
\end{figure}

The comparison is illustrated in Fig.~\ref{fig:Z_COMP_app}. The large-$N$ results of~\cite{Castagnini:2015ejr} are obtained from the so-called Rome--Southampton method (indicated as RI-MOM for shortness). In addition, we present two large-$N$ TEK determinations of $\ZP/\ZS$ and $\ZA$:
\begin{itemize}
\item This work: $\ZP/\ZS$ is obtained from the low-lying Dirac eigenvectors. Then, $\ZA$ is obtained from $(\ZP/\ZS) \times (1/Z)$, with $Z=\ZP/(\ZS\ZA)$ is obtained from the chiral scaling of the PCAC quark mass.
\item Ref.~\cite{Perez:2020vbn}: $\ZA$ is obtained from the ratio of $F_\pi/\ZA$ (obtained from Wilson fermions) and $F_\pi$ (obtained from twisted mass Wilson fermions). Then, $\ZP/\ZS$ is obtained as $\ZA \times Z$, where again $Z=\ZP/(\ZS\ZA)$ is obtained from the chiral scaling of the PCAC quark mass.
\end{itemize}
As earlier outlined, different methods to obtain renormalization constants correspond to different lattice discretizations, thus in general they will not give exactly the same results. However, the observed differences among different determinations are $\sim 6\%$ at most. These differences are reduced as the lattice spacing is decreased: for example, at the finer lattice spacing $a\sqrt{\sigma}\simeq 0.1573$, the value $\ZA=0.810(10)$ was found in Ref.~\cite{Perez:2020vbn} from twisted mass, while in this work we find $\ZA=0.831(10)$ from the spectral method at the same lattice spacing. They are compatible within 1.5 standard deviations.

\providecommand{\href}[2]{#2}\begingroup\raggedright\endgroup

\end{document}